\documentclass[rmp,aps,twocolumn,nofootinbib,floatfix,a4paper,hyperref]{revtex4}

\usepackage{amsmath,amssymb,amsfonts}
\usepackage{bm}\let\vec\bm
\usepackage{graphicx}\graphicspath{{fig/}}
\usepackage{mathrsfs}
\usepackage{longtable}
\usepackage{dcolumn}

\usepackage{color}
\definecolor{citations}{rgb}{0.196,0.196,0.392}
\definecolor{links}{rgb}{0.392,0,0}

\begin{document}

\title{Scanning tunneling spectroscopy of high-temperature superconductors}

\author{\O ystein Fischer}
\email{oystein.fischer@physics.unige.ch}
\author{Martin Kugler}
\author{Ivan Maggio-Aprile}
\author{Christophe Berthod}
\affiliation{MaNEP-DPMC,\footnote{National Center of Competence in Research on
``Materials with Novel Electronic Properties'' (MaNEP) and D\'epartement de
Physique de la Mati\`ere Condens\'ee (DPMC)} Universit\'e de Gen\`eve, 24 quai
Ernest-Ansermet, 1211 Gen\`eve 4, Switzerland}
\author{Christoph Renner}
\affiliation{London Center for Nanotechnology}
\affiliation{Department of Physics and Astronomy, University College London,
Gower Street, London WC1E 6BT, UK}

\begin{abstract}

Tunneling spectroscopy played a central role in the experimental verification of
the microscopic theory of superconductivity in the classical superconductors.
Initial attempts to apply the same approach to high-temperature superconductors
were hampered by various problems related to the complexity of these materials.
The use of scanning tunneling microscopy/spectroscopy (STM/STS) on these
compounds allowed to overcome the main difficulties. This success motivated a
rapidly growing scientific community to apply this technique to high-temperature
superconductors.  This paper reviews the experimental highlights obtained over
the last decade. We first recall the crucial efforts to gain control over the
technique and to obtain reproducible results. We then discuss how the STM/STS
technique has contributed to the study of some of the most unusual and
remarkable properties of high-temperature superconductors: the unusual large gap
values and the absence of scaling with the critical temperature; the pseudogap
and its relation to superconductivity; the unprecedented small size of the
vortex cores and its influence on vortex matter; the unexpected electronic
properties of the vortex cores; the combination of atomic resolution and
spectroscopy leading to the observation of periodic local density of states
modulations in the superconducting and pseudogap states, and in the vortex
cores.

\end{abstract}


\date{\today}

\vspace*{0.01em}
\maketitle

\tableofcontents

\section{Introduction}
\label{sect_introduction}

Superconductivity is one of the most remarkable phenomena observed in physics in
the 20th century. Discovered in 1911 by \citeauthor*{Kamerlingh-Onnes-1911},
nearly 50 years passed before this phenomenon was explained by a microscopic
quantum mechanical theory. The theory published in 1957 by
\citeauthor*{Bardeen-1957} (BCS) became a universal basis for describing
superconductors. Among the experimental techniques which allowed testing this
theory, quasiparticle tunneling introduced by \citet{Giaever-1960} played a
central role. Giaever showed that a planar junction composed of a
superconducting film and a normal metal separated by a nanometer thin insulator,
has striking current voltage characteristics. He showed that the
derivative---the tunneling conductance---has a functional dependence on voltage
which reflects the BCS quasiparticle density of states.
\citeauthor{Giaever-1960}'s finding was subsequently put on a firm theoretical
basis by \citet{Bardeen-1961}. This was the beginning of tunneling spectroscopy,
which was later used by \citet*{McMillan-1965} to establish a quantitative
confirmation of the BCS theory and its extension by \citet{Eliashberg-1960}.

In the 1980s, two major events took place. First, the invention of the scanning
tunneling microscope (STM) by \citet*{Binnig-1982a} opened a new world of
possibilities for tunneling spectroscopy. With the technique of scanning
tunneling spectroscopy (STS), it is possible to carry out spectroscopic studies
with a spatial resolution down to the atomic scale, something no other technique
can do. A beautiful demonstration of the possibilities to study superconductors
with this new instrument was realized when \citet{Hess-1989} observed the vortex
lattice in NbSe$_2$, and showed how the electronic structure of the vortex core
can be explored in detail. Second, the discovery of high-temperature
superconductivity in cuprates by \citet*{Bednorz-1986} initiated a burst of new
activity in the field of superconductivity. It was soon realized that the
superconductivity in these materials might be quite different from the one
observed in most low-temperature superconductors. As a result, numerous
investigations, using many different techniques with an unprecedented level of
sophistication and precision, have been carried out to study these
unconventional compounds. In spite of this wealth of results we still do not
have a consensus on the mechanism leading to superconductivity in the cuprates.

In this quest for an understanding of high-temperature superconductors (HTS), it
was obvious from the early days that tunneling spectroscopy could turn out to be
a key experimental technique. However, tunneling experiments using various
techniques struggled for a long time with what appeared to be irreproducible
results. In this context the availability of the STM became an important asset
for tunneling spectroscopy. The difficulty of obtaining reproducible data on the
cuprates was partly due to a bad control of the tunnel barrier and partly due to
material inhomogeneities. Using the STM it was possible to gain control over
these difficulties, to demonstrate reproducible spectra, and to identify the
essential intrinsic features of tunneling spectra on high-temperature
superconductors \cite{Renner-1994, Renner-1995}. Although for the study of HTS
there have been very important contributions from several experimental methods,
STM/STS has over the past 10 years made remarkable progress and greatly
contributed to the world-wide effort toward an understanding of the underlying
mechanism. With its unrivaled spatial and energy resolution, it is complementary
to other techniques like optical spectroscopy and ARPES which offer $k$-space
resolution. The STM/STS technique holds promise to shine new light on the key
questions in the future.

In this article we review the STM/STS investigations on high-temperature
superconductors which have been reported over the last decade. In
Sec.~\ref{sect_technique} we address experimental aspects and in
Sec.~\ref{sect_theory} we review the tunneling theory used for the
interpretation of various STS measurements. Surface characterization is
essential in this field, and these aspects are the focus of
Sec.~\ref{sect_characterization}. The following sections then discuss the main
results obtained by STS, many of which are significantly new as compared to
other methods. In Sec.~\ref{sect_gap_spectroscopy} we report the main results on
low-temperature spectroscopy, including the studies of impurities. We then
address in Sec.~\ref{sect_pseudogap} the temperature dependence of the spectra
and the pseudogap as seen by STS. Intrinsic spatial variations of the tunneling
spectra are discussed in Secs.~\ref{sect_vortices} and \ref{sect_modulations}.
In the former, we focus on vortex matter and vortex core spectroscopy, whereas
in the latter we turn our attention to the recently observed spatial periodic
modulations of the tunneling conductance. Each section contains an introductory
paragraph and a summary highlighting the main results. The appendices provide
additional elements of tunneling theory and a list of the superconducting gap
values observed by STS in high-temperature superconductors.

\section{Experimental aspects of STM and STS}
\label{sect_technique}

The invention of scanning tunneling microscopy (STM) by \citet{Binnig-1982b} set
a new milestone in the world of experimental physics. The STM measures the
tunneling current which flows between a sharp metallic tip and a conducting
sample separated by a thin insulating barrier, generally vacuum. The major
technological breakthrough occurred when the control of the spacing between the
tip and the sample, and of the lateral position of the tip reached picometric
precision using piezoelectric transducers. This ability added another dimension
to the well-established tunneling spectroscopy. Beside imaging the surface
topography with atomic-scale resolution, it allows to probe the \emph{local}
electron density of states (LDOS) with exceptional spatial resolution and well
controlled tunneling barriers. This technical accomplishment gave rise to the
development of a variety of scanning probe microscopes, which turned out to be
very powerful tools to investigate materials. Scanning Tunneling Spectroscopy
(STS) provides direct insight into fundamental properties of superconductors,
such as the superconducting gap, as remarkably illustrated by the real-space
imaging of the vortex lattice \cite{Hess-1989}.

In this section, we briefly recall the basic experimental principle of STM and
STS. We shall in particular describe the general configuration and operating
mode of the instrument, point out some specific technical challenges and give an
overview of the various acquisition methods used to investigate superconductors.
For a more detailed description of scanning tunneling microscopy, as well as
other scanning probe techniques, we refer the reader to \citet{Chen-1993,
Stroscio-1993, Guntherodt-1994}.

\subsection{The basic principle of STM}
\label{sect_STMprinciple}

\begin{figure}[tb]
\includegraphics[width=8.6cm]{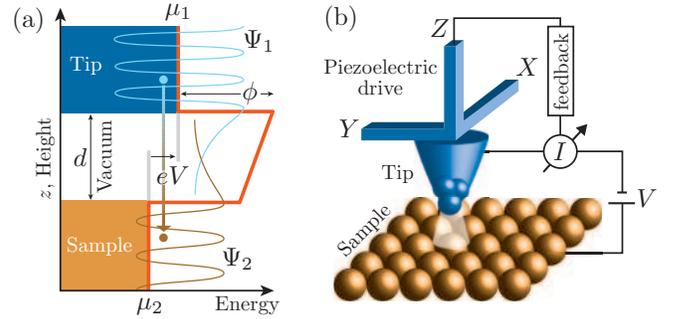}
\caption{\label{fig_STMprinciple}
(a) Tunneling process between the tip and the sample across a vacuum barrier of
width $d$ and height $\phi$ (for simplicity, the tip and the sample are assumed
to have the same work function $\phi$). The electron wave functions $\Psi$ decay
exponentially into vacuum with a small overlap, allowing the electrons to tunnel
from one electrode to the other. With a positive bias voltage $V$ applied to the
sample the electrons tunnel preferentially from the tip into unoccupied sample
states. (b) Schematic view of the scanning tunneling microscope.
}
\end{figure}

The phenomenon behind scanning tunneling microscopy is the quantum tunneling of
electrons between two electrodes separated by a thin potential barrier
(Fig.~\ref{fig_STMprinciple}a). This phenomenon was known since the early days
of quantum mechanics, and its application to study the superconducting gap was
first demonstrated in superconductor/insulator/normal metal (SIN) planar
junctions \cite{Giaever-1960} and point-contact junctions
\cite{Levinstein-1966}. No spatially resolved tunneling was possible in these
rigid electrode configurations. In 1981, \citeauthor{Binnig-1987} developed the
scanning tunneling microscope. Their successful idea was to mount a sharp
metallic tip, which acts as a local probe, on a three dimensional piezoelectric
drive (Fig.~\ref{fig_STMprinciple}b). The tip is scanned in the $xy$-plane above
the sample using the $X$ and $Y$ actuators, while its height is controlled using
the $Z$ actuator. Applying a bias voltage between the metallic tip and the
conducting sample, and approaching the tip within a few \AA ngstr\"oms of the
sample surface, results in a measurable tunneling current. An electronic
feedback loop is used to maintain this current constant by permanently adjusting
the tip height.

The most striking feature of this instrument is the remarkable spatial
resolution it can achieve. The key for reaching a vertical resolution of a few
hundredths of an \AA ngstr\"om is the exponential dependence of the tunneling
current, $I$, on the tip-to-sample spacing, $d$:
	\begin{equation}\label{eq_ExpDecay}
		I \sim e^{-2\kappa d},\quad
		\kappa=\sqrt{\frac{2m\phi}{\hbar^2}}\approx0.5\sqrt{\phi~\text{[eV]}}
		\text{~\AA}^{-1}.
	\end{equation}
For a typical metal ($\phi\sim 5$~eV) the current will decrease by about one
order of magnitude for every \AA ngstr\"om increase in the electrode spacing.
The lateral resolution mainly depends on the apex geometry and electronic
orbitals of the scanning tip, which confine the tunneling electrons into a
narrow channel, offering the unique opportunity to perform real-space imaging
down to atomic length scales. Sec.~\ref{sect_theory} provides a more detailed
discussion of the theoretical aspects of electron tunneling.

\def\fnlt#1#2{\multicolumn{6}{l}{$~^#1$\footnotesize #2}\\ [-0.1em]}
\begin{table*}[bt]
\caption{\label{tab_LTsystems}
Selection of home-built low-temperature STM listing the main experimental
conditions (temperature range, vacuum condition, magnetic field) and specific
design features. This is not an exhaustive table, but rather a guide for further
reading. The references are classified chronologically for each refrigeration
type. Abbreviations are RT: room temperature, ExGas: exchange gas, CryVac:
cryogenic vacuum, UHV: ultra-high vacuum, L$^3$He: liquid $^3$He, and ULT:
ultra-low temperature. \\ [-2em]
}
\begin{tabular*}{\textwidth}[t]{l@{\extracolsep{\fill}}c@{\extracolsep{\fill}}
								c@{\extracolsep{\fill}}c@{\extracolsep{\fill}}
								l@{\extracolsep{\fill}}l}
\hline\hline \\ [-1em]
Description &\multicolumn{3}{c}{Experimental conditions}& Specificities & Reference\\
\hline \\ [-0.5em]
$^4$He STM$^{ab}$ & 1.6/4.2~K  & ExGas  & 8/10~T  & cleavage and tip/sample exchange in UHV&\citet{Kent-1992}\\
$^4$He STM$^a$    & 35~K--RT   & UHV    & --      & \textit{in situ} cleaving&\citet{Ikeda-1993}\\
$^4$He STM$^c$    & 11--400~K  & UHV    & --      & \textit{in situ} tip/sample conditioning, $^4$He-flow cryostat&\citet{Horch-1994}\\
$^4$He STM$^{ac}$ & 7~K--?     & UHV    & 8~T     & \textit{in situ} tip/sample exchange, magnet lifted by bellow&\citet{Schulz-1994a}\\
$^4$He STM$^a$    & 1.5~K--RT  & ExGas  & 8~T     & \textit{in situ} tip/sample exchange&\citet{Tessmer-1994}\\
$^4$He STM$^a$    & 4.2~K      & ExGas  & --      & cold cleaving, magnetic coarse approach&\citet{Hancotte-1995}\\
$^4$He STM$^c$    & 15~K--RT   & UHV    & --      & \textit{in situ} sample exchange&\citet{Meyer-1996}\\
$^4$He STM$^d$    & 8~K--RT    & UHV    & 7~T     & split-coil, STM body pressed against conus&\citet{Wittneven-1997}\\
$^4$He STM$^a$    & 11~K--RT   & UHV    & --      & \textit{in situ} tip/sample condition., top-loading with bellow&\citet{Foley-2000}\\
$^4$He STM$^d$    & 16~K       & UHV    & 2.5~T   & split-coil, \textit{in situ} tip/sample exchange&\citet{Pietzsch-2000}\\
$^4$He STM        & 4--300~K   & UHV    & 8~T     & rotatable field, STM cooled by superfluid $^4$He&\citet{Kondo-2001}\\
$^4$He STM$^d$    & 6.5~K--RT  & UHV    & --      & \textit{in situ} tip/sample conditioning, $^4$He-flow cryostat&\citet{Zhang-2001}\\
\hline \\ [-0.5em]
$^3$He STM$^d$    & 240~mK--RT & CryVac & 7~T     & cold cleaving, \textit{in situ} sample exchange&\citet{Pan-1999}\\
$^3$He STM$^{ab}$ & 275~mK--RT & UHV    & 12/14~T & \textit{in situ} tip/sample exchange, bottom-loading cryostat&\citet{Kugler-2000a}\\
$^3$He STM$^{ac}$ & 260~mK--RT & L$^3$He& --      & top-loading, thermally compensated STM&\citet{Urazhdin-2000}\\
$^3$He STM$^a$    & 500~mK--?  & UHV    & 7~T     & \textit{in situ} facilities, Joule-Thomson refrigerator&\citet{Heinrich-2003}\\
$^3$He STM$^{ad}$ & 315~mK--RT & UHV    & 12/14~T & \textit{in situ} tip/sample conditioning, spin-polarized STM&\citet{Wiebe-2004}\\
\hline \\ [-0.5em]
ULT STM           & 90~mK--RT  & CryVac & --      & \textit{ex situ} sample preparation&\citet{Fukuyama-1996}\\
ULT STM           & 20~mK      & UHV    & 6~T     & \textit{in situ} tip/sample conditioning, bottom-load. cryostat&\citet{Matsui-2000}\\
ULT STM$^b$       & 60~mK      & CryVac & --      & very compact design, rapid cool down&\citet{Moussy-2001}\\
ULT STM$^b$       & 70~mK      & CryVac & 12~T    & home-built STM integrated into Oxford Kelvinox 100&\citet{Upward-2001}\\ [0.3em]
\hline\hline \\ [-0.6em]
\fnlt{a}{System developed for the study of superconductors}
\fnlt{b}{Based on coaxial inertial slider or `slip-stick' design \cite{Lyding-1988, Renner-1990b}}
\fnlt{c}{Based on Besocke or `beetle' design \cite{Besocke-1987}}
\fnlt{d}{Based on `friction motor' design \cite{Pan-1999}}
\end{tabular*}
\end{table*}

The tunneling regime is defined by a set of three interdependent parameters: the
electrode spacing $d$ (typically 5--10~\AA), the tunneling current $I$
(typically 0.01--10~nA), and the bias voltage $V$ (typically 0.01--2~V). The
parameters $I$ and $V$ are generally chosen to set the tunneling resistance
$R_t=V/I$ in the G$\Omega$ range. It is worth mentioning that the absolute
electrode spacing $d$ is not readily accessible by experiment: only relative
variations can be measured. Another important point is that the measured
tunneling current is a convolution of the electron densities of states of both
the tip and the sample (Sec.~\ref{sect_theory}). To study intrinsic sample
properties, it is therefore preferable to use tips with a featureless density of
states and a well-defined Fermi surface (ideally spherical). The metals most
commonly used for the tip are Au, W, Ir, and PtIr.\footnote{A selection of tip
preparation techniques is discussed by \citet{Ekvall-1999}.}

\subsection{Technical challenges of low-temperature STM}
\label{sect_TechIssues}

A successful investigation of superconductors by STM relies essentially on three
prerequisites: (i) a suitable sample surface; (ii) a scanning tunneling
microscope allowing atomic resolution and stable spectroscopy; and (iii) an
experimental environment featuring vacuum conditions, low and variable
temperatures, as well as magnetic fields. This is a highly delicate endeavor,
which requires the mastering of many technical challenges. In the following we
briefly outline the major technical difficulties, and review some of the STM
designs used to study superconductors at low temperatures (see
Table.~\ref{tab_LTsystems}).

Controlling the sample surface quality is essential. Contamination in ambient
atmosphere may rapidly degrade the sample top layer, often preventing stable and
reproducible tunnel junctions and the investigation of intrinsic properties.
This issue is non trivial and different for each compound, depending on its
crystallographic structure and surface nature. The most suitable surfaces for
STM/STS are those prepared \textit{in situ}. Ideally, the top layers are
mechanically removed by cleaving the sample in ultra-high vacuum (UHV), either
at room temperature before cooling or at low temperature. This procedure works
best for the more anisotropic layered compounds with a natural cleaving plane,
such as Bi2212 \cite{Renner-1994, DeWilde-1998, Pan-2001}, Bi2201
\cite{Kugler-2001, Shan-2003}, Bi2223 \cite{Kugler-2006}, and Hg-compounds
\cite{Rossel-1994, Wei-1998}.

Successful STM studies have also been reported for HTS which do not offer the
advantage of a natural cleaving plane, such as Y123, Nd123 and the
electron-doped superconductor NCCO \cite{Alff-1998}. In these cases, besides
cleaving in UHV \cite{Edwards-1992, Pan-1999, Nishiyama-2002}, a number of
successful experiments were carried out on as grown surfaces
\cite{Maggio-Aprile-1995} and surfaces prepared \textit{ex-situ} by chemical
etching \cite{Wei-1998, Shibata-2003b} or by cutting using a razor blade
\cite{Hayashi-1998b}.

Most low-temperature STMs described in this section are home-built and generally
based on one of the following designs: the Besocke or `beetle' setup
\cite{Besocke-1987}, the coaxial inertial slider or `slip-stick' design
\cite{Lyding-1988, Renner-1990b} and the `friction motor' \cite{Pan-1999}. A
more complete description of the devices would go beyond the scope of this
review: the interested reader may consult, for example, \citet{Guntherodt-1994}.

The modern low-temperature STMs used for spectroscopic studies of oxide
superconductors are generally mounted in a UHV chamber for the reasons mentioned
above. The need for magnetic fields and ample cryogenic hold-time considerably
increases the system complexity. The three main design challenges are (i)
ensuring optimal vibration isolation to achieve high resolution and stable
tunneling conditions, (ii) proper thermalization of the instrument to achieve
low and stable temperatures, and (iii) confining the instrument inside the small
bore of a superconducting coil. Table~\ref{tab_LTsystems} provides a list of
references for guiding the reader through the various existing low-temperature
STM configurations developed to deal with these constraints.

The early low-temperature STMs were mostly based on conventional $^4$He
cryostats working at $T\geqslant4.2$~K (in a few cases down to 1.5~K via a
pumped $^4$He pot), generally cooled by exchange gas. Combining low temperatures
and UHV leads to more complex systems, with the difficult task to properly
thermalize the STM without coupling to mechanical vibrations. Optimal
thermalization becomes even more important when sub-Kelvin temperatures and a
variable temperature range are required. Moreover, for maximal flexibility and
short turn-around times, an easy \textit{in-situ} access to the STM for tip and
sample exchange, as well as \textit{in-situ} tip and sample conditioning, are
necessary. Magnetic fields impose further considerable size and materials
constraints: the STM has to be non-magnetic and has to fit into the coil.

In order to study ultra-low temperature phenomena with enhanced energy
resolution, a few groups have developed sub-Kelvin systems, inserting the STM
into $^3$He or dilution refrigerators. First STS studies in small fields below
1~K were achieved by \citeauthor{Hess-1990}, with the historical imaging of
individual Abrikosov vortex cores in NbSe$_2$, first at 300~mK using a $^3$He
refrigerator \cite{Hess-1990} and later at 50~mK in a dilution fridge
\cite{Hess-1991}. Other groups followed and added higher magnetic fields and/or
UHV conditions with \textit{in situ} facilities. Compared to $^3$He
refrigeration, dilution refrigeration offers the advantage of reaching
temperatures of the order of 50~mK. However, the thermalization of the sample
and STM below 100~mK is difficult, making the operation in a dilution
refrigerator much more complex.

\subsection{Operating modes}
\label{sect_STMmodes}

A scanning tunneling microscope allows to collect topographic \emph{and}
spectroscopic data on a local scale. In the topographic mode, the surface is
mapped via the dependence of the tunneling current upon the tip-to-sample
distance. In spectroscopy, the LDOS of the material is extracted through
measurements of the tunneling conductance.

\subsubsection{Topography}
\label{sect_TopoIm}

\begin{figure}[tb]
\includegraphics[width=8.6cm]{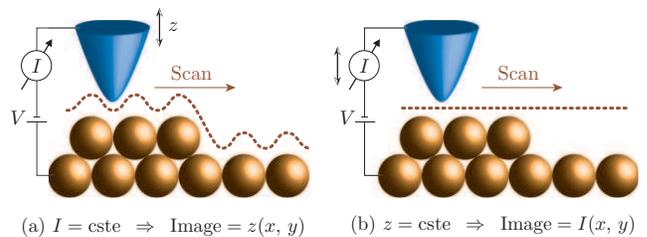}
\caption{\label{fig_STMmode}
Generic STM operating modes: (a) constant-current and (b) constant-height
imaging.
}
\end{figure}

\paragraph{Constant-current imaging}

(Fig.~\ref{fig_STMmode}a) In this standard mode, the tunneling current $I$ is
kept constant by continuously feedback-adjusting the tip vertical position
during the scan. Since the tunneling current integrates over all states above or
below $E_{\text{F}}$, up to an energy equal to the tunnel voltage
(Sec.~\ref{sect_theory}), the constant-current mapping corresponds to a profile
of constant integrated electron density of states (DOS). If the LDOS is
homogeneous over the mapped area, this profile corresponds to constant
tip-to-sample spacing, and recording the height of the tip as a function of
position gives a three dimensional image of the surface $z=z(x,\,y)$. Because
the tip follows the corrugations of the surface at a constant spacing, the scan
speed is limited by the feedback loop bandwidth, which is typically in the
kilohertz range.

\paragraph{Constant-height imaging}

(Fig.~\ref{fig_STMmode}b) In this mode the tip is scanned over the sample
surface while maintaining the tip at a constant absolute height (feedback loop
turned off). For ideal tip and sample, modulations of the tunneling current
$I(x,\,y)$ are only due to variations in the tip-to-sample spacing, and
recording the current as a function of position will reflect the surface
topography. This mode allows fast scanning, but is restricted to surface areas
where the corrugations do not exceed a few \AA ngstr\"oms, to avoid tip
collisions with large surface protrusions. According to Eq.~(\ref{eq_ExpDecay}),
the image corrugation depends on the local work function $\phi$ as
$d(x,\,y)\sim\ln I(x,\,y)/\sqrt{\phi}$. Thus, unless the actual local value of
$\phi$ is known, quantitative characterizations of topographic features are
difficult to achieve.

\subsubsection{Local tunneling spectroscopy}
\label{sect_STS}

Locally resolved electron spectroscopy is probably the most sophisticated
application of the STM. The electronic density of states can be accessed by
recording the tunneling current $I(V)$ while the bias voltage is swept with the
tip held at a fixed vertical position. If a positive bias voltage $V$ is applied
to the sample, electrons will tunnel into \emph{unoccupied} sample states,
whereas at negative bias they will tunnel out of \emph{occupied} sample states
(Fig.~\ref{fig_STMprinciple}a). Although the interpretation of spectra can be
quite complex, it can be shown that in ideal conditions the tunneling
conductance $dI/dV(V)$ provides a valid measurement of the sample LDOS [see
Eqs.~(\ref{eq_tunneling_DOS}) and (\ref{eq_conductance_Tersoff})]. This
straightforward way to interpret the experiments is used in most STM/STS
studies.

$dI/dV$ spectra can either be obtained by numerical differentiation of $I(V)$
curves or by a lock-in amplifier technique. In the latter case, a small
ac-voltage modulation $V_{\text{ac}}\cos(\omega t)$ is superimposed to the
sample bias $V$, and the corresponding modulation in the tunneling current is
measured. Expanding the tunneling current into a Taylor series,
	\[
		I=I(V)+\left(\frac{dI}{dV}\right)V_{\text{ac}}\cos(\omega t)
		+\mathcal{O}(V_{\text{ac}}^2),
	\]
one finds that the component at frequency $\omega$ is proportional to
$dI/dV(V)$. This statement is only valid if $V_{\text{ac}}\ll V$ and if $I(V)$
is sufficiently smooth. For optimal energy resolution, $V_{\text{ac}}$ should
not exceed $k_{\text{B}}T$, and typical values are in the few hundred
$\mu\text{V}$ range. The advantage offered by the lock-in technique is that the
sampling frequency $\omega$ can be selected outside the typical frequency
domains of mechanical vibrations or electronic noise, considerably enhancing the
measurement sensitivity.

\subsubsection{Spectroscopic imaging}
\label{sect_STSIm}

Most STS experiments use `current-imaging tunneling spectroscopy' (CITS),
introduced by \citet{Hamers-1986}. A CITS image is based on a regular matrix of
points distributed over the surface. The tip is scanned over the sample surface
with a fixed tunneling resistance $R_t=V_t/I$, recording the topographic
information. At each point of the CITS array, the scan and the feedback are
interrupted to freeze the tip position ($x$, $y$, and $z$). This allows the
voltage to be swept to measure $I(V)$ and/or $dI/dV$, either at a single bias
value or over an extended voltage range. The bias voltage is then set back to
$V_t$, the feedback is turned on and the scanning resumed. The result is a
topographic image measured at $V_t$, and simultaneous spectroscopic images
reconstructed from the $I(V)$ and/or $dI/dV$ data. Because the feedback loop is
interrupted, $V$ can take any value, even those where $I(V)=0$. This technique
provides a very rich set of information.

\begin{figure}[tb]
\includegraphics[width=8.6cm]{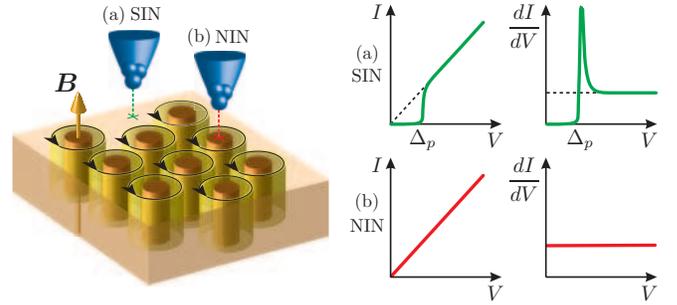}
\caption{\label{fig_Vimaging}
Illustration of the vortex-lattice imaging by STM: (a) Local SIN junction with
typical BCS $s$-wave characteristics when the tip is sitting between vortices.
(b) Local NIN junction with a constant conductance (for a dirty BCS
superconductor) when the tip is positioned over a vortex core.
}
\end{figure}

For best contrast in the spectroscopic images it is important to select the
energy where maximal variations in the tunneling conductance occur. In the case
of vortex imaging (Fig.~\ref{fig_Vimaging}) the mapping energy is usually
selected at the position of the coherence peaks (superconducting gap), or close
to zero energy where the amplitude of the localized core states is largest
(Sec.~\ref{sect_vortices}). This technique also enables to reconstruct
spectroscopic maps acquired simultaneously at different energies. An example of
such analysis was provided by \citet{Hess-1991} who measured the energy
dependence of the star-shaped vortex core structure in NbSe$_2$, and by
\citet{Pan-2000b} who used Fourier transforms of such maps on Bi2212 to reveal
periodic modulations of the LDOS in real space (Sec.~\ref{sect_modulations}).
From the energy dependence of these modulations, energy dispersion curves could
be extracted, opening to STS the door of reciprocal-space spectroscopies.

\section{The theory of electron tunneling}
\label{sect_theory}

The theory of tunneling began in the first years of quantum mechanics, and now
covers a large variety of experimental situations.\footnote{See for example
\citet{Duke-1969}, \citet{Wolf-1985}, and the series edited by
\citet{Wiesendanger-1993}.} The interpretation of most tunneling measurements on
high-$T_c$ superconductors is based on the tunneling-Hamiltonian formalism,
which we shortly describe in Sec.~\ref{sect_tunneling_Hamiltonian} and present
in more detail in Appendix~\ref{app_tunneling_theory_r}. In
Sec.~\ref{sect_tunneling_other_approaches} we mention some of the more
sophisticated approaches to electron tunneling which have been developed over
the years. Sec.~\ref{sect_matrix_element} is primarily dedicated to the often
overlooked question of the tunneling matrix element, with a particular emphasis
on the case of local vacuum tunneling (STM junction). Finally, in
Sec.~\ref{sect_tunneling_HTS} we formulate some issues in the theoretical
interpretation of the tunneling experiments on HTS.

\subsection{The tunneling Hamiltonian formalism}
\label{sect_tunneling_Hamiltonian}

Like any transport process, the tunneling of electrons across a vacuum or
insulating barrier is a non-equilibrium phenomenon, and the injected
quasiparticles are not \textit{a priori\/} in thermal equilibrium with the
lattice. In the STM experiments, however, the amplitude of the current is so low
that the time between two tunneling events is much longer than the typical
quasiparticle relaxation time. In this situation it is appropriate to use an
equilibrium theory; it is also reasonable to assume that the correlations play
no role in the barrier and that the elementary process involved is the tunneling
of a {\em single\/} electron. When the two metals are superconductors, the
coherent tunneling of a Cooper pair can also take place, leading to the
celebrated Josephson effect \cite{Josephson-1962}.

\begin{figure}[tb]
\includegraphics[width=7cm]{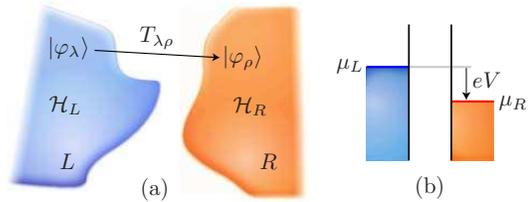}
\caption{\label{fig_junction}
Schematics of a tunnel junction. (a) Geometrical view: $\mathcal{H}_L$ and
$\mathcal{H}_R$ are the Hamiltonians of the isolated left and right materials,
respectively, and $T_{\lambda\rho}$ is the probability of tunneling from a state
$|\varphi_{\lambda}\rangle$ on the left side to a state
$|\varphi_{\rho}\rangle$ on the right side. (b) Energy diagram: a potential
difference $V$ is applied to the junction, resulting in a relative shift of the
chemical potentials $\mu_L$ and $\mu_R$ in both materials.
}
\end{figure}

The tunneling-Hamiltonian formalism \cite{Bardeen-1961, Cohen-1962,
Bardeen-1962} provides a framework to understand both the single-particle and
the pair tunneling phenomena. The two materials forming the tunnel junction are
considered as two independent systems (Fig.~\ref{fig_junction}). The transfer of
particles across the barrier is described by the phenomenological `tunneling
Hamiltonian'
	\begin{equation}\label{eq_transfer_Hamiltonian}
			\mathcal{H}_T = \sum_{\lambda\rho} T_{\lambda\rho}
			c^{\dagger}_{\rho}c^{\phantom{\dagger}}_{\lambda}+\text{h.c.}
	\end{equation}
We use the index $\lambda$ to label the single-particle states on the left side
of the junction, and the index $\rho$ for the states on the right side. The
operator $c_{\lambda}$ destroys a particle in the state $\varphi_{\lambda}$ and
the operator $c^{\dagger}_{\rho}$ creates a particle in the state
$\varphi_{\rho}$. The quantity $T_{\lambda\rho}$ is known as the `tunneling
matrix element'. It mainly depends upon the geometry of the tunnel junction, but
also on the details of the electronic states on both sides. The proper
definition and the explicit evaluation of $T_{\lambda\rho}$ are difficult
problems which have stimulated a lot of work, especially in the case of the STM
junction. We shall discuss the tunneling matrix element further in
Sec.~\ref{sect_matrix_element}.

If the bias $V$ applied across the junction is small, the current can be
calculated using linear-response theory (see
Appendix~\ref{app_tunneling_Hamiltonian}). The total current turns out to be the
sum of two contributions, $I_s$ and $I_J$. The first part is the single-particle
current and is given by
	\begin{multline}\label{eq_tunneling_current}
	I_s=\frac{2\pi e}{\hbar}\int d\omega\, \left[f(\omega-eV)
		-f(\omega)\right]\times\\
		\sum_{\lambda\rho}|T_{\lambda\rho}|^2
		A_{\lambda}(\omega-eV)A_{\rho}(\omega),
	\end{multline}
where $A_{\lambda}(\omega)$ and $A_{\rho}(\omega)$ are the single-particle
spectral functions of the (isolated) left and right materials, respectively, and
$f(\omega)$ is the Fermi function. Eq.~(\ref{eq_tunneling_current}) applies to
any type of tunnel junction. The most common cases, planar and STM junctions,
will be discussed shortly. The meaning of Eq.~(\ref{eq_tunneling_current}) is
obvious: the Fermi factors select the energy window where occupied states in one
material, say on the left, can be aligned with empty states on the right; within
that window, an electron in a state $\lambda$ can tunnel into an empty state
$\rho$ if the two states are connected by a non vanishing matrix element, and if
they are aligned in energy by the applied bias. Eq.~(\ref{eq_tunneling_current})
can also be rationalized within the simple `semiconductor model'
\cite{Nicol-1960, Tinkham-1996}. We shall not consider in this review the second
contribution to the current, $I_J$, which describes the coherent tunneling of
electron pairs.\footnote{The observation of the Josephson current by STM using a
superconducting tip is challenging because it requires a relatively low junction
resistance. This type of experiment has not been reported so far for the HTS,
but was recently achieved with conventional superconductors \cite{Naaman-2001,
Naaman-2003, Naaman-2004, Suderow-2002, Martinez-Samper-2003, Rodrigo-2004}.}

The main strength of Eq.~(\ref{eq_tunneling_current}) is its generality: it
applies to any tunnel junction, provided that the tunneling takes place at
energies lower than the barrier height, and that high-order effects in
$\mathcal{H}_T$ (such as interferences occurring within the barrier) are not
important. Surface and geometric effects are in principle embodied in the
spectral functions and in the matrix element. In practice one often assumes that
the surface disruption does not change significantly the electronic properties,
and that the spectral functions entering Eq.~(\ref{eq_tunneling_current}) can be
identified with the \emph{bulk} ones.

The explicit appearance of the spectral functions in
Eq.~(\ref{eq_tunneling_current}) is another appealing feature of this formalism,
which opens the door for tunneling spectroscopy. Consider a junction between a
simple metal with a structureless DOS on the left---the probe---and a given
material on the right---the sample. In general, the tunneling matrix element
depends upon the electron quantum numbers (see Sec.~\ref{sect_matrix_element}),
but for the sake of the argument we assume here that it is a constant:
$T_{\lambda\rho}\equiv T$. Then at zero temperature
Eq.~(\ref{eq_tunneling_current}) leads to the following expression for the
tunneling conductance:
	\begin{equation}\label{eq_tunneling_DOS}
		\sigma(V)=\frac{dI_s}{dV}=\frac{2\pi e^2}{\hbar}|T|^2N_L(0)N_R(eV)
	\end{equation}
where $L$ and $R$ denote `left' and `right' and $N(\omega)$ is the DOS measured
from the Fermi energy. This simple formula shows the essence of tunneling
spectroscopy: the bias dependence of the tunneling conductance directly probes
the DOS of the sample. This result is not valid in general, however, as
discussed at length in Sec.~\ref{sect_matrix_element}. The correspondence
between the tunneling conductance and the sample DOS, first postulated by
\citet{Giaever-1960}, was the initial motivation for the \citet{Bardeen-1961}
tunneling theory.\footnote{ For SIS junctions, \textit{i.e.} junctions involving
two superconductors separated by a thin insulating barrier,
Eq.~(\ref{eq_tunneling_current}) and the assumption $T_{\lambda\rho}\equiv T$
yield a partial convolution of the materials' DOSs instead of the DOS itself,
that is $I_s\propto\int_0^{eV}d\omega\,N_L(\omega-eV)N_R(\omega)$ at zero
temperature.}

For understanding \emph{local} probes like the STM, it is helpful to express
Eq.~(\ref{eq_tunneling_current}) in terms of the real-space spectral function,
which is closely related to the LDOS (Appendix~\ref{app_tunneling_Hamiltonian}).
\citet{Tersoff-1983} performed the first calculation of the tunneling matrix
element for the STM junction, and found an explicit relation between the
tunneling current and the sample LDOS $N_{\text{sample}}(\vec{x},\,\omega)$:
	\begin{multline}\label{eq_current_Tersoff}
		I_s\propto\int d\omega\,\left[f(\omega-eV)-f(\omega)\right]\times\\
		N_{\text{tip}}(\omega-eV)N_{\text{sample}}(\vec{x},\,\omega)
	\end{multline}
where $\vec{x}$ denotes the tip center of curvature
(Appendix~\ref{app_Tersoff}). Assuming a structureless tip DOS,
$N_{\text{tip}}(\omega-eV)=\text{const.}$, Eq.~(\ref{eq_current_Tersoff}) gives
	\begin{equation}\label{eq_conductance_Tersoff}
		\sigma(\vec{x},\,V)\propto \int d\omega\,[-f'(\omega-eV)]
		N_{\text{sample}}(\vec{x},\,\omega),
	\end{equation}
where $f'$ is the derivative of the Fermi function. Thus the interpretation of
scanning tunneling spectroscopy becomes remarkably simple: the voltage
dependence of the tunneling conductance measures the thermally smeared LDOS of
the sample at the position of the tip. Because the local tunneling matrix
element is properly taken into account, Eq.~(\ref{eq_conductance_Tersoff})
describes the STM much better than Eq.~(\ref{eq_tunneling_DOS}) describes planar
junctions.

\subsection{Other approaches to the tunneling problem}
\label{sect_tunneling_other_approaches}

The success of the tunneling-Hamiltonian formalism in describing many
experimental results, in particular the Josephson effect, led several authors to
investigate this approach in depth. One of the first questions raised concerned
the primary assumption of the theory, namely the possibility to represent the
junction as two independent systems coupled by a term like $\mathcal{H}_T$. Not
surprisingly it was found that this assumption is not valid in general
\cite{Prange-1963, Zawadowski-1967}; however the associated error turned out to
be small. The effect of interactions in the electrodes and in the barrier was
also investigated \cite{Schrieffer-1964, Scalapino-1966, Appelbaum-1969a,
Appelbaum-1969b, Duke-1969}; this allowed, in particular, to explain the fine
structure of the conductance in BCS superconductors in terms of the phonon
spectrum \cite{McMillan-1965}. The most significant improvements on the original
theory are perhaps the incorporation of non-equilibrium effects by means of the
Keldysh formalism \cite{Caroli-1971a, Feuchtwang-1974a}, as well as the
non-perturbative treatments \cite{Noguera-1990, Sacks-1991}, which gave useful
information on the accuracy of the perturbative approach. The resulting theories
are rather complicated, and were not yet used in the context of the high-$T_c$
superconductors.

In order to avoid difficulties inherent to the tunneling-Hamiltonian formalism,
and for treating the regime where the interaction between the two materials
cannot be neglected, several authors proposed approaches based on the scattering
theory \cite{Lucas-1988, Doyen-1993, Cuevas-1996, Ness-1997, Carminati-2000}, in
the spirit of the \citet{Buttiker-1985} transport theory. For superconducting
junctions, the most popular method is due to \citet{Blonder-1982}. Initially
developed for \emph{planar} junctions involving a BCS $s$-wave \cite[or
$d$-wave,][]{Tanaka-1995} superconductor, this approach was nevertheless
occasionally used to interpret STM spectra on HTS materials.

Another approach was explored by \citet{Lang-1985}, who represented both tip and
sample by a single atom adsorbed on a jellium surface, and calculated the
tunneling current using a first-principles method. This study illustrated the
atomic character of the STM junction, which in effect can be viewed as an
atom-to-atom contact \cite[see also][]{Tsukada-1993}, and initiated many
subsequent works based on the \textit{ab initio} methods \cite[see,
\textit{e.g.}][]{Ciraci-1987, Tersoff-1989, Hofer-2001, Hofer-2003}.

\subsection{The tunneling matrix element}
\label{sect_matrix_element}

In the \citet{Bardeen-1961} theory the matrix element is the expectation value
of the single-particle current in the direction $z$ normal to the plane of the
junction, through a surface $S$ lying entirely in the barrier region:
	\begin{equation}\label{eq_T_Bardeen}
		T_{\lambda\rho}=-\frac{\hbar^2}{2m}\int_S dS
		\left(\varphi_{\rho}^*\,\frac{\partial\varphi_{\lambda}}{\partial z}
		-\varphi_{\lambda}\,\frac{\partial\varphi_{\rho}^*}{\partial z}\right).
	\end{equation}
\citet{Duke-1969} and \citet{Wolf-1985} give comprehensive reviews of the
various models for $T_{\lambda\rho}$ developed prior to the invention of the
STM. Hereafter we describe the most popular models for planar and STM junctions.

\subsubsection{The planar junction}

$T_{\lambda\rho}$ has a particularly simple form at ideal planar junctions, due
to conservation of momentum in the plane parallel to the interface
\cite{Harrison-1961, Duke-1969, Wolf-1985}. The exponential decay of the wave
functions in the barrier leads to an exponential dependence of
$|T_{\lambda\rho}|^2$ on the barrier thickness $d$, which reduces to
$\exp(-2\kappa d)$ in the limit of a wide and square barrier;
$\kappa=[\frac{2m}{\hbar^2}(U-\varepsilon_z)]^{\frac{1}{2}}$ is the electron
momentum in the barrier and $\varepsilon_z$ is the energy for the motion normal
to the interface, measured from the top $U$ of the barrier. This exponential
factor suggests that the tunneling current is dominated by states with a large
velocity along the $z$ direction, and consequently a small momentum in the
$(x,\,y)$ plane, leading to a momentum selectivity often referred to as the
`tunneling cone' \cite{Beuermann-1981}. The matrix element is also proportional
to the group velocity along $z$ \cite{Harrison-1961}. In
Eq.~(\ref{eq_tunneling_current}), however, this velocity factor
$\partial\xi_{\vec{k}}/\partial k_z$ is canceled by an equivalent DOS factor. A
counterintuitive consequence of this cancellation is the possibility to tunnel
along the $c$ axis into a quasi two-dimensional material (without dispersion
along $k_z$), in spite of the low conductivity in the $z$ direction. This latter
property would imply that $k_z$-related DOS features should not show up at all
in the planar tunneling spectra \cite{Harrison-1961}, in sharp contrast with the
simple result of Eq.~(\ref{eq_tunneling_DOS}), and with the experimental
observation of the superconducting DOS by \citeauthor{Giaever-1960}.
\citet{Bardeen-1961} attributed this discrepancy to the fact that the electrons
are not paired in the barrier region; thus the velocity entering
$T_{\lambda\rho}$ would be the velocity of the bare electrons rather than the
quasiparticle velocity $\partial E_{\vec{k}}/\partial k_z$, and the
superconducting gap would show up in the tunneling spectra, although the
normal-state DOS would not.

It should be noted that these arguments apply to materials with spherical Fermi
surfaces and isotropic superconducting gaps, and must be reconsidered when
dealing with the cuprate superconductors. As pointed out by \citet{Wei-1998},
the tunneling cone ``flattens out'' for $c$-axis tunneling into the $ab$ plane
of the HTS, because $\varepsilon_z$ is almost independent of the in-plane
momentum. Therefore one expects to measure the $ab$-plane band structure in the
planar $c$-axis tunneling spectra of the HTS.

\subsubsection{The STM junction}

Because of the lateral confinement, the one-electron states of the STM tip
cannot be characterized by a single momentum in the $(x,\,y)$ plane, which makes
the calculation of the tunneling matrix element more difficult than in the
planar junction case. Soon after the invention of the STM, \citet{Tersoff-1983,
Tersoff-1985} proposed an expression for $T_{\lambda\rho}$, which is still
widely used to interpret STM images. They represented the tip apex by a
spherical potential well and found that Bardeen's matrix element
Eq.~(\ref{eq_T_Bardeen}) is proportional to the sample wave function at the
center $\vec{x}$ of the tip apex:
	\begin{equation}\label{eq_T_Tersoff}
		|T_{\lambda\rho}|^2\propto|\varphi_{\rho}(\vec{x})|^2.
	\end{equation}
Using Eq.~(\ref{eq_T_Tersoff}) they were able to explain the measured
corrugations for the various superstructures observed on the reconstructed
Au~(110) surface. The case of a non-spherical tip and a rough sample surface was
later treated along the same lines \cite{Sacks-1988, Sestovic-1995}. In these
models the atomic structure of the tip is not taken into account. This issue was
investigated by \citet{Chen-1988, Chen-1990a, Chen-1990b}, who assumed instead
that a \emph{single} atom at the tip apex is responsible for the interaction
with the sample surface. \citeauthor{Chen-1988} summarized his results in a
`derivative rule': $T_{\lambda\rho}$ is proportional to a derivative of the
sample wave function, which derivative depends upon the orbital state of the
apex atom. Eq.~(\ref{eq_T_Tersoff}) can be regarded as a particular case
appropriate for $s$-wave tips. The unexpectedly large corrugation amplitudes
observed on HTS surfaces could possibly be explained using this formalism.

According to Eq.~(\ref{eq_T_Tersoff}) and the resulting
Eq.~(\ref{eq_current_Tersoff}), the $c$-axis STM tunneling into a nearly
two-dimensional material is possible, like for the planar junction, but for
different reasons. On the one hand the local nature of the STM tip explains its
ability to measure all momenta in the $ab$ plane, thus preventing any momentum
selectivity. On the other hand $T_{\lambda\rho}$ in Eq.~(\ref{eq_T_Tersoff}) is
not proportional to the group velocity along $z$, in contrast to the planar
junction case. The complete absence of a velocity factor in
Eq.~(\ref{eq_T_Tersoff}) results from specific approximations made by
\citeauthor{Tersoff-1983}, such that having a spherical tip wave function and
identical work functions in the sample and the tip; the relaxation of these
assumptions would yield some dependence upon the group velocity, but is not
expected to change the picture qualitatively.

\subsection{Interpretation of STM experiments on HTS}
\label{sect_tunneling_HTS}

A consistent interpretation of the STM experiments on HTS must address two
different questions. The first concerns the electronic nature of the materials
themselves, and the second is the coupling of the material surfaces with the STM
tip. Until now theorists have focused  on the former problem, and made
considerable progress; several excellent reviews of this effort were given
recently.\footnote{\citet{Dagotto-1994, Anderson-1997, Orenstein-2000,
Chubukov-2003, Norman-2003, Sachdev-2003, Kivelson-2003, Carlson-2004, Lee-2004,
Demler-2004}. This is not an exhaustive list.} In this paragraph we leave this
question aside, and we address some issues related to the tunneling process
itself, concentrating on the Bi2212 compound.

Generally the conditions of the STM measurements on HTS are consistent with the
basic assumptions of the tunneling-Hamiltonian theory. The absence of a
substantial interaction between tip and sample was evidenced by the
insensitivity of the tunneling conductance to the tip-sample distance
\cite{Renner-1995}. The interval between two tunneling events is typically
$10^{-10}$~s for a current of $1$~nA, which is long compared to the relaxation
times $\lesssim 10^{-12}$~s in the materials; therefore non-equilibrium effects
are not expected to play a role in STM spectra.

\subsubsection{The role of the BiO surface layer}

Many STM experiments on Bi2212 have been successfully interpreted using
Eq.~(\ref{eq_conductance_Tersoff}). It is commonly believed that the measured
LDOS originates from the CuO$_2$ bilayer (see Fig.~\ref{fig_crystals}).
According to Eq.~(\ref{eq_conductance_Tersoff}), though, the LDOS is not
measured directly on a CuO$_2$ plane, but at some point $\vec{x}$ typically
10~\AA\ above that plane, since Bi2212 cleaves between the weakly bonded BiO
sheets. The CuO$_2$ layer lies 4.5~\AA\ beneath the surface BiO layer, raising
the question of the role played by the BiO in the tunneling process from the STM
tip to the CuO$_2$. That the BiO indeed plays a significant role is obvious from
the topographic images, which show the lattice of Bi atoms on the surface (see
Sec.~\ref{sect_surface_Bi}).

The band calculations indicate that the BiO layer is
metallic.\footnote{\citet{Hybertsen-1988, Massidda-1988, Krakauer-1988,
Herman-1988, Mattheiss-1988, Szpunar-1992, Jarlborg-2000}} This result
contradicts the common experimentalist's wisdom that the BiO layer is
insulating. The metallic nature of BiO in these calculations is due to a pair of
conduction Bi($6p$)-O($2p$) hybrids, which disperse below the Fermi energy in a
small region around the $(\pi,\,0)$ point of the Brillouin zone. It is well
known that correlation effects on the Cu($3d$) orbitals lead to a failure of the
LDA approaches in the antiferromagnetic phase; such effects are not expected in
the BiO layer. Therefore, a Bi-O band at the Fermi surface would give rise to a
sharp peak in the spectral function near $E_{\text{F}}$. This peak is not seen
in ARPES experiments \cite{Damascelli-2003, Campuzano-2004}, suggesting that the
Bi-O band indeed lies above $E_{\text{F}}$. \citet{Singh-1995} found that in
Bi2201 the calculated Bi-O bands move upward by $\sim 400$~meV when the
distortion of the BiO planes is taken into account. Very recently,
\citet{Lin-2006} found that a similar shift occurs in Bi2212 when lead or oxygen
doping is introduced in the calculation.

If the BiO surface layer is not conducting, then at sufficiently low energies
the STM would necessarily measure the spectral properties of the buried CuO$_2$
bilayer. The observation of the Bi lattice in the topographic images could then
be due to \textit{e.g.} (i) a modulation of the shape of the tunnel barrier by
the Bi atoms, (ii) a diffraction of the tunneling electron by the BiO (and SrO)
intermediate layers, (iii) the admixture of some Bi($6p$) character in the
Cu($3d$)-O($2p$) states of the CuO$_2$ bilayer, (iv) the fact that the bias
applied to record the topography (typically $\gtrsim0.5~$V) would lie within the
Bi-O bands.

\subsubsection{Momentum dependence of the matrix element}
\label{sect_Mk}

The matrix element Eq.~(\ref{eq_T_Tersoff}) is local in space [see also
Eq.~(\ref{eq_T_Tersoff_r})]. In some instances, especially in the context of STM
tunneling on impurities (Sec.~\ref{sect_impurities}), it has been claimed that
the matrix element could be non local, \textit{i.e.} could depend upon the
sample wave function in some vicinity of the tip position $\vec{x}$. A
phenomenological generalization of Eq.~(\ref{eq_T_Tersoff}) taking into account
this possibility would be
	\begin{equation}\label{eq_M_r}
		T_{\lambda\rho}\propto\int d\vec{r}\,M(\vec{r}-\vec{x})
		\varphi_{\rho}^*(\vec{r}).
	\end{equation}
The function $M$ weights the various contributions of the sample wave function
around the tip, and reduces to $\delta(\vec{r}-\vec{x})$ for a local matrix
element. For a translationally invariant system, Eq.~(\ref{eq_M_r}) leads to
(Appendix~\ref{app_Mk}):
	\begin{equation}\label{eq_conductance_Mk}
		\sigma(V)\propto\int d\omega\,[-f'(\omega-eV)]
		\sum_{\vec{k}}|M_{\vec{k}}|^2A(\vec{k},\,\omega),
	\end{equation}
where $M_{\vec{k}}$ is the Fourier transform of $M(\vec{x})$ and
$A(\vec{k},\,\omega)$ is the sample spectral function.\footnote{The assumption
of translational invariance excludes from the outset any atomic-scale dependence
of the tunneling conductance. Hence Eq.~(\ref{eq_conductance_Mk}), unlike
Eq.~(\ref{eq_conductance_Tersoff}), cannot address the question of the atomic
resolution or describe inhomogeneous systems like vortex cores.
Eq.~(\ref{eq_conductance_matrix_element}) provides the suitable generalization
of Eq.~(\ref{eq_conductance_Tersoff}) to account for a non-local matrix
element.} The vector $\vec{k}$ in Eq.~(\ref{eq_conductance_Mk}) belongs to the
three-dimensional Brillouin zone of the sample. In the case of the HTS, though,
the $k_z$ dependence of the spectral function can be to first approximation
neglected, and the $k_z$ sum does not contribute to the $V$ dependence of
$\sigma(V)$. Anisotropic matrix elements $M_{\vec{k}}$ have been invoked for the
interpretation of various experimental results as discussed in
Secs.~\ref{sect_CuO2}, \ref{sect_impurities}, \ref{sect_CoreHTS}, and
\ref{sect_Dopplershift}.

The STM measurements on zinc impurities in Bi2212 \cite[][see
Sec.~\ref{sect_impurities}]{Pan-2000a} revealed a sharp maximum in the LDOS at
the impurity site. Similar spectra were observed on surface defects
\cite{Yazdani-1999}. These observations contrast with the prediction of the BCS
theory in the unitary limit, namely a vanishing LDOS on the impurity and
resonances on the neighboring sites \cite{Salkola-1996}. \citet{Martin-2002}
argued that the tunneling into the planar Cu($3d$) orbitals occurs indirectly
through Cu or Bi orbitals extending out of the surface and having zero in-plane
orbital momentum. As a result the STM would not probe the wave function at the
Zn site, but at the four neighboring Cu sites (see Fig.~\ref{fig_Pan-2000}). The
weighting function in Eq.~(\ref{eq_M_r}) would then take the form
$M(\vec{x})\propto \delta(\vec{x}-\vec{a}_0)+ \delta(\vec{x}+\vec{a}_0)
-\delta(\vec{x}-\vec{b}_0)- \delta(\vec{x}+\vec{b}_0)$, where $\vec{a}_0$ and
$\vec{b}_0$ are the basis vectors of the Cu lattice. The opposite signs along
the $a$ and $b$ directions reflect the $d_{x^2-y^2}$ symmetry of the CuO$_2$
orbitals near $E_{\text{F}}$. In the reciprocal space this corresponds to
$|M_{\vec{k}}|^2\propto(\cos k_xa_0-\cos k_yb_0)^2$. The absence of a large
zero-bias conductance peak in the vortex cores was also tentatively attributed
to matrix-element effects \cite[][see Sec.~\ref{sect_CoreHTS}]{Wu-2000}. In the
STM literature, this $d_{x^2-y^2}$ matrix element is often attributed to
\citet{Chakravarty-1993}, although these authors introduced it  for describing
the coherent tunneling of Cooper pairs in bilayer compounds. \citet{Franz-1999}
obtained the same result for a planar-junction, after summing over the momenta
along the $c$ axis, and argued that this matrix element would prevent the STM
from seeing the Doppler-shift induced zero-energy DOS in the mixed state
\cite[][see Sec.~\ref{sect_Dopplershift}]{Volovik-1993}. Finally a similar form
for $|M_{\vec{k}}|^2$ (with a plus sign instead of the minus sign) was also
suggested by arguing that direct tunneling into the planar CuO$_2$ orbitals is
blocked by the atomic cores in the intermediate SrO and BiO layers
\cite{Zhu-2000}.

The anisotropic $d_{x^2-y^2}$ matrix element would prevent the electrons from
tunneling into states with momenta along the nodal directions, and would
therefore highlight the $(\pi,0)$ region of the Brillouin zone. For a
homogeneous $d$-wave superconductor the low-energy conductance resulting from
Eq.~(\ref{eq_conductance_Mk}) with an anisotropic $M_{\vec{k}}$ would show a
rounded U-shape instead of the V-shape characteristic of nodal quasiparticles.
However, after a detailed fit of the Bi2212 data, \citet{Hoogenboom-2003b}
concluded that the low-energy conductance is incompatible with an anisotropic
matrix element. The recent observation that the STM spectra measured directly on
the CuO$_2$ layer exhibit a U-shape, while the spectra taken on the BiO layer
have the $d$-wave V-shape \cite[][see Sec.~\ref{sect_CuO2}]{Misra-2002a}, was
attributed by the authors to a different matrix element in the two tunneling
configurations.

The actual form of the matrix element for STM tunneling into HTS remains largely
controversial.\footnote{Some authors analyzed tunneling spectra on HTS using a
matrix element analogous to the one derived by \citeauthor{Harrison-1961} for
planar junctions, $|M_{\vec{k}}|^2\propto v_g$ \cite{Kouznetsov-1996} or
$|M_{\vec{k}}|^2\propto v_gD(\vec{k})$ \cite{Yusof-1998}. The main motivation
was to explain the absence of some sharp structures, such as van-Hove
singularities, in the experimental tunneling spectra, and the observed asymmetry
in the background conductance (Sec.~\ref{sect_gap_spectroscopy}) as well as
specific features observed in point-contact tunneling experiments. Here $v_g$ is
the group velocity and $D(\vec{k})$ is a function selecting the momenta close to
the tunneling direction, which was introduced by \citet{Ledvij-1995} as a
generic model for the Josephson coupling between planar superconductors. The
validity of these planar-junction matrix elements to account for the STM and
point-contact geometries was not addressed, however.} Since no direct
experimental determination of $T_{\lambda\rho}$ (or $M_{\vec{k}}$) has been
possible so far, and no microscopic calculation of the tip/surface system has
been reported, most assessments about the matrix element rely upon
phenomenological or heuristic arguments. The V-shaped spectra acquired on the
BiO layer support the idea of a \citeauthor{Tersoff-1983}-like matrix element,
\textit{i.e.} $M_{\vec{k}}\equiv1$. Although there is no compelling experimental
evidence in favor of an anisotropic $M_{\vec{k}}$, the measurements on zinc
impurities and possibly also on vortices could be interpreted in the framework
of the BCS theory assuming a more complicated matrix element like
$M_{\vec{k}}\propto\cos k_x a_0-\cos k_yb_0$. However, alternate explanations
which emphasize the non-BCS character of the HTS have also been proposed.

\subsection{Summary}

Most STM spectroscopic measurements on HTS have been successfully analyzed in
the framework of the tunneling-Hamiltonian formalism. Complemented with the
\citeauthor{Tersoff-1983} treatment of the STM junction, this formalism provides
a straightforward interpretation of the tunneling data,
Eq.~(\ref{eq_conductance_Tersoff}), in terms of the sample LDOS. Although there
is a broad consensus regarding this interpretation of the data, a definitive
assessment requires to fix the issue of the tunneling matrix element.

\section{Crystal structure and surface characterization}
\label{sect_characterization}

\begin{figure}[tb]
\includegraphics[width=7.5cm]{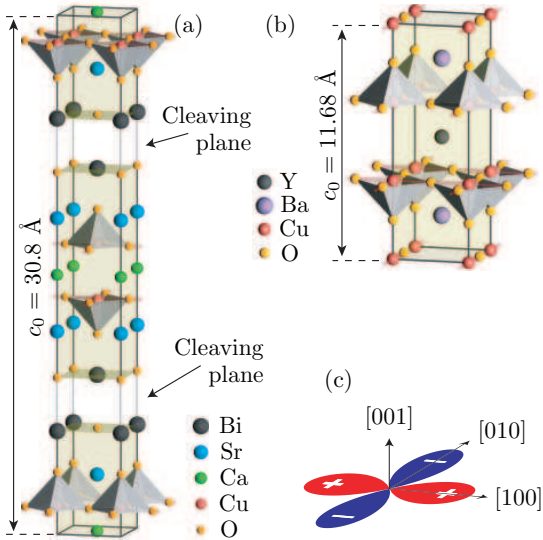}
\caption{\label{fig_crystals}
(a) Tetragonal unit cell of Bi$_2$Sr$_2$CaCu$_2$O$_{8+\delta}$ with
$a=b=5.4$~\AA\ defined along $[110]$ and $[\bar{1}10]$, respectively. (b)
Orthorhombic unit cell of YBa$_2$Cu$_3$O$_{6+\delta}$ with $a_0=3.82$~\AA\ along
$[100]$ and $b_0=3.89$~\AA\ along $[010]$. (c) Schematics of the $d_{x^2-y^2}$
superconducting gap in the unit-cell coordinate system. Note that $a_0$ and
$b_0$ designate the lattice constants along the Cu-O-Cu bonds ($Pmmm$ space
group for Y123), whereas $a$ and $b$ designate the lattice constants in larger
unit cells ($Cmmm$ space group for Bi2212).
}
\end{figure}

The superconducting gap of  standard isotropic BCS superconductors is
independent on position in real space and on momentum along the Fermi surface in
$k$-space ($s$-wave gap symmetry). Therefore, the tunneling spectra of  a
homogeneous sample neither depend on the tunneling direction, nor on the
position along the surface, nor on the chosen surface. Under those circumstances
the measured gap does not depend upon the above-mentioned experimental
configuration, allowing for an unambiguous interpretation of tunneling
experiments. The situation is very different in layered HTS cuprates, owing to
their very anisotropic structural, electronic, and superconducting properties.

The HTS cuprates consist of one or more superconducting copper oxide (CuO$_2$)
sheets sandwiched between non-metallic charge reservoir layers
(Figs.~\ref{fig_crystals}a,b). Because of the alternate stacking of
superconducting and non-superconducting layers along the crystallographic
$[001]$ direction, the tunneling spectra of $(001)$ surfaces are likely to
depend on the termination layer. As a result of the $d$-wave symmetry
(Fig.~\ref{fig_crystals}c) of the superconducting gap, the spectra may also
strongly depend on which momentum states contribute to the tunneling current
(see Sec.~\ref{sect_Mk}). Finally, the sign change of the $d$-wave gap function
at the nodal points along $(\pi,\,\pi)$ induces quasiparticle bound states at
surfaces or step edges oriented perpendicular to the nodal directions. The
spectroscopic signature of these bound states is an enhanced conductance---so
called zero-bias conductance peak---instead of a gap at $E_{\text{F}}$
\cite{Kashiwaya-2000}. For all these reasons, tunneling spectra measured on HTS
are likely to depend on the exposed surface layer, its orientation, and the
details of the tunneling matrix element. Hence, a range of different tunneling
spectra is expected, even for perfect vacuum tunnel junctions, making their
interpretation very challenging.

On layered HTS, such as Bi$_2$Sr$_2$CaCu$_2$O$_{8+\delta}$, a further key issue
is to identify which states are being probed by STM. Does the tunneling current
involve pure BiO or CuO$_2$ derived states, or some hybridized ones? This
question, which we already touched upon in Sec.~\ref{sect_theory}, is of
fundamental importance. Superconductivity is believed to reside in the CuO$_2$
plane, but most tunneling measurements of the superconducting $d$-wave gap were
obtained on the BiO surface. Ascribing the measured gap to the superconducting
state of HTS, implies that tunneling into the BiO surface allows to measure the
superconducting gap of the CuO$_2$ layer situated several \AA ngstr\"oms beneath
the surface. This statement is far from obvious because of the short coherence
length in HTS and the non-metallic layers sandwiching the superconducting
CuO$_2$ sheets. The possibility to sense electronic contributions from
sub-surface atomic layers is well known in highly ordered pyrolytic graphite, a
very simple layered conductor. STM images of this material reveal two non
equivalent carbon sites instead of a regular honeycomb lattice
\cite{Wiesendanger-1992}, due to site-dependent hybridization of carbon orbitals
from the neighboring graphene sheet.

A careful assessment of the probed surface and of the tunnel junction is
prerequisite to lift the above ambiguities and enable a meaningful
interpretation of the tunneling spectra measured on HTS. Identification of the
local surface layer is best done through atomic-scale imaging. However, such
optimal resolution remains elusive on most HTS cuprates, and in
Secs.~\ref{sect_surface_Bi} and \ref{sect_surface_Y} we restrict the discussion
to the archetypical Bi- and Y-compounds, respectively. The quality of the tunnel
junction can be inferred from the reproducibility of the spectra as a function
of position along the surface and as a function of tip-to-sample distance as
discussed in Sec.~\ref{sect_tip_sample_distance}.

\subsection{Surfaces of Bi-based cuprates}
\label{sect_surface_Bi}

Bi$_2$Sr$_2$CaCu$_2$O$_{8+\delta}$ (Bi2212) is the most widely studied HTS using
STM for a simple reason: it is straightforward to prepare atomically flat and
clean surfaces by cleaving. Cleaving is most likely to occur between the weakly
van der Waals coupled adjacent BiO layers (Fig.~\ref{fig_crystals}a), and with
few exceptions, STM studies have focused on BiO terminated surfaces. It was the
first surface of any HTS to yield STM images with atomic-scale topographic
resolution. Following the pioneering work by \citet{Kirk-1988b}, the resolution
steadily improved to reveal the tetragonal unit cell and the $\sim5\,b$
incommensurate supermodulation along $[\bar{1}10]$ with stunning sharpness.

Cleaving predominantly exposes the BiO layer to the surface. However, other
surface terminations were occasionally observed by STM \cite{Murakami-1995,
Pan-1998b, Misra-2002a}. In the following paragraphs, we discuss STM
investigations of the BiO and CuO$_2$ surfaces in greater details.

\begin{figure}[tb]
\includegraphics[width=8.6cm]{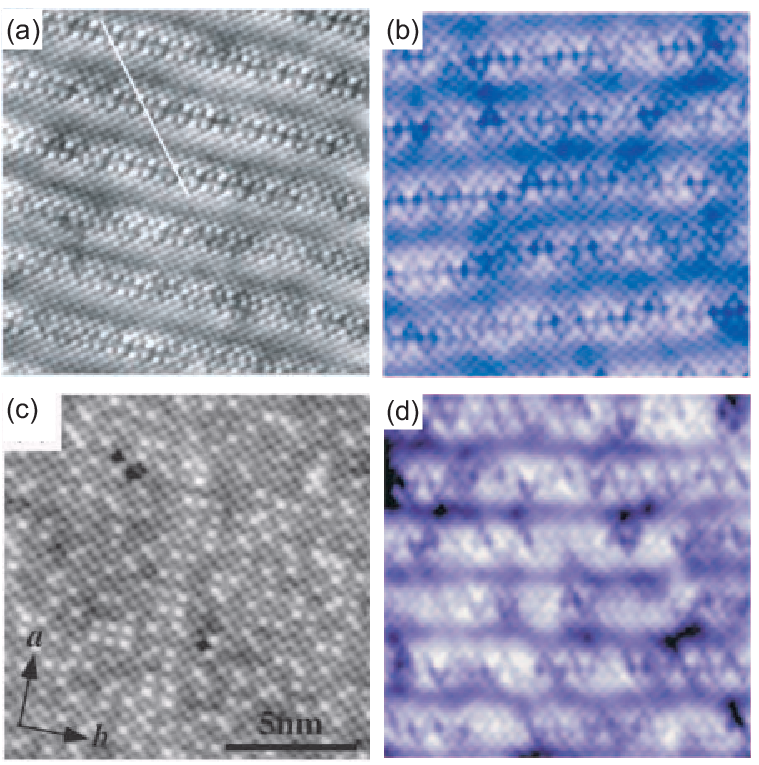}
\caption{\label{fig_stmbisurface}
STM images of the BiO surface of Bi-based HTS cuprates. (a)
$13.1\times13.1$~nm$^2$ area of Bi$_2$Sr$_2$CuO$_6$ at 4.6~K \cite{Shan-2003}.
(b) $15\times15$~nm$^2$ area of Bi$_2$Sr$_2$CaCu$_2$O$_8$ at 4.2~K
\cite{Pan-2001}. (c) Lead-doped Bi$_2$Sr$_2$CaCu$_2$O$_8$ at 4.3~K
\cite{Kinoda-2003}. The bright lattice sites correspond to lead dopant atoms.
Note the absence of the supermodulation and dark-atom rows. (d)
$15\times15$~nm$^2$ area of zinc-doped Bi$_2$Sr$_2$CaCu$_2$O$_8$ at 4.3~K
\cite{Pan-2000a}.
}
\end{figure}

\subsubsection{BiO surface}
\label{sect_BiO}

The structural characteristics of the BiO surface, as seen by STM, are similar
in all parent Bi-based HTS cuprates (Figs.~\ref{fig_stmbisurface}a,b).
Distinctive features include a nearly commensurate supermodulation along
$[\bar{1}10]$ and concomitant dark-atom rows running along the supermodulation
ridges. The supermodulation develops throughout the bulk BiO layers, and is not
a surface effect \cite{Heinrich-1994, Gladyshevskii-1996}. Note that the
supermodulation along $[\bar{1}10]$ in Bi2212 is mostly structural. No
associated spatial variation of the superconducting gap amplitude has been
reported. The dark-atom rows are observed on all pure BiO surfaces except in few
instances on Bi2201 \cite{Inoue-1995} and Bi2212 \cite{Shih-1991}.

The origin of the supermodulation and the dark-atom rows is still matter of
debate. Initially interpreted as rows of missing atoms \cite{Kirk-1988b}, the
latter were recently found to be more consistent with depressed rather than
vacant lattice sites. \citet{Inoue-1994} attributed them to spatial ordering of
non equivalent Bi atoms, whereas \citet{Zandbergen-1998} ascribed them to extra
oxygen in the BiO layers. The supermodulation could result from a structural
mismatch among the constituent layers of Bi2212. Based on X-ray refinement,
\citet{Gladyshevskii-2004} propose an alternative model where it stems from a
rotation of Bi-O trimer in the BiO plane. In this scenario, the dark-atom rows
reflect additional oxygen atoms necessary to register this rotation to the BiO
lattice. \citeauthor{Gladyshevskii-2004} further observe that this additional
oxygen site is quenched in lead-doped Bi2212, suppressing both the
supermodulation and the dark-atom rows, consistent with STM topographs of
lead-doped Bi2212 (Fig.~\ref{fig_stmbisurface}c).

The BiO layers contain two atomic species, but atomic-scale STM images reveal
only one lattice site. \citet{Shih-1991}  found no dependence  on bias polarity
or amplitude as would be expected if the BiO layer was an ionic insulator,
suggesting a single atomic species is contributing to the contrast. Strong
indications that STM is imaging the Bi lattice come from studies of Pb
\cite{Kinoda-2003}, Zn \cite{Pan-2000a}, and Ni \cite{Hudson-2001} substituted
single crystals. In the case of lead doping, images of the BiO surface reveal
some brighter lattice sites due to the dopant atoms
(Fig.~\ref{fig_stmbisurface}c). Lead substitutes for bismuth in the BiO layer,
hence the STM imaging contrast shows the bismuth lattice. Zinc and nickel
substitute for copper in the CuO$_2$ layer. Unlike lead dopant atoms, the latter
are not directly resolved in topographic images (Fig.~\ref{fig_stmbisurface}d).
However, they are clearly seen at the bismuth lattice sites in spectroscopic
images of the surface (\textit{i.e.} spatially resolved maps of the local
tunneling conductance, see Figs~\ref{fig_Pan-2000} and \ref{fig_Hudson-2001}).
Since the bismuth and copper lattices are aligned along the $[001]$ direction,
the implication is again that STM topographs of the BiO surface do reveal the
bismuth lattice. Moreover, the latter  results show that tunneling into the BiO
surface feels the underlying CuO$_2$ plane, supporting the idea that the
superconducting gap of the CuO$_2$ layer can be measured by tunneling into the
BiO surface.

\subsubsection{CuO$_2$ surface}
\label{sect_CuO2}

\begin{figure}[tb]
\includegraphics[width=8.6cm]{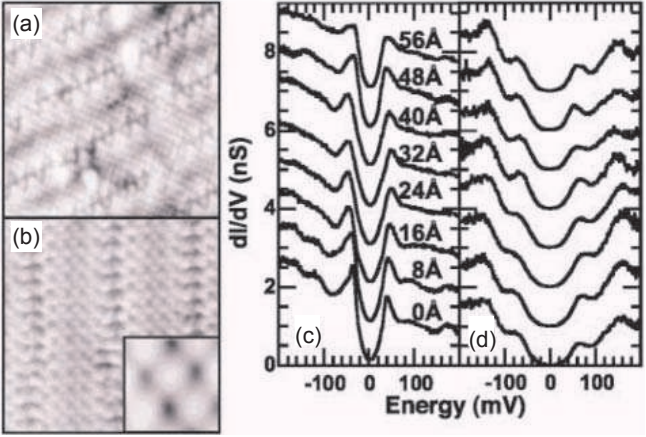}
\caption{\label{fig_Misra2002}
STM topography and spectroscopy of a cleaved Bi$_2$Sr$_2$CaCu$_2$O$_8$ thin
film. (a) $10\times10$~nm$^2$ topographic image of a BiO terminated surface. (b)
$6.4\times6.4$~nm$^2$ topographic image of a CuO$_2$ terminated surface (inset
$0.8\times0.8$~nm$^2$). (c) Tunneling spectra taken at the indicated distances
along a straight line on the BiO plane and (d) on the CuO$_2$ plane. Spectra are
offset for clarity, and all data were taken with $V=200$~mV and $I=200$~pA. From
\citet{Misra-2002a}.
}
\end{figure}

The most comprehensive topographic and spectroscopic STM study of the CuO$_2$
surface to date was carried out by \citet{Misra-2002a} on Bi2212 thin films.
They clearly resolved the tetragonal lattice of the CuO$_2$ and the BiO $(001)$
surfaces (Fig.~\ref{fig_Misra2002}). Both layers host a supermodulation of
similar periodicity with one characteristic difference: the in-plane concomitant
lattice modulation is much weaker in CuO$_2$ layers than in BiO layers
\cite{Gladyshevskii-1996}. While they obtain tunneling spectra of the BiO
surface consistent with a $d$-wave superconducting gap
(Fig.\ref{fig_Misra2002}c),  CuO$_2$ terminated surfaces show a wider gap
($60\pm10$~meV) with a strongly suppressed conductance near the Fermi energy
(Fig.\ref{fig_Misra2002}d).  This U-shaped energy dependence of the CuO$_2$
tunneling conductance close to $E_{\text{F}}$ (Fig.\ref{fig_Misra2002}d) is
unexpected for a $d$-wave superconductor, raising the question of the electronic
nature of a bare CuO$_2$ layer. \citet{Kitazawa-1996} already noted the tendency
of CuO$_2$ surfaces to yield spectra with a U-shaped energy dependence at low
bias (expected for a $s$-wave BCS gap), whereas other surface terminations of
HTS cuprates show a V-shaped energy dependence near $E_{\text{F}}$ (expected for
a $d$-wave BCS gap).

The electronic structure of the BiO layer is likely to be the same at the
surface and in the bulk, because cleaving between adjacent BiO layers does not
break any strong bond. Exposing the CuO$_2$ layer to the surface is
significantly more disruptive, hence its structural and electronic properties
may be very different from the bulk. \citet{Misra-2002a} ruled out any surface
reconstruction to explain the non $d$-wave CuO$_2$ spectra, because the measured
surface structure (Fig.~\ref{fig_Misra2002}b) matches the bulk structure
\cite{Heinrich-1994, Gladyshevskii-1996}. The large gap revealed by spectroscopy
may signal an insulating CuO$_2$ surface. But \citeauthor{Misra-2002a} contend
it would have to be a very homogeneous insulator, with an electron-hole
symmetric low energy DOS and $E_{\text{F}}$ lying exactly half way between the
empty and filled bands (Fig.~\ref{fig_Misra2002}d).

\citet{Misra-2002a} argue that the $s$-wave looking spectra they measured are
consistent with $d$-wave superconductivity in the CuO$_2$ planes, provided the
tunneling matrix element is of the anisotropic form $|M_{\vec{k}}|^2
\propto(\cos k_xa_0-\cos k_yb_0)^2$ (see Sec.~\ref{sect_Mk}). This particular
choice of $M_{\vec{k}}$ enables them to reproduce the gap structure and the
conductance peaks below $\pm100$~meV. Their fit yields a gap amplitude
consistent with a reduced doping, possibly due to oxygen loss and asymmetric
doping by the BiO layer underneath: $60$~meV is indeed close to the gap measured
in highly underdoped, yet superconducting samples \cite{Zasadzinski-2001}. The
analysis by \citeauthor{Misra-2002a} implies that the bare CuO$_2$ layer at the
surface is superconducting, which is an important finding. However, their model
only partially reproduces the experimental spectra of CuO$_2$. In particular, it
does not account for the higher energy peak structures, and does not exclude the
possibility that the CuO$_2$ spectra reflect some signature of the pseudogap
(Sec.~\ref{sect_pseudogap}). Finally, they do not answer the question as to why
$M_{\vec{k}}$ of BiO and CuO$_2$ surfaces would be so different, leaving the
nature of the CuO$_2$ surface open to further investigations.

\subsection{Surfaces of YBa$_2$Cu$_3$O$_7$}
\label{sect_surface_Y}

YBa$_2$Cu$_3$O$_7$ (Y123) is the second most studied HTS by STM. Y123 is much
less anisotropic than Bi2212 and offers no natural cleaving plane. STM
experiments were carried out on as grown and chemically etched surfaces, as well
as on single crystals cleaved at low temperature ($\sim 20$~K). Y123 cleaves
between the BaO and CuO-chain layers, and  both were imaged with atomic
resolution STM on low-temperature cleaved samples \cite{Edwards-1994, Pan-1999}.
The problem with cleaved specimens is that STM \cite{Edwards-1992} and ARPES
\cite{Campuzano-1990} experiments degrade if the surfaces are not kept below
70~K, indicating they are unstable and possibly different from the bulk due to
mechanical strain and unbalanced charges.

On the other hand, reproducible temperature-dependent STM spectroscopy was
demonstrated on as grown surfaces of single crystal grown in BaZrO$_3$ crucibles
\cite{Maggio-Aprile-2000}. These surfaces could sustain repeated thermal cycling
between 4~K and 300~K. They also allowed the first successful imaging of the
Abrikosov vortex lattice in a HTS \cite{Maggio-Aprile-1995}. Similar results
have been achieved on chemically etched surfaces \cite{Shibata-2003b,
Wei-1998b}. Atomic resolution on as-grown Y123 surfaces appears very difficult
to achieve and has been reported only on thin films
\cite{Lang-1991,Nantoh-1995}.  A reliable recipe allowing both atomic-resolution
imaging and reproducible spectroscopy over a wide temperature range on Y123 has
yet to be found.

\begin{figure}[tb]
\includegraphics[width=7.5cm]{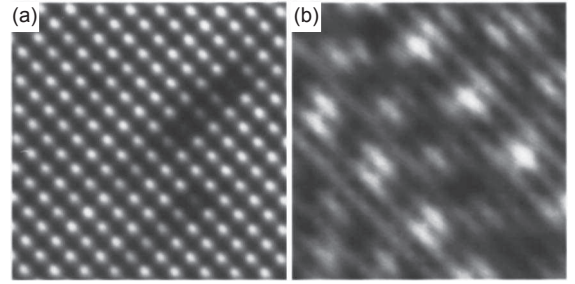}
\caption{\label{fig_Pan-1999}
$5\times5$~nm$^2$ STM images of the (a) BaO and (b) CuO-chain surfaces of Y123.
From \citet{Pan-1999}.
}
\end{figure}

The orthorhombic lattice of BaO terminated low-temperature cleaved surfaces has
been clearly resolved by STM (Fig.~\ref{fig_Pan-1999}a), with a few single
atomic defects attributed to oxygen vacancies \cite{Pan-1999}. Similar
micrographs were obtained by \citet{Edwards-1994} who found them to be
independent of bias amplitude and polarity, although a systematic energy
dependence  is not available yet.

CuO-chain terminated surfaces have been investigated in greater details. They
show a more complex structure  with a large charge modulation superimposed on
the atomic lattice (Fig.~\ref{fig_Pan-1999}b). The copper and oxygen lattices
are resolved as oval and faint round  features, respectively. In contrast to
BaO, STM images of the CuO-chain surface do depend on energy and bias polarity,
which led \citet{Edwards-1994, Edwards-1995} to attribute the spatial modulation
of the DOS to a charge density wave (CDW). More recently, \citet{Derro-2002}
identified a series of strong dispersive resonances at energies below the
superconducting gap $\Delta_p=25$~meV.  In their view, these resonances are
hallmarks of a predominantly 1D character of the CuO-chain layer DOS, and the
charge modulation observed in the topography is more consistent with
superconducting quasiparticle scattering than with a CDW. In this scenario, the
CuO-chains would be superconducting via proximity coupling to the CuO$_2$
planes. Note that the presence of these two superconducting layers could provide
an explanation of the multiple peak structure in Y123 tunneling spectra
discussed in Sec.~\ref{sect_Y_compounds}.

\subsection{Tip-to-sample distance dependence}
\label{sect_tip_sample_distance}

The tip-to-sample distance dependence of the tunneling spectra is a key
experimental test to appraise the quality of the STM vacuum tunnel junction. If
only the states from a single surface layer were contributing to the tunneling
process, the spectra should be independent of tip-to-sample distance. On the
other hand, if the junction is contaminated or if different decoupled layers
were contributing to the tunneling process, one might expect some tip-to-sample
distance dependence.

\begin{figure}[tb]
\includegraphics[width=8.6cm]{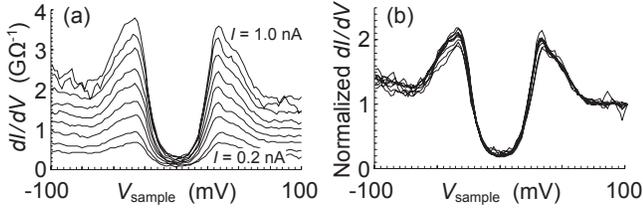}
\caption{\label{fig_tip_distance}
Ideal STM vacuum tunnel junction between a Au tip and Bi2212 at $4.2$~K. (a)
Tunneling spectra measured as a function of increasing tip-to-sample distance.
The current is changed from $I=1.0$~nA (short distance, top curve) to $I=0.2$~nA
(large distance, bottom curve) in 0.1~nA increments at constant sample voltage
$V=0.4$~V. (b) Same data plotted on a normalized conductance scale to emphasize
the distance independence. Adapted from \citet{Renner-1995}.
}
\end{figure}

Large tip-to-sample distance dependencies of $c$-axis imaging and spectroscopy
were common features of early STM experiments on HTS cuprates. Only
low-resistance tunnel junctions ($R_t\sim\text{M}\Omega$), \textit{i.e.} when
the tip is in very close proximity to the surface, allowed to measure
superconducting gap features on Bi2212 and Y123 \cite{Hasegawa-1991a,
Hasegawa-1992, Nantoh-1994b, Aleszkiewicz-1997}. Upon increasing the junction
resistance ($R_t\sim\text{G}\Omega$) by pulling the tip away from the surface,
these superconducting features vanished and the tunneling spectra evolved into
semiconducting-like line shapes. Likewise, some groups claimed that topographic
imaging of Bi2212 was depending on tip-to-sample distance \cite{Nishiyama-1996,
Oda-1996}, the BiO layer being imaged at large and the CuO$_2$ layer at small
tip-to-sample distance. More recently, \citet{Sugita-2000} reported exactly the
opposite distance dependence in experiments where the topographic contrast was
seen to switch randomly between BiO and CuO$_2$.

A better control over the tunnel junctions became possible through progress in
HTS single crystal growth and STM instrumentation, and the above distance
dependencies have not been confirmed on ideal junctions. As shown in
Fig.~\ref{fig_tip_distance}, these improvements enabled to achieve tip-to-sample
distance independent STM tunneling spectroscopy on Bi2212 \cite{Renner-1994,
Renner-1995} and Y123 \cite{Maggio-Aprile-1996}. Recently, \citet{Sakata-2003}
confirmed this distance independence in the entire range from vacuum tunneling
to point contact on Bi2212. The only minute change \citeauthor{Sakata-2003}
found was a slight reduction in the gap amplitude at very short electrode
spacing, which they reckon  might result from a change in the local DOS due to
the pressure applied by the STM tip.

The best tunnel junctions are not only characterized by tip-to-sample distance
independent STM imaging and spectroscopy, they also yield the largest work
function\footnote{The work function can be estimated from the
distance-dependence of the tunneling current \cite{Pan-1998}.} ($\phi>1$~eV) and
tunneling spectra with the sharpest superconducting DOS structures. On Bi2212, a
couple of other distinctive spectral features enable to readily identify the
best junctions: (i) The background conductance is flat or even slightly
decreasing at energies $eV<500$~meV \cite[see][and references
therein]{Renner-1995}. (ii) The conductance peaks at the gap edges are most
intense and sharp in spectra  with a flat background conductance  below
$\pm500$~meV. Smeared coherence peaks and a parabolic background conductance
both signal a contaminated tunnel junction. Indeed, the energy dependence of the
tunneling probability expected in low work function tunnel junctions leads to a
background conductance increasing with energy, and quasiparticle scattering
causes damping of the coherence peaks.

The degree of quasiparticle scattering and the energy dependence of the
background conductance are prone to vary, depending on the quality of the local
tunnel junction. As a result, the surface will appear very non-uniform from a
spectroscopic point of view, with $dI/dV$ spectra varying from parabolic curves
with only a weak or no gap structure in degraded surface areas, to flat
characteristics with a very sharp gap structure in clean surface areas. Such
inhomogeneities were first observed by \citet{Wolf-1994} in spatially resolved
tunneling spectroscopy of Bi2212. They are strictly related to the (poor)
quality of the surface, and are to be clearly distinguished from the
inhomogeneities discussed in Secs.~\ref{sect_spatial_homogeneity} and
\ref{sect_modulations}.

\subsection{Summary}
\label{sect_summary_characterization}

The early days of STM spectroscopy on HTS cuprates were plagued by a very wide
spread in tunneling line shapes, as pointed out by \citet{Kitazawa-1996}. Under
those circumstances, it was extremely challenging to delineate the intrinsic
features of the measured local DOS. It appears now that much of the spread in
experimental data was a consequence of ill-defined tunnel junctions and sample
surfaces. Atomic resolution STM imaging, although not sufficient to ensure true
vacuum tunneling spectroscopy as shown by \citet{Renner-1995}, plays an
important part in assessing the surface.  The more stringent tests of the
junction quality is a reasonably large work function ($\phi\gtrsim 1$~eV) and
tunneling spectra which are independent of tip-to-sample distance.

\section{Low-temperature tunneling spectroscopy}
\label{sect_gap_spectroscopy}

Since the pioneering experiments of \citet{Giaever-1960}, electron tunneling
spectroscopy has become the technique of choice to probe the superconducting
quasiparticle density of states (DOS). It is probably the most successful and
sensitive way to measure the gap in the electronic excitation spectrum at the
Fermi energy, one of the hallmarks of superconductivity (see textbooks by
\citet{Tinkham-1996} and \citet{Wolf-1985}). Tunneling spectroscopy was
instrumental in validating the BCS theory for conventional low-$T_c$
superconductors. Naturally, when superconductivity was discovered at
unprecedented high temperatures in copper oxide perovskites \cite{Bednorz-1986},
large expectations were put on tunneling experiments to unravel the underlying
physics.

The focus of this section is on low-temperature tunneling spectroscopy of HTS
cuprates performed by STM in zero external magnetic field. We shall review the
differential tunneling conductance spectra $dI/dV$ measured on selected HTS,
with the aim to identify generic features in the electronic DOS related to their
extraordinary superconducting properties. We discuss in particular the gap
structure at the Fermi energy, the background conductance, the dip-hump feature
above the gap, the zero-bias conductance peak at the Fermi energy (ZBCP), and
the doping dependence and spatial (in-)homogeneity of the tunneling spectra.

The overwhelming majority of experiments probe $(001)$ surfaces with the STM tip
perpendicular to the CuO$_2$ planes. Hence, we shall mostly discuss this
experimental configuration. Cross-sectional tunneling, \textit{i.e.} tunneling
into $(hk0)$ surfaces with the STM tip parallel to the CuO$_2$ planes, is
briefly discussed in Sec.~\ref{sect_cross_section}.

\subsection{Gap spectroscopy}
\label{sect_low_T_spectro}

A series of typical SIN tunneling characteristics measured by STM on selected
superconductors at low temperature is presented in Fig.~\ref{fig_fig5_gaps}.
Niobium (Fig.~\ref{fig_fig5_gaps}a) exhibits all the classic features expected
for a conventional low-$T_c$ BCS superconductor: (i) a completely developed gap
centered on $E_{\text{F}}$ ($\Delta=1.0$~meV) characterized by a U-shaped
$dI/dV$ with zero conductivity at $E_{\text{F}}$, and (ii) two symmetric square
root singularities at the gap edges. Such spectra are fully described by the BCS
theory \cite{Tinkham-1996}. In contrast, tunneling spectra of HTS cuprates
deviate in a number of remarkable ways from the BCS predictions as described in
the following paragraphs.

\subsubsection{Bi-compounds}
\label{sect_Bi_compounds}

\begin{figure}[tb]
\includegraphics[width=8.6cm]{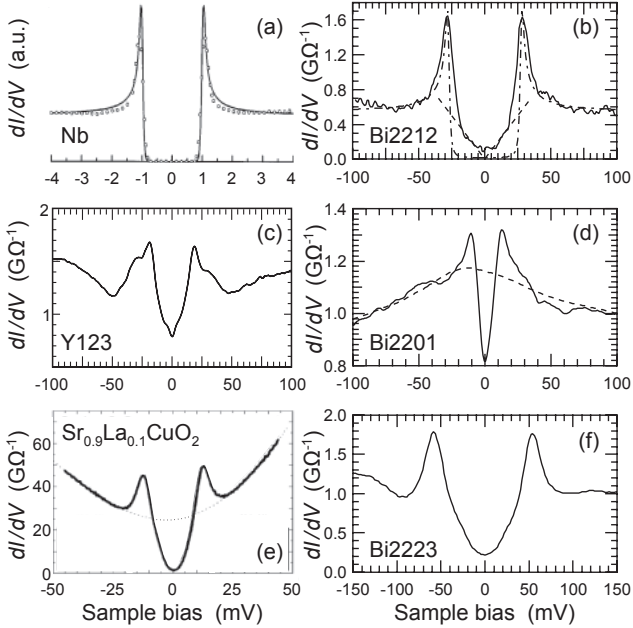}
\caption{\label{fig_fig5_gaps}
$c$-axis SIN vacuum tunneling conductance spectra of selected superconductors
(single crystals) measured by STM. (a) Nb at $335$~mK (circles); BCS fit with
$\Delta=1.0$~meV (solid line); from \citet{Pan-1998}. (b) Optimally-doped Bi2212
($T_c=92$~K) at $4.8$~K (solid line); $s$-wave BCS fit with $\Delta=27.5$~meV
and pair-breaking parameter $\Gamma=0.7$~meV (dash-dotted line), and low-energy
V-shaped $d$-wave BCS conductance (dashed line); from \citet{Renner-1995}. (c)
Y123 at $4.2$~K; from \citet{Maggio-Aprile-1995}. (d) Overdoped Bi2201
($T_c=10$~K) at $2.5$~K (solid line) and $82$~K (dashed line); adapted from
\citet{Kugler-2001}. (e) Optimally-doped Sr$_{0.9}$La$_{0.1}$CuO$_2$ at $4.2$~K;
adapted from \citet{Yeh-2002}. (f) Underdoped Bi2223 ($T_c=111$~K) at $4.2$~K;
from \citet{Kugler-2006}.
}
\end{figure}

A characteristic $dI/dV$ spectrum measured on a cleaved Bi2212 single crystal is
shown in Fig.~\ref{fig_fig5_gaps}b \cite{Renner-1994, Renner-1995, DeWilde-1998,
Pan-2001}. Spectra with similar line shapes were obtained on Bi2223
\cite[Fig.~\ref{fig_fig5_gaps}f,][]{Kugler-2006} and Hg-based cuprates
\cite{Rossel-1994, Wei-1998}, though the latter exhibit a very different
(increasing) energy dependence of the background conductance. The prominent
low-energy features of the Bi2212 spectra are two large conductance peaks
defining a clear gap centered on $E_{\text{F}}$ and a conductance increasing
linearly with energy near $E_{\text{F}}$. For convenience, the superconducting
gap is often defined as half the energy separating the two conductance peaks
($\Delta_p$). The gap obtained this way is somewhat larger than the value
$\Delta$ calculated from a proper fit of the spectra \cite{Hoogenboom-2003b}. In
HTS, the reduced gap defined as $2\Delta_p/k_{\text{B}}T_c$ is far in excess of
3.5 or 4.3, the values expected for a weak coupling $s$-wave or $d$-wave BCS
superconductor, respectively \cite{Won-1994}.

Two of the most striking features of the $dI/dV$ spectra of Bi2212 are the
V-shaped energy dependence near $E_{\text{F}}$ and the very large peaks at the
gap edges. The V-shaped low-bias conductance (Fig.~\ref{fig_fig5_gaps}b) is
indicative of nodes in the gap. This is consistent with the $d$-wave symmetry
deduced from tricrystal experiments \cite{Tsuei-2000} and with the angular
dependence observed in ARPES measurements \cite{Damascelli-2003,
Campuzano-2004}. However, this simple model fails to reproduce the large
spectral weight in the conductance peaks at the gap edges. \citet{Wei-1998} and
\citet{Hoogenboom-2003b} showed that both characteristics can be simulated
simultaneously by taking into account the band structure, and especially the van
Hove singularity (vHs) near the saddle points at $(\pi,\,0)$ and $(0,\,\pi)$ in
the Brillouin zone (see Sec.~\ref{sect_dip_hump}).

The simulated spectral weight in the peaks could also be enhanced by assuming an
anisotropic tunneling matrix element, which would filter the tunneling
probability in favor of the anti-nodal states near $(\pi,\,0)$, and suppress the
tunneling probability near the nodal states along $(\pi,\,\pi)$
\cite{Yusof-1998, Franz-1999}. However, this simple model does not only enhance
the DOS in the coherence peaks in agreement with experiment, it also reduces the
DOS below the gap too much, producing more $s$-wave looking U-shaped spectra
inconsistent with experiments \cite{Hoogenboom-2003b}.

The tunneling spectra of the three layer Bi-compound
Bi$_2$Sr$_2$Ca$_2$Cu$_3$O$_{10+\delta}$ (Bi2223) look very similar to those of
Bi2212, except that all features are shifted to higher energies
(Fig.~\ref{fig_fig5_gaps}f). On the contrary, the single layer Bi-compound
Bi$_2$Sr$_2$CuO$_6$ (Bi2201) yields substantially different spectra
(Fig.~\ref{fig_fig5_gaps}d) with a finite zero-bias conductance and a much
larger reduced gap, \textit{e.g.} $2\Delta_p/k_{\text{B}}T_c\sim28$ in overdoped
Bi2201 \cite{Kugler-2001}. \citet{Mashima-2003} measured similar looking spectra
on lead-doped Bi2201. The considerable difference between Bi2201 and Bi2212 may
stem from the more pronounced two-dimensional nature of Bi2201.
Superconductivity in Bi2201 resides in a single CuO$_2$ layer, resulting in
larger fluctuations effectively reducing $T_c$. The much wider temperature range
of the pseudogap phase in Bi2201 (see Sec.~\ref{sect_pseudogap}) could be a
direct consequence of such fluctuations. The finite zero-bias conductance and
the broad peaks at the gap edges remain open issues.

\begin{figure}[tb]
\includegraphics[width=6cm]{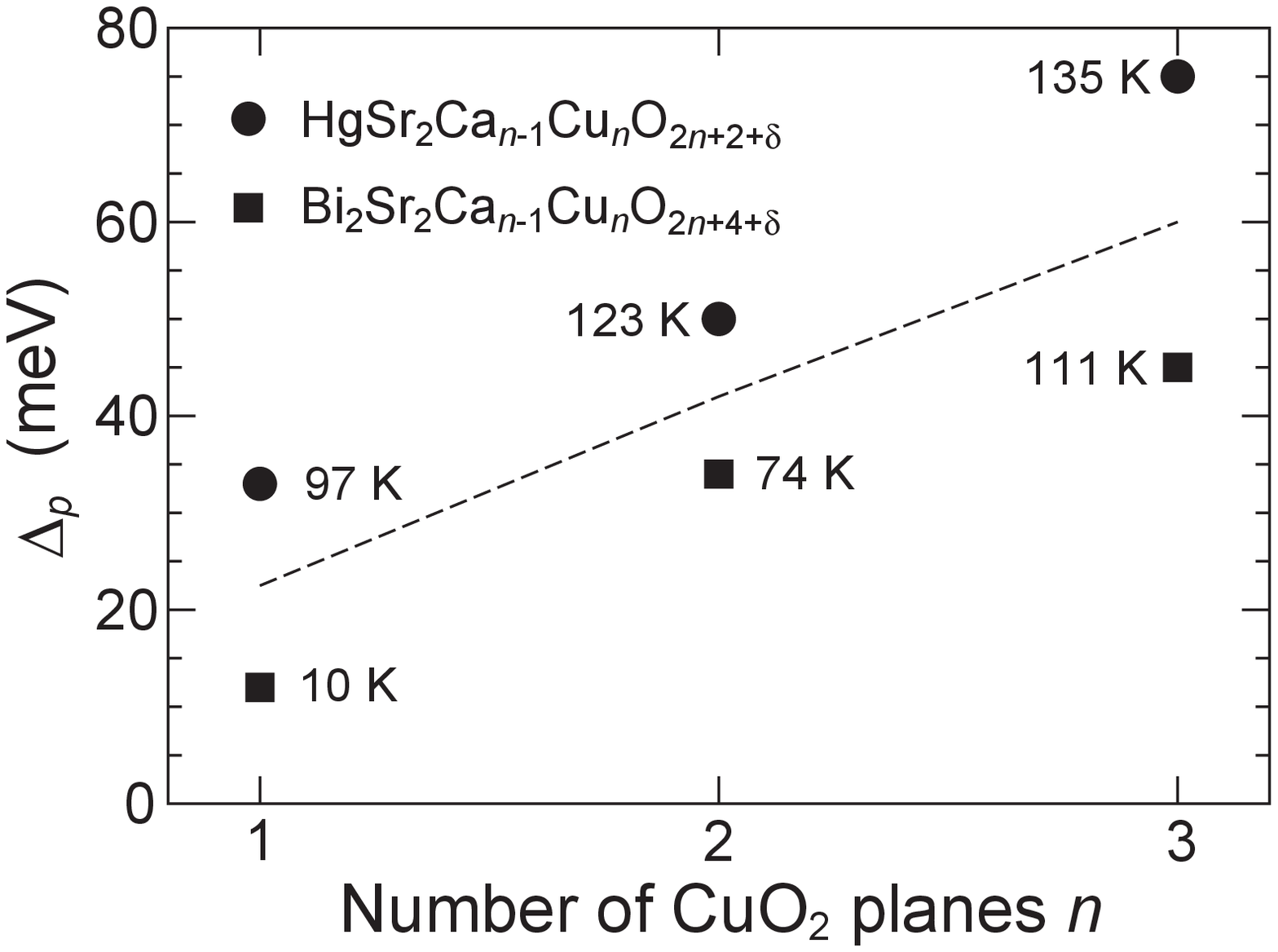}
\caption{\label{fig_fig5_gap-planes}
Gap amplitude $\Delta_p$ as a function of the number $n$ of CuO$_2$ planes in
the bismuth and mercury series. Optimally-doped
HgSr$_2$Ca$_{n-1}$Cu$_n$O$_{2n+2+\delta}$ (circles); gap values from
\citet{Wei-1998}. Slightly overdoped
Bi$_2$Sr$_2$Ca$_{n-1}$Cu$_n$O$_{2n+4+\delta}$ (squares); gap values for $n=1$
from \citet{Kugler-2001}, for $n=2$ from \citet{Renner-1998a}, and for $n=3$
from \citet{Kugler-2006}. Each compound's $T_c$ is indicated next to the symbol.
The dashed line is a guide to the eye.
}
\end{figure}

The gap amplitude of hole-doped HTS cuprates series appears to increase with the
number $n$ of CuO$_2$ planes per unit cell, at least up to $n=3$ for which
tunneling data exist. This dependence is depicted in
Fig.~\ref{fig_fig5_gap-planes} for two series with very different gap
amplitudes: the bismuth series Bi$_2$Sr$_2$Ca$_{n-1}$Cu$_n$O$_{2n+4+\delta}$ and
the mercury series HgSr$_2$Ca$_{n-1}$Cu$_n$O$_{2n+2+\delta}$. The increase of
the gap with $n$ in the Bi-compounds is illustrated graphically in
Figs.~\ref{fig_fig5_gaps}d, b, and f showing the tunneling spectra for $n=1$, 2,
and 3, respectively. (Note that the dependence on $n$ appears accentuated in
Fig.~\ref{fig_fig5_gaps} due to the spread of doping levels of that Bi-series.)
A similar behavior is observed for $T_c^{\text{max}}$: it increases with
increasing $n$ up to $n=3$ ($T_c^{\text{max}}$ decreases upon further increasing
$n$), and can be very different in distinct homogeneous series. Understanding
the dependence of the gap and of $T_c^{\text{max}}$ on the number of CuO$_2$
planes may turn out to be one important key to unravel the mechanism of HTS
\cite[see \textit{e.g.}][]{Pavarini-2001, Chen-2004d}.

\subsubsection{Y-compounds}
\label{sect_Y_compounds}

A characteristic SIN $dI/dV$ spectrum measured on an as grown, fully oxygenated
YBa$_2$Cu$_3$O$_7$ (Y123) single crystal ($T_c=91$~K) is shown in
Fig.~\ref{fig_fig5_gaps}c. The spectrum reveals a number of remarkable
differences compared to Bi2212 with similar $T_c$: (i) a smaller reduced gap
($\sim4.9$ in the sample shown here), (ii) a large finite conductance at $V=0$,
and (iii) multiple broad coherence peaks at the gap edges \cite{Miller-1993,
Maggio-Aprile-1995}. The main coherence peaks define a gap of $\sim20\pm2$~meV,
much closer to the value expected for a BCS $d$-wave superconductor than the gap
in the Bi-based cuprates discussed above. Additional structures of interest are
weak shoulders flanking the main peaks at higher energy, and two weak features
often developing below the gap at about $\pm6$~meV.

The multiple peak structure does not correspond to the simple $d$-wave
expectation. \citet{Tachiki-1990} and \citet{Miller-1993} ascribed the features
inside the gap near $\pm6$~meV to weak proximity-induced superconductivity in
the BaO and CuO-chain planes. \citet{Yeh-2001} identified the peaks at
$\pm20$~meV and $\pm6$~meV with a mixed $(d_{x^2-y^2}+s)$-wave gap in the
CuO$_2$ plane. In this scenario, two sets of conductance peaks are indeed
expected, one at $\pm(\Delta_d-\Delta_s)$ and the other at
$\pm(\Delta_d+\Delta_s)$, $\Delta_d$ and $\Delta_s$ being the maximum amplitude
of the $d$-wave and $s$-wave component of the gap, respectively. However, this
would imply a strong $s$-wave admixture in the Y123 ground state, which is
opposed by other experiments, like tricrystal experiments \cite{Tsuei-2000},
showing a pure $d$-wave. Thus the exact nature of the Y123 ground state and the
corresponding interpretation of the tunneling data needs further investigation.

Different ideas have been developed to understand the shoulders outside the main
coherence peaks. Such structures have been modeled in terms of the band
structure van-Hove singularity \cite{Tachiki-1990, Hoogenboom-2003b}. Similar,
though much weaker, shoulders have been measured on Bi2212 and were ascribed to
the onset of a strong coupling effect \cite{Zasadzinski-2003}. One aspect of
Y123, which certainly complicates the analysis of the spectra, is the presence
of CuO chains along with the CuO$_2$ planes, and a completely satisfactory
description has yet to be devised.

Spectra with a line shape similar to those obtained on Y123 were measured on
La$_{1.84}$Sr$_{0.16}$CuO$_4$ (LSCO) by \citet{Kato-2003}. Like Y123, their
background conductance is increasing with energy, the DOS at $E_{\text{F}}$ does
not vanish, and the reduced gap amounts to $\sim4.8$. However, unlike Y123, they
do not show any multiple peak structure.

\subsubsection{Electron-doped HTS}
\label{sect_electron_HTS}

Over the past two decades, some consensus has emerged that the gap symmetry in
hole-doped HTS is $d_{x^2-y^2}$. In electron-doped HTS, a majority of tunneling
experiments point at a more conventional $s$-wave gap, although some do look
more consistent with a $d$-wave symmetry. This controversy in electron-doped HTS
is illustrated in Nd$_{1.85}$Ce$_{0.15}$CuO$_{4-y}$ (NCCO), one of the few
electron-doped HTS which has been investigated by STM. \citet{Kashiwaya-1998}
and \citet{Alff-1998} contend that NCCO is an $s$-wave superconductor. Their
conclusion is based on the absence of a zero-bias conductance peak (ZBCP) on
surfaces perpendicular to the nodal lines (see Sec.~\ref{sect_cross_section}),
and on the good fit of the spectra obtained with an anisotropic $s$-wave BCS
model. A different picture emerges from STM studies by \citet{Hayashi-1998} who
conclude that NCCO is a $d$-wave superconductor based on the ZBCP which appears
in some of their low-resistance tunnel junction spectra.

Magnetic and non-magnetic impurities provides an indirect probe of the gap
symmetry (see Sec.~\ref{sect_impurities}). Non-magnetic impurity scattering has
no effect on an $s$-wave superconductor, whereas it will cause pair breaking in
a $d$-wave superconductor. Magnetic impurities have the opposite effect and do
cause pair breaking in conventional $s$-wave BCS superconductors. The gap
measured on Sr$_{0.9}$La$_{0.1}$CuO$_2$ (SLCO, Fig.~\ref{fig_fig5_gaps}e), an
infinite layer electron-doped HTS, is in good agreement with an isotropic
$s$-wave BCS model \cite{Yeh-2002}. \citeauthor{Yeh-2002} report that doping
SLCO with nickel, a magnetic impurity, has a large effect on the line shape of
the LSCO spectra, whereas doping with Zn, a non-magnetic impurity, has very
little effect as expected for an $s$-wave superconductor. In Bi2212, a $d$-wave
hole-doped HTS, the opposite seems to happen, where zinc doping is most
detrimental to superconductivity \cite{Hudson-2001}---Note that in both
compounds, the impurity atom is substituting for the copper atom in the CuO$_2$
plane. Such studies of atomic scale impurities in HTS are discussed further in
Sec.~\ref{sect_impurities}.

\subsection{Dip-hump structure}
\label{sect_dip_hump}

\begin{figure}[tb]
\includegraphics[width=6.5cm]{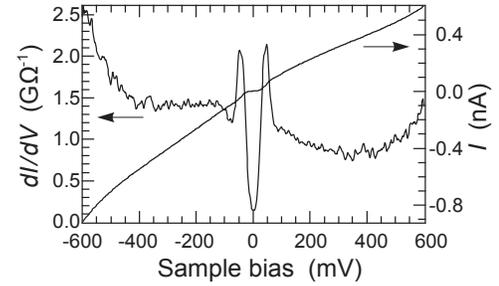}
\caption{\label{fig_fig5_largespec}
Large energy scale SIN tunneling characteristics of Bi2212 (Ch. Renner and \O.
Fischer, unpublished).
}
\end{figure}

Sec.~\ref{sect_low_T_spectro} was devoted to the region of the tunneling
conductance curves below the superconducting gap. Here, we extend the discussion
to include the conductance background at higher energy beyond the gap.
Figure~\ref{fig_fig5_largespec} shows a typical large energy range spectrum of
Bi2212 exhibiting a sizeable electron-hole asymmetry in the background
conductance \cite[see also][]{Sugita-2000}. A very similar background was
measured on Bi2223 \cite{Kugler-2006}. The increasing conductance at high energy
beyond $400$~meV may be a signature of the valence bands below $E_\text{F}$.

A remarkable dip structure develops outside the superconducting gap on top of a
smooth broad maximum in the background conductance near $E_\text{F}$. In early
STM experiments, this dip was mostly obscured by an increasing (linear or
parabolic) background conductance \citep[\textit{e.g.}][]{Liu-1994}. The dip
became clearly apparent in spectra obtained with the best STM junctions, as
defined in Sec.~\ref{sect_tip_sample_distance}. In Bi2212, the dip is
systematically seen at negative sample bias \cite{Renner-1994, Renner-1995}. A
symmetric dip is present at positive bias in some exceptional cases
\cite{DeWilde-1998, Hudson-1999}. The dip structure, often accompanied by a hump
at higher energy, is at the core of a heated debate. One key question is whether
it is a band structure or a strong coupling effect.

Band-structure effects would be a natural possibility to explain the dip. ARPES
spectra, which first showed the same dip feature beyond the gap
\cite{Dessau-1991, Hwu-1991}, have been interpreted along those lines. However,
this interpretation is not widely accepted. \citet{Zasadzinski-2003} exclude in
particular the possibility that the dip is due to a vHs or that it results from
the superconducting gap opening in a pseudogap background.

\begin{figure}[tb]
\includegraphics[width=8.6cm]{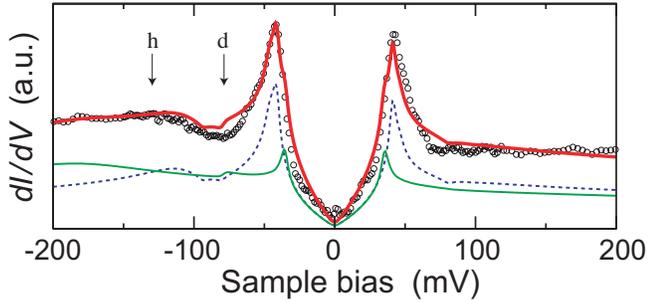}
\caption{\label{fig_fig5_fitspec}
Fit of the differential tunneling conductance measured on optimally-doped
Bi2212. The fit (red line) to be compared with the experimental data (circles)
is the sum of the contributions from anti-bonding (dotted line) and bonding
(green line) bands describing the CuO$_2$ bilayer splitting. Adapted from
\citet{Hoogenboom-2003b}.
}
\end{figure}

Strong coupling effects are another possibility to explain the dip. They are
known to have similar effects on tunneling spectra in conventional
superconductors. Tunnel junctions on lead, for example, reveal some fine
structure in the differential conductance above the gap, which was shown to
correspond to peaks in the phonon density of states $F(\omega)$. The fine
structure was directly linked to the electron-phonon mechanism of
superconductivity, and the electron-phonon spectral function $\alpha^2F(\omega)$
could be quantitatively obtained through inversion of the spectra using strong
coupling Eliashberg theory \cite{Schrieffer-1963, McMillan-1965}. This
eventuality has been addressed by \citet{Zasadzinski-2003} to fit a particular
symmetric Bi2212 spectrum normalized to an arbitrary state-conserving
normal-state background. Using a modified Eliashberg model to account for the
$d$-wave symmetry, \citet{Zasadzinski-2003} self consistently reproduce the
overall shape, position and magnitude of the dip as well as the superconducting
gap structure with a single narrow bosonic mode centered at $36.5$~meV.

In the strong coupling scenario, the dip could be due to a phonon mode (as in
low-temperature superconductors) or to a collective electronic mode. A candidate
electronic mode was observed by neutron scattering near $(\pi,\,\pi)$ at
$41$~meV in Y123 \cite{Fong-1995}, and subsequently in Bi2212 \cite{Fong-1999}.
This mode was determined to scale with $T_c$ \cite{He-2001}, suggesting a close
relation with the superconducting state. Specific models were developed to
describe the coupling of the electrons to this mode \cite{Eschrig-2000}.
\citet{Hoogenboom-2003a} used this model to calculate the spectra of Bi2212
using doping-dependent band structure parameters from ARPES, including the
bilayer splitting, and the electronic mode measured by neutrons. The result is
plotted as the red curve in Fig.~\ref{fig_fig5_fitspec}. Note the importance of
the bilayer splitting to obtain the proper background conductance intensity
relative to the peaks height. In this model the asymmetry of the dip with
respect to $E_{\text{F}}$ is not a consequence of the nature of the mode
(\textit{e.g.} phonon or spin wave) but is due to the presence of the vHs at
negative energy.

\subsection{Doping dependence of the superconducting gap}
\label{sect_doping_dependence}

\begin{figure}[t]
\includegraphics[width=6.5cm]{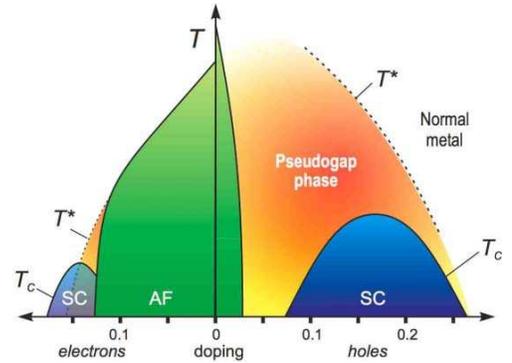}
\caption{\label{fig_fig5_phasediagram}
Schematic doping phase diagram of electron- and hole-doped high-$T_c$
superconductors showing, in particular, the superconducting (SC) and
antiferromagnetic (AF) phases.
}
\end{figure}

\begin{figure}[b]
\includegraphics[width=8.2cm]{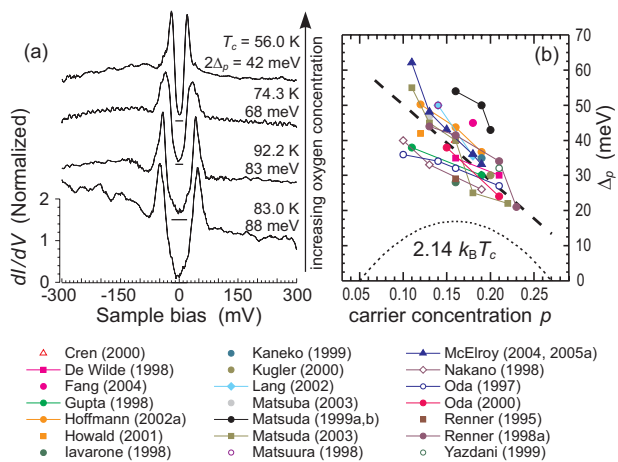}
\hphantom{
\cite{Cren-2000, DeWilde-1998, Fang-2004, Gupta-1998, Hoffman-2002, Howald-2001,
Iavarone-1998, Kaneko-1999, Kugler-2000, Lang-2002, Matsuda-1999a,
Matsuda-1999b, Matsuda-2003, Matsuura-1998, McElroy-2004a, McElroy-2005a,
Nakano-1998, Oda-1997, Oda-2000, Renner-1995, Renner-1998a, Yazdani-1999}
}
\vspace*{-2em}
\caption{\label{fig_fig5_doping}
(a) Doping dependence of the STM tunneling spectra measured on Bi2212
\cite{Renner-1998a}. (b) Compilation of the gap of Bi2212 measured by various
techniques in a wide doping range. The numeric values are listed in
Appendix~\ref{app_gap_tables}. The dashed line is a linear fit of the average gap
at a given doping. Note the absence of scaling between the gap and $T_c$.
}
\end{figure}

A generic doping phase diagram of HTS is shown in
Fig.~\ref{fig_fig5_phasediagram}. Undoped cuprates are antiferromagnetic (AF)
Mott insulators with one electron per copper site in the CuO$_2$ plane. Upon
adding either holes or electrons, they eventually become superconducting (SC)
and $T_c$ increases to reach a maximum at optimal doping before vanishing again
at higher hole or electron concentration.

A series of unusual properties characterize hole-doped cuprates, including a
pseudogap, non Fermi liquid behavior, and an unusual superconducting state.
These properties depend on doping, and are most anomalous in underdoped
compounds (\textit{i.e.} at doping below optimal).

In conventional BCS superconductors, $\Delta$ is proportional to the
superconducting transition temperature $T_c$, and the reduced gap
$2\Delta/k_{\text{B}}T_c$ lies in the range $3.5$ to $4.3$. A surprising result
was that for most HTS compounds, STM tunneling spectroscopy and related methods
(planar, break, and point-contact tunnel junctions) as well as ARPES
spectroscopy, found a very large and doping-dependent reduced gap
(Fig.~\ref{fig_fig5_doping}). This result reflects the absence of scaling of the
superconducting gap with $T_c$ (Fig.~\ref{fig_fig5_doping}b). In overdoped
Bi2212, the gap decreases with decreasing $T_c$, as expected. But in underdoped
Bi2212, the gap increases with decreasing $T_c$ \cite{Renner-1998a,
Miyakawa-1999}. The reduced gap defined in the standard way ranges from 4.3 to
values as high as 28 \cite{Kugler-2001}. This wide range suggests that $T_c$,
the temperature where phase coherence and a macroscopic superconducting state
are established, is not the appropriate energy scale to describe the gap.
Replacing $T_c$ by $T^*$, the temperature where the pseudogap appears, enables
to recover a constant reduced gap with a more conventional value close to $4.3$
\cite{Kugler-2001, Dipasupil-2002, Nakano-2003} expected for a $d$-wave BCS
superconductor (see Sec.~\ref{sect_pseudogap}). Other spectroscopic features
also show a characteristic doping dependence in Bi2212. The dip-hump feature
discussed in Sec.~\ref{sect_dip_hump} becomes weaker and shifts closer to the
coherence peaks in underdoped \emph{and} overdoped compounds
\cite{Zasadzinski-2001}. The background conductance shows a systematic
evolution, becoming gradually more asymmetric with decreasing hole concentration
(Fig.~\ref{fig_fig5_doping}a). Likewise, the coherence peaks are most asymmetric
in highly underdoped crystals \cite{Miyakawa-1999}. The change in the background
conductance can be partly explained in terms of the vHs shifting away from
$E_{\text{F}}$ in underdoped Bi2212 \cite{Hoogenboom-2003b}. However, the more
pronounced asymmetry observed in heavily underdoped samples has been ascribed to
strong-correlations effects \cite{Rantner-2000, Anderson-2006}. The smeared
coherence peaks at the gap edges in underdoped Bi2212 may be a consequence of
the very short quasiparticle lifetime due to the proximity of a Mott insulator.

The doping dependence of the gap of Y123 is much less documented than Bi2212.
\citet{Deutscher-1999} and \citet{Yeh-2001} report that unlike in Bi2212, the
gap in Y123 does scale with $T_c$ as a function of doping, and the reduced gap
does not reach as high values as in Bi2212. $2\Delta_p/k_{\text{B}}T_c$ remains
much closer to conventional superconductors, ranging from 7.8 (underdoped) to
4.5 (overdoped). \citet{Kohen-2000} and \citet{Yeh-2001} report another
interesting observation in Y123, namely that the symmetry of the gap is changing
from pure $d$-wave to $(d+s)$-wave or $s$-wave above optimal doping (see
discussion in Sec.~\ref{sect_Y_compounds}). \citet{Vobornik-1999} claimed to
observe a similar crossover from a $d$-wave to a less anisotropic gap in
overdoped Bi2212 in ARPES experiments. However, this result is in contradiction
with STM spectroscopy, which shows the same linear energy dependence of the
conductance near $E_{\text{F}}$ throughout the entire superconducting range
\cite{Renner-1998a, Miyakawa-1999}.

\subsection{Spatial gap (in-)homogeneity}
\label{sect_spatial_homogeneity}

The question of spatial homogeneity of the superconducting gap has stirred up a
lot of interest. This ongoing excitement is driven by models where electronic
phase separation is at the core of the high superconducting transition
temperature \cite{Phillips-2003}. Two types of spatial inhomogeneities of the
local DOS emerge from STS studies of Bi2212: (i) spatial inhomogeneity of the
superconducting gap amplitude, and (ii) spatial inhomogeneity of the low-energy
DOS, deep inside the gap. The latter is seen as a $\sim4a_0\times4a_0$ periodic
electronic modulation, where $a_0$ is the unit-cell lattice parameter
\cite{Howald-2003, Hoffman-2002}. Recently, a similar electronically modulated
phase was also found to develop in the pseudogap state of Bi2212 above $T_c$
\cite{Vershinin-2004} as well as in superconducting and non-superconducting
Ca$_{2-x}$Na$_x$CuO$_2$Cl$_2$ \cite{Hanaguri-2004}. These DOS modulations appear
intimately linked to the HTS state, and are the focus of
Sec.~\ref{sect_modulations}. Here, we shall discuss the notorious spatial
inhomogeneities of the gap amplitude observed in Bi2212.

\begin{figure}[tb]
\includegraphics[width=8.6cm]{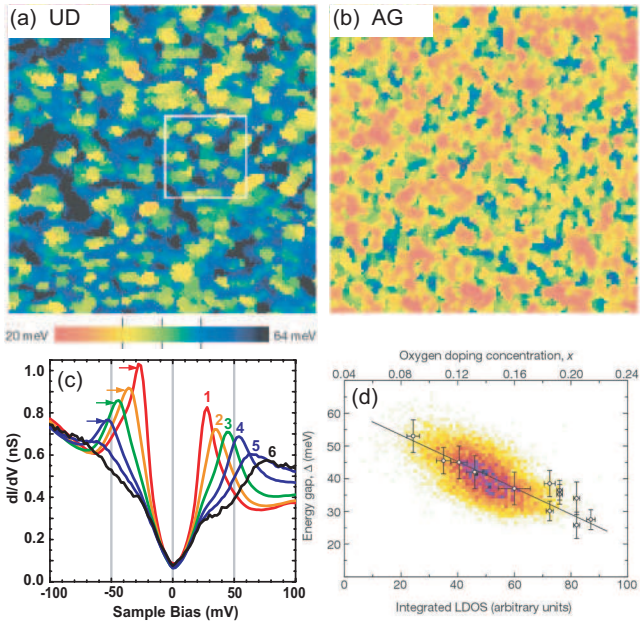}
\caption{\label{fig_fig5_inhomogene}
$56\times56$~nm$^2$ maps of the spatial gap distribution in (a) underdoped and
(b) as grown Bi2212 single crystals; adapted from \citet{Lang-2002}. (c) Range
of tunneling conductance spectra measured on inhomogeneous superconducting
Bi2212; from \citet{McElroy-2005a}. (d) Scatter plot of the superconducting gap
versus the integrated local DOS in optimally-doped Bi2212; from
\citet{Pan-2001}.
}
\end{figure}

Inhomogeneous superconductivity in Bi2212 was reported in a number of STM
studies \cite{Cren-2000, Howald-2001, Pan-2001, Lang-2002, Kinoda-2003}.
Real-space STS maps of the local superconducting gap shown in
Figs.~\ref{fig_fig5_inhomogene}a,b reveal a network of nanometer-scale islands,
on average $3$--$5$~nm wide, with different gap amplitudes.\footnote{Degraded or
contaminated surfaces also lead to spatial variations of the tunneling spectra,
but of a very different kind (see Sec.~\ref{sect_pseudogap_PGinDisorder}).}

\begin{figure}[tb]
\includegraphics[width=8.6cm]{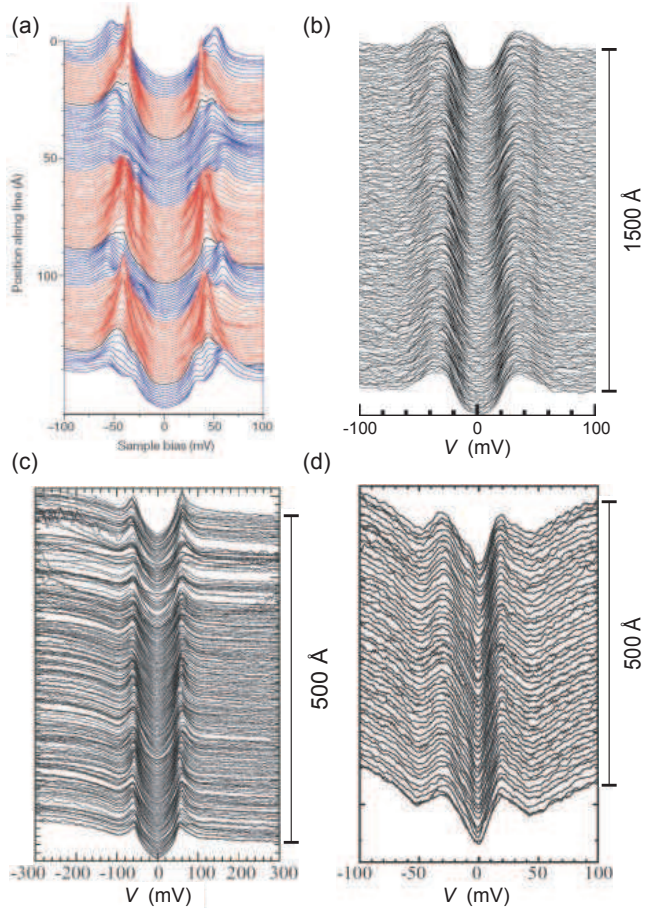}
\caption{\label{fig_fig5_traces}
Spatially resolved STM tunneling characteristics measured along different traces
on (a) Bi2212 \cite{Lang-2002}, (b) Bi2212 \cite[adapted from][]{Renner-1998a},
(c) Bi2223 \cite{Kugler-2006}, and (d) Y123 \cite{Maggio-Aprile-1996}. Note the
high degree of homogeneity in (b--d).
}
\end{figure}

The key question is to what extent this inhomogeneity is an intrinsic phenomenon
at the core of high temperature superconductivity? By \emph{intrinsic} we mean
an electronic effect, possibly related to the mechanism for HTS, taking place in
an otherwise homogeneous system. Electronic phase separation would be one
candidate phenomenon \cite{Phillips-2003}. This is in contrast to an
\emph{extrinsic} origin of the spectral inhomogeneities, by which we mean a
phenomenon unrelated to HTS, resulting from stoichiometric inhomogeneities such
as the distribution of atoms and dopants. \citet{Howald-2001} argued strongly in
favor of an intrinsic inhomogeneity. They pointed out that the boundaries
between the domains are always of the order of the coherence length and that
there was no obvious correlation with structural inhomogeneities of the sample.
Their conclusion was that this disorder represents a phase separation phenomenon
into ``good and bad superconducting regions''. Such an interpretation has also
been put forward by \citet{Pan-2001} and \citet{Lang-2002}. On the contrary,
\citet{Renner-1994a} noted the existence of such inhomogeneities and concluded
that they are related to chemical inhomogeneities. They further found that
homogeneous tunneling spectra can be obtained, and they demonstrated atomic
resolution under these conditions \cite{Renner-1995}. \citet{Hoogenboom-2003a}
showed that homogeneous samples could be obtained by adequate annealing, at
least in the overdoped case. They concluded that inhomogeneity is most likely
due to inhomogeneous oxygen distribution and that it is not an essential part of
the mechanism for HTS. Thus, in recent years, there has been a controversy about
the interpretation of these inhomogeneities. Although this question is not
definitively settled, the authors of this review are of the opinion that there
is more and more experimental evidence for the extrinsic scenario. In the
following, we shall develop further arguments on which this statement is based.

The first piece of evidence comes from a series of experiments showing
reproducible spectra with very little spread in gap amplitude in Bi2212, Bi2223,
Y123 (Fig.~\ref{fig_fig5_traces}b--d) and NdBa$_2$Cu$_3$O$_{7-\delta}$
\cite{Nishiyama-2002} single crystals. These measurements were made in
conditions where concurrent atomic resolution and/or STS vortex-core imaging
could be achieved, indicating that, if present, changes in the spectral
signature, even over sub-nanometer distances, could be detected. Such distances
are much shorter than the typical size of the domains seen in inhomogeneous
samples (Figs.~\ref{fig_fig5_inhomogene}a,b), and the observed homogeneity can
not be ascribed to a broad STM tip.

A second piece of evidence showing that inhomogeneous superconductivity is not
essential for a high $T_c$ comes from STM studies by \citet{Matsuba-2003}. They
observe that the spread in gap amplitude can be very different in distinct
regions on a given Bi2212 surface, although the average gap value, and possibly
$T_c$, are the same. In another systematic study of slightly overdoped Bi2212,
\citet{Hoogenboom-2003a} demonstrated that spatial gap inhomogeneities as shown
in Figs.~\ref{fig_fig5_inhomogene}a,b and \ref{fig_fig5_traces}a were intimately
associated with a broad superconducting transition, as measured by
ac-susceptibility on the same samples. Conversely, single crystals with narrow
superconducting transitions ($\Delta T_c<0.5$~K) yield very homogeneous
tunneling spectroscopy as shown in Fig.~\ref{fig_fig5_traces}b. This
correlation between a wide gap distribution and the ac-susceptibility transition
width was also observed in Bi2223 \cite{Kugler-2006}, and suggests that the
nanometer length scale gap inhomogeneities stem from \emph{extrinsic}
stoichiometric disorder rather than \emph{intrinsic} electronic phase
separation.

A very recent study provides further compelling experimental evidence that the
gap inhomogeneity is indeed an \emph{extrinsic} stoichiometric effect, as
inferred from the above experiments. Combining high resolution topographic and
spectroscopic STM maps, \citet{McElroy-2005} captured the direct correlation
between the gap amplitude and oxygen impurity distributions in real space. This
result is in agreement with the spread of spectra measured on inhomogeneous
samples \cite{McElroy-2005a}, whose line shapes vary from typical for
optimally-doped samples (Fig.~\ref{fig_fig5_inhomogene}c, spectrum 1) to typical
for heavily underdoped samples (Fig.~\ref{fig_fig5_inhomogene}c, spectrum 6).
Note however that the detailed relation between the local gap and the local
oxygen impurity distribution remains unclear and requires further
investigations.

In summary, the spatial inhomogeneities in the gap amplitude appear to be driven
by a property which transition metal perovskite oxides share with all oxides,
namely their propensity to high concentration of defects, especially oxygen
inhomogeneities. Note that lead impurities and disorder in the Sr and Ca layers
were found to have no effect on the gap inhomogeneity \cite{Kinoda-2003}. These
results all point at oxygen disorder as the main cause for the inhomogeneities
found by spectroscopy. Generally, as-grown samples tend to be inhomogeneous and
this tendency increases as doping is reduced. However, these inhomogeneities are
not consistent with simple charge segregation in a perfect lattice. Special
post-synthesis treatments allow to make more homogeneous samples with equally
high $T_c$, as has been clearly demonstrated in overdoped Bi2212
\cite{Hoogenboom-2003a}. Most importantly, the experimental evidence is that an
inhomogeneous gap distribution is not a prerequisite to the occurrence of high
temperature superconductivity, since there is no correlation between such a
distribution and the value of $T_c$. However, note that the oxygen concentration
and distribution are probably not the only differences between samples from
different sources and batches. Therefore in a detailed comparison between
samples other parameters have to be taken into account as well.

\subsection{STM on impurities}
\label{sect_impurities}

Impurities in superconductors have been a subject of research since the BCS
theory was established. Whereas non-magnetic impurities were found to have
little or no effect on $T_c$, magnetic impurities turned out to suppress
superconductivity. Indeed, for $s$-wave superconductors the Anderson theorem
\cite{Anderson-1959} states that non-magnetic impurity scattering does not reduce
$T_c$, whereas \citet{Abrikosov-1960} showed how magnetic impurities reduce
$T_c$ to zero through exchange scattering at impurity concentrations of the
order of one percent. Later on \citet{Yu-1965}, \citet{Shiba-1968}, and
\citet{Rusinov-1969} independently found that a magnetic impurity can induce
quasiparticle sub-gap bound states. The first direct observation of such
behavior by STM was reported by \citet{Yazdani-1997} on a Nb surface with Mn, Gd
or Ag adatoms. Ag, believed to be non-magnetic, has no effect on the
superconducting $s$-wave state. However when Mn or Gd are deposited on the
surface of Nb the zero-bias conductance is enhanced when tunneling through
magnetic Mn and Gd. The spatial extension of the impurity effect is about
10~\AA\ and the resulting spectra were found to be asymmetric.

In a $d$-wave superconductor Anderson's theorem does not hold and it is expected
that non-magnetic impurities have a profound effect on the superconducting
properties. This has been investigated by several groups \cite{Lee-1993,
Byers-1993, Balatsky-1995}. These theories predict local signatures of $d$-wave
pairing, and variations in the LDOS around the impurities. A review of these
theories has been given by \citet{Balatsky-2006}.

First attempts to study impurities in HTS by STM were carried out by
\citet{Yazdani-1999} and by \citet{Hudson-1999}. The former studied Bi2212
surfaces onto which nanoscale Au dots were deposited from the STM gold tip. It
was found that strong zero-bias enhancements occurred in $dI/dV$ when the tip
was placed just above these particles, demonstrating the existence of a strong
resonance as expected for a $d$-wave superconductor. In the latter paper an
undoped Bi2212 crystal was studied. Their approach was to make zero-bias
conductance maps similar to \citeauthor{Yazdani-1999}. These maps showed
nanoscale areas with an unusually high zero-bias conductance. Analyzing these
unveiled a zero-bias anomaly corresponding to what would be expected from
scattering of a $d$-wave superconductor on an impurity. However, the nature of
these impurities was not identified in this work.

\begin{figure}
\includegraphics[width=8cm]{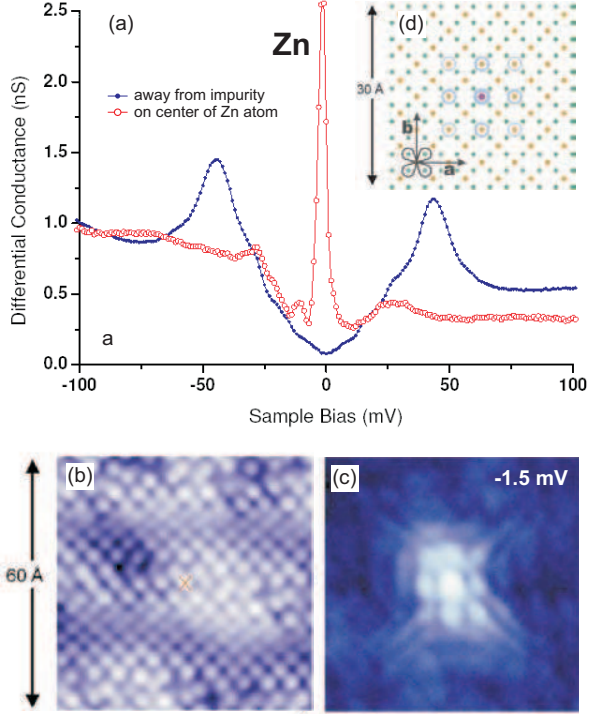}
\caption{\label{fig_Pan-2000}
Spectroscopy above a Zn atom in
Bi$_2$Sr$_2$Ca(Cu$_{1-x}$Zn$_x$)$_2$O$_{8+\delta}$ with $x=0.6$\% and
$T_c=84$~K. (a) Spectra taken at the center and away from the Zn impurity, (b)
$60\times60$~\AA$^2$ topography and (c) conductance map at $-1.5$~meV
(logarithmic color scale). (d) $30\times30$~\AA$^2$ representation of the square
CuO$_2$ lattice (Cu in orange, O in green) lying below the BiO surface layer.
The position of the Zn atom is indicated in red and the conductance maxima
observed in (c) in grey. Adapted from \citet{Pan-2000a} and
\citet{Balatsky-2006}.
}
\end{figure}

\begin{figure}
\includegraphics[width=8cm]{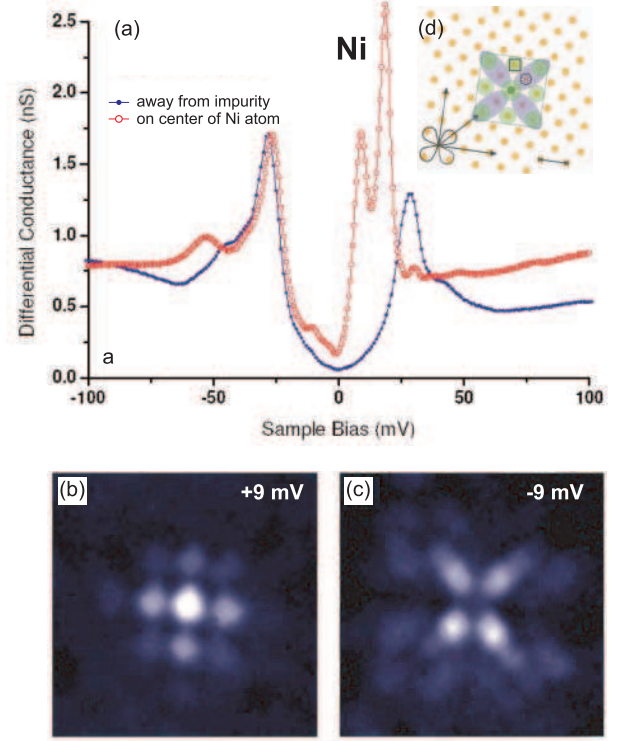}
\caption{\label{fig_Hudson-2001}
Spectroscopy above a Ni atom in
Bi$_2$Sr$_2$Ca(Cu$_{1-x}$Ni$_x$)$_2$O$_{8+\delta}$ with $x\approx0.2$\% and
$T_c=85$~K. (a) Spectra taken at the center and away from the Ni impurity. (b,
c) $35\times35$~\AA$^2$ conductance maps at positive ($+9$~mV) and negative
($-9$~mV) bias. (d) Position of the Ni atom (green solid circle) with respect to
the Cu atoms (solid orange circles) in the invisible CuO$_2$ plane, and
schematics (purple) of the $d_{x^2-y^2}$ order parameter. Adapted from
\citet{Hudson-2001} and \citet{Balatsky-2006}.
}
\end{figure}

Following this \citet{Pan-2000a} investigated Zn-doped Bi2212. They found an
intense quasiparticle scattering resonance at $-1.5$~meV at many sites on the
surface of the sample. Since Zn substitutes for Cu, two layers below the
surface, it is not seen in topography. However the effect of the resonant
scattering can be seen at low energies. The density of such sites corresponds to
about one third of the nominal concentration of Zn atoms and it is therefore
reasonable to assume that the strong zero-bias sites reflect the effect of Zn
doping. Fig.~\ref{fig_Pan-2000}a shows the spectra away from the impurity and on
top of it, the latter displaying a strong zero bias anomaly coincident with a
strong suppression of superconductivity within a 15~\AA\ radius of the
scattering sites. Imaging of the spatial dependence of the quasiparticle density
of states in the vicinity of the impurity atoms directly reveals the four-fold
symmetric quasiparticle `cloud' aligned with the nodes of the $d$-wave
superconducting gap (Figs.~\ref{fig_Pan-2000}b--d). Whether this structure
reflects the symmetry of the superconducting state or the symmetry of the atomic
lattice remains to be determined.

Particularly interesting in this context is the comparison of the effects of
magnetic and non-magnetic impurities substituted on equivalent Cu sites. In
comparison to Zn impurities, magnetic Ni impurities lead to a very different
behavior. This was studied by \citet{Hudson-2001}. At each Ni site they observe
two $d$-wave impurity states of apparently opposite spin polarization, whose
existence indicates that Ni retains a magnetic moment in the superconducting
state (Fig.~\ref{fig_Hudson-2001}a). However, analysis of the impurity-state
energies shows that quasiparticle scattering at Ni is predominantly
non-magnetic. The spatial distribution of the conductance maxima around the
impurity is strongly asymmetric with respect to the Fermi level
(Figs.~\ref{fig_Hudson-2001}b and c). Furthermore, they show that the
superconducting energy gap is unimpaired at Ni. This is in strong contrast to
the effects of non-magnetic Zn impurities, which locally destroy
superconductivity.

Motivated by the STM observations, several microscopic calculations of the
electronic structure near Zn or Ni impurities have been reported. Most of these
calculations are based on the BCS $d$-wave theory, and they differ mainly by the
model employed for the impurity, and the treatment of the superconducting order
parameter modification around the impurity site. In short, impurity potentials
in a BCS $d$-wave state induce resonances, either below or above the Fermi
energy depending upon the sign of the potential. In the unitary limit the
resonance approaches the Fermi energy. The measured LDOS around Zn impurities,
however, turns out to be more difficult to understand in this class of models.
Indeed for strong point-like impurities the spectral weight is expelled from the
impurity to the neighboring sites, and the LDOS at the impurity site is nearly
zero, in contrast to experiment. Various solutions to this problem have been put
forward. \citet{Zhu-2000}, \citet{Zhu-2001d}, and \citet{Martin-2002} invoked an
anisotropic tunneling matrix element, which would filter the tunnel electrons in
such a way that the current measured on top of the impurity site in effect
originates from the neighboring sites. Other authors claimed that correlation
effects and the formation of a local moment at the Zn site are important
\cite{Polkovnikov-2001}. Recently \citet{Tang-2004} were able to give a fairly
complete account of the experimental Zn impurity LDOS by assuming that the Zn
potential has a finite spatial extension, and by including a modification of the
pairing potential and hopping energies in the neighborhood of the Zn site. Which
one of these interpretations of the STM data is the right one is still unclear.

\subsection{Cross-sectional tunneling}
\label{sect_cross_section}

Most published STM experiments were carried out with the STM tip positioned
perpendicular to the perovskite $(001)$ surface. In this $c$-axis tunneling
configuration, tunneling samples an angular average over the $ab$-plane density
of states (see Sec.~\ref{sect_theory}). Here, we briefly review the few attempts
to do cross-sectional tunneling, \textit{i.e.} tunneling into an arbitrary plane
parallel to the $[001]$ direction with the STM tip perpendicular to the $[001]$
direction. This geometry is attractive for several reasons. It should enable a
direct probe of the alternate stacking of insulating and superconducting layers
along $[001]$. It further should allow to probe the $d_{x^2-y^2}$ symmetry of
the gap in the CuO$_2$ sheets by measuring the gap amplitude as a function of
the $(hk0)$ surface orientation and by detecting the zero-bias conductance peak
(ZBCP) expected on $(110)$ and $(\bar{1}10)$ surfaces.\footnote{The main
theoretical expectations for the electronic structure of superconductors near
surfaces or interfaces have been reviewed by \citet{Lofwander-2001} and
\citet{Deutscher-2005}.}

Cross-sectional tunneling is a very challenging endeavor, primarily due to the
difficulty in preparing a suitable surface for STM. The first experiments were
performed on Bi2212, resulting in topographic images with atomic scale
resolution of the $(100)$ surface \cite{Hasegawa-1990, Hasegawa-1991b}.
\citeauthor{Hasegawa-1991b} also found spectroscopic signatures of the alternate
stacking along $[001]$ of gapped BiO layers and metallic CuO$_2$ layers.

\begin{figure}[tb]
\includegraphics[width=5.5cm]{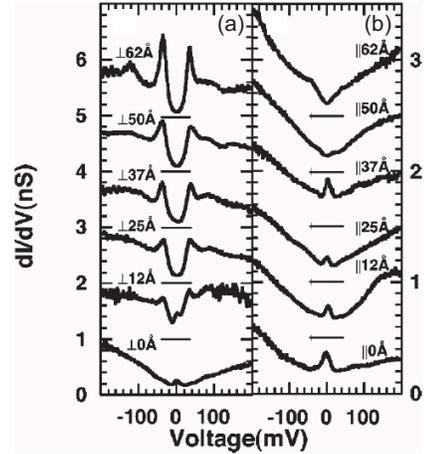}
\caption{\label{fig_fig5_bi2212stepedge}
STM tunneling spectra measured relative to a $[\bar{1}10]$ step edge
($45^{\circ}$ boundary) on Bi2212 at $4.2$~K. (a) Differential conductance
spectra measured at increasing distances in the direction perpendicular to the
step edge. (b) Differential conductance spectra measured along the step edge.
The ZBCP is only measured for tunnel junctions close to the step edge. From
\citet{Misra-2002b}.
}
\end{figure}

The potential to gather experimental evidence of the $d_{x^2-y^2}$ symmetry of
the superconducting gap has compelled several groups to attempt cross-sectional
tunneling experiments. \citet{Suzuki-1999} cut thin Bi2212 single crystals along
different $(hk0)$ planes and probed the exposed surfaces by STM. The $(110)$
surface revealed a conductance peak at $E_{\text{F}}$. All other $(hk0)$
surfaces were gapped, with the largest gap on the $(100)$ surface. This is
precisely what is expected for a $d_{x^2-y^2}$ gap in the $(001)$ plane.
\citet{Kane-1996} did a similar experiment, tunneling into $(hk0)$ edges of a
Bi2212 single crystal using a wire aligned parallel to the $c$-axis as the
``tip'' electrode. These experiments yield a very different result: all surfaces
show a gapped spectrum (no ZBCP), and the largest gap is measured on $(110)$
surfaces. This angular dependence of the gap is suggesting $d_{xy}$ symmetry.
However, the absence of any ZBCP makes this experiment inconclusive.

\citet{Misra-2002b} circumvent the difficulty of preparing cross-sectional
surfaces by measuring tunneling characteristics along a $[\bar{1}10]$ oriented
step edge of a Bi2212 thin film. The step should give rise to a ZBCP in a
superconductor with a $d_{x^2-y^2}$ gap in the $(001)$ plane \cite{Hu-1994}.
Indeed, they do find a clear signature of a conductance peak at $E_{\text{F}}$
in the vicinity of the step (Fig.~\ref{fig_fig5_bi2212stepedge}). Another
interesting result reported by \citet{Misra-2002b} is the absence of a ZBCP near
a $90^{\circ}$ twin boundary in Bi2212, indicating that such crystallographic
defects are not a strong perturbation for the superconducting state---unlike in
Y123, where STM measurements in a magnetic field show that they strongly perturb
the superconducting condensate \cite{Maggio-Aprile-1997}.

\begin{figure}[tb]
\includegraphics[width=8.6cm]{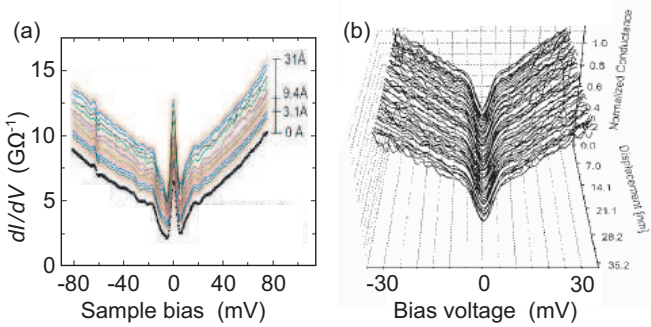}
\caption{\label{fig_fig5_zbcp}
Spatially resolved cross-sectional STM tunneling spectroscopy at $4.2$~K. (a)
$31$~nm trace (running parallel to $[001]$) on Y123 showing the presence of a
ZBCP on the $(110)$ surface of a hole-doped HTS; adapted from \citet{Yeh-2001}.
(b) $35$~nm trace on NCCO showing the systematic absence of the ZBCP on the
$(110)$ surface of an electron-doped HTS; adapted from \citet{Kashiwaya-1998}.
}
\end{figure}

In Y123, \citet{Yeh-2001} were able to measure the ZBCP as a function of
position along the $(110)$ surface of an optimally-doped single crystal
(Fig.~\ref{fig_fig5_zbcp}a). The same experiment along the $(100)$ surface
showed a large gap at $E_{\text{F}}$, in agreement with $d_{x^2-y^2}$ symmetry.
Y123 can be grown as $[110]$ oriented thin films, thus offering an alternative
approach to measuring the ZBCP. \citet{Yeh-2001} measured the expected ZBCP on
such films, but could not achieve atomic-resolution imaging. Conversely,
\citet{Nantoh-1995} achieved atomic-resolution imaging of $[110]$ oriented thin
films, but failed to observe a ZBCP.

\citet{Tanaka-1995} measured spatially homogeneous and tip-to-sample distance
independent tunneling spectra on the $(100)$ surface of
La$_{1.85}$Sr$_{0.15}$CuO$_4$ (La214). However, the spectra show a ZBCP, which
is not consistent with $d_{x^2-y^2}$ symmetry, but indicates $d_{xy}$ symmetry.
\citet{Tanaka-1994} also found a $d_{xy}$ gap by cross-sectional spectroscopy in
Y123. Introducing momentum-dependent tunneling currents to take the shape of the
electron wave function of the surface oxygen atoms into account for the
evaluation of the tunneling matrix elements, \citet{Tanaka-1995} argue they can
reconcile their data with a $d_{x^2-y^2}$ symmetry of the gap.

The ZBCP appears to be a characteristic property of $(110)$ and $(\bar{1}10)$
surfaces of hole-doped HTS, in agreement with the $d$-wave symmetry of the gap.
Most STM experiments favor $s$-wave superconductivity in electron-doped HTS. One
argument is the absence of a ZBCP on $(110)$ surfaces \cite{Kashiwaya-1998,
Alff-1998} as shown for Nd$_{1.85}$Ce$_{0.15}$CuO$_{4-y}$ (NCCO) in
Fig.~\ref{fig_fig5_zbcp}b. \citet{Hayashi-1998} measure a ZBCP on the $(100)$
surface of NCCO, suggesting a $d_{xy}$ symmetry. However, this result seems less
systematic, as the ZBCP is only observed in some of their low-resistance
junctions. The relation of these results to other experiments showing $d$-wave
symmetry needs further clarification.

The ZBCP has been more extensively investigated using planar and point contact
tunnel junctions, including the influence of disorder and magnetic field
\cite[see \textit{e.g.}][]{Aprili-1998, Aprili-1999}. A very comprehensive
discussion of probing the ZBCP by tunneling can be found in two recent reviews
by \citet{Kashiwaya-2000} and \citet{Deutscher-2005}.

\subsection{Summary}
\label{sect_summary-gap}

The very unusual nature of tunneling spectra measured on HTS was recognized from
the first successful tunneling experiments, and caught immediate attention. Most
striking were the unconventional line shape in hole-doped cuprates, with a
finite density of states in the gap, and the large energy scale of the
characteristic spectral features. The low-energy line shape has since been
explained in terms of the $d$-wave symmetry of the gap. Higher-energy spectral
features above the gap are still matter of discussion---although some very good
theoretical fits have been worked out---in particular the asymmetric and doping
dependent background and dip-hump.

The energy scale of the gap feature in HTS is way off the expectations of the
conventional theory for superconductivity. This is in particular the case for
the bismuth compounds (Bi2201, Bi2212, Bi2223). In these materials the gap
depends on the doping and number of CuO$_2$ planes in the unit cell, and grows
with reducing hole concentration and increasing number of planes. This leads in
certain cases to exceptionally high values of $2\Delta_p/k_{\text{B}}T_c$. This
surprising behavior has to be viewed in relation to the properties of the
pseudogap, to be discussed in the next section.

Spatial inhomogeneities of the superconducting gap have stirred up considerable
interest, fuelled in part by models where phase separation plays a key role for
high-temperature superconductivity. However, experimental evidence strongly
suggests that such inhomogeneities are an extrinsic effect, stemming from
stoichiometric variations. Most importantly, they are correlated with the
superconducting transition width, but not with the maximum value of the bulk
transition temperature. Magnetic and non-magnetic atomic impurities
intentionally introduced into the CuO$_2$ planes result in a particular kind of
local inhomogeneity, providing an original local test of the pairing state.

\section{The pseudogap}
\label{sect_pseudogap}

Many experimental techniques provided evidence for an unconventional normal
state common to all HTS and characterized by the opening of a gap in the
electronic excitation spectrum at a temperature $T^*$ above the critical
temperature $T_c$, the so-called `pseudogap' (PG). This observation initiated an
intense debate about its origin, since the answer to this question may turn out
to be essential for the understanding of high-$T_c$ superconductivity \cite[for
a review see][]{Timusk-1999}.

The various theoretical interpretations of the pseudogap phenomenon may be
roughly classified in two groups: (i) The pseudogap is the manifestation of some
order, static or fluctuating, generally of magnetic origin, but unrelated to
and/or in competition with the superconducting order.\footnote{Most of these
theories view the superconducting state of the cuprates as a consequence of
doping a Mott insulator. \citet{Lee-2004} give a comprehensive review of this
type of approach. In the resonating valence bond (RVB) theory
\citet{Anderson-1987} argued that the ground state of a doped antiferromagnet
would be a liquid of singlet spin pairs. In this framework the pseudogap appears
as the spin gap associated with the breaking of the RVB singlets
\cite{Paramekanti-2001, Anderson-2004c}. Several authors have considered the
possibility that doping the cuprates has the effect of weakening some order
present in the parent compound. The pseudogap would thus be a consequence of
this order which progressively disappears upon doping as superconductivity sets
in, and would vanish at a quantum critical point (QCP) presumably below the
superconducting dome. This (possibly dynamical) order could be the AFM order
itself \cite{Kampf-1990}, a state of microscopic orbital currents known as
staggered-flux state \cite{Affleck-1988} or $d$-density wave (DDW) state
\cite{Schulz-1989, Wen-1996, Varma-1997, Chakravarty-2001}, a stripe state in
which the doped holes order into one-dimensional patterns (\citet{Zaanen-1989,
Emery-1990,White-1998}; the field was recently reviewed by \citet{Kivelson-2003}
and \citet{Carlson-2004}), a spin-density wave (SDW) and/or charge-density wave
(CDW) state \cite{Chubukov-1996, Varlamov-1997, Vojta-1999}. In some of these
theories the two orders appear as two faces of the same coin \cite{Zhang-1997},
or they turn out to be dual one to another \cite{Franz-2001b,Wang-2004}.} (ii)
The pseudogap is the precursor of the superconducting gap, and reflects pair
fluctuations above $T_c$.\footnote{The short coherence length of the cuprate
superconductors suggests that they lie somewhere between the BCS limit of very
large momentum-space pairs and the opposite case of small real-space pairs
undergoing a Bose-Einstein condensation (BEC) \cite[see the review
of][]{Chen-2005}. In this picture the pseudogap is the binding energy of the
preformed pairs above $T_c$ \cite{Geshkenbein-1997}. The BCS-BEC crossover
scenario has been studied within the boson-fermion model \cite{Ranninger-1995}
and attractive Hubbard model \cite{Moreo-1992, Randeria-1992, Micnas-1995}.
Other approaches emphasize the weak phase stiffness in the underdoped cuprates
resulting from the low superfluid density, which would lead to a suppression of
$T_c$ by phase fluctuations \cite{Emery-1995, Schmalian-1997, Franz-1998a,
Giovannini-2001, Eckl-2002b, Lammert-2001}.} At present, a definitive answer is
still lacking, although the STS results favor the latter, as we shall see, or
more generally a situation where both phenomena have the same origin.

Closely related to this fundamental issue is the understanding of the HTS doping
phase diagram, schematically shown in Fig.~\ref{fig_fig5_phasediagram}. How does
the pseudogap vary with doping, temperature, and magnetic field? How does the
pseudogap scale with $T_c$ and the superconducting gap? Is there a universal HTS
doping phase diagram?

In this section we review the key STS results related to the pseudogap under
different experimental conditions. In Sec.~\ref{sect_pseudogap_Tdep} we describe
the $T$-dependence of the LDOS of Bi2212, Bi2201, and Y123 and we analyze the
scaling relations between the pseudogap and the superconducting gap. In
Sec.~\ref{sect_pseudogap_LTPG} we address the observation of the pseudogap at
low temperatures inside vortex cores and on disordered surfaces. In
Sec.~\ref{sect_pseudogap_DPD} we discuss the doping phase diagram which results
from the tunneling experiments reviewed here and finally, in
Sec.~\ref{sect_pseudogap_summary}, we highlight the key results of this section.
The connection between the pseudogap and the recently reported experiments
revealing spatial modulations in the LDOS are discussed separately in
Sec.~\ref{sect_modulations}.

\subsection{Temperature dependence of the local DOS}
\label{sect_pseudogap_Tdep}

Already by the end of the eighties bulk techniques revealed an unconventional
normal state behavior above $T_c$, but it took about a decade to obtain
sufficient sample and surface quality to evidence the PG spectral signature
directly in the DOS using ARPES \cite{Ding-1996, Loeser-1996} and STS
\cite{Renner-1998a}.

Measuring the $T$-dependence of the LDOS by STM is a challenge. The difficulty
is to avoid the tip from shifting relative to the sample when the temperature is
changed, due to uncompensated thermal expansion coefficients in the experimental
setup. If the tip does shift, as is usually the case, variations in the $I(V)$
characteristics may not only be due to temperature, but also to the different
tunneling locations, especially if the spectroscopic properties are not
homogeneous over sufficiently large areas as discussed in
Sec.~\ref{sect_gap_spectroscopy}. Not surprisingly, there are only few relevant
$T$-dependent STM studies of HTS. The first successful STM experiments where
obtained on Bi2212 \cite{Renner-1998a, Matsuda-1999a}, followed by Y123
\cite{Maggio-Aprile-2000} and Bi2201 \cite{Kugler-2001}.

\subsubsection{The case of Bi2212}
\label{sect_pseudogap_TdepBi2212}

Historically, the first tunneling results revealing the pseudogap (PG) were
reported on Bi2212 by \citet{Tao-1997} on planar junctions and by
\citet{Renner-1998a} using an STM junction. These experiments showed
unambiguously that HTS do not follow what the BCS theory predicts, namely that
the superconducting gap closes at $T_c$. In this theory, indeed, both the
pairing and the phase coherence are lost simultaneously at $T_c$. The origin of
this behavior is illustrated by the reduced gap value
$2\Delta/k_{\text{B}}T_c=3.5$ for an $s$-wave BCS superconductor, telling that
at the mean field critical temperature $T_c$, thermal fluctuations are strong
enough to overcome the pairing energy $2\Delta$. In the case of UD Bi2212,
reduced gap values as high as 20 have been reported \cite{Miyakawa-1999},
providing a strong hint that the $T$-dependence of the superconducting gap of
HTS might be very different from the BCS mean-field behavior.

\begin{figure}[tb]
\includegraphics[width=8.5cm]{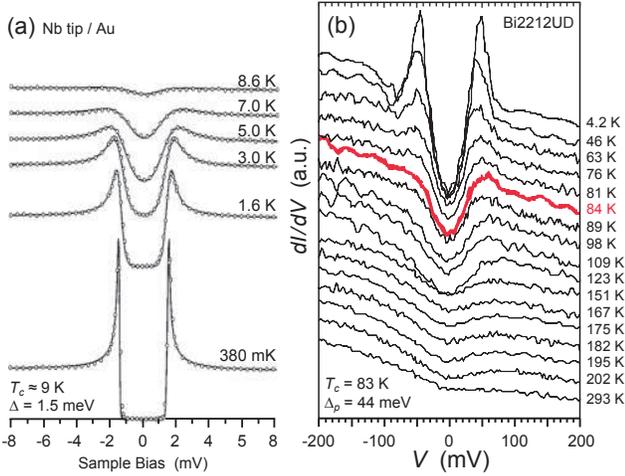}
\caption{\label{fig_Tdep-Nb-Bi2212UD}
$T$-dependencies of the DOS measured by STM. (a) Junction between a gold sample
and a Niobium tip with $T_c\approx9$~K, $\Delta_p=1.5$~meV; adapted from
\citet{Pan-1998}. (b) Junction between an Iridium tip and UD Bi2212 with
$T_c=83$~K, $\Delta_p=44$~meV, and $T^*$ near room temperature; adapted from
\citet{Renner-1998a}.
}
\end{figure}

Figure~\ref{fig_Tdep-Nb-Bi2212UD} illustrates the $T$-dependence of the
quasiparticle DOS of Niobium \cite{Pan-1998},\footnote{Note the different gap
amplitude compared to Fig.~\ref{fig_fig5_gaps}a for the same experimental
configuration, which \citet{Pan-1998} tentatively ascribe to the detailed
geometry, structure, and composition of the Nb tip apex.} a conventional BCS
superconductor, and of UD Bi2212 \cite{Renner-1998a}. The most striking
difference with Nb is the existence in Bi2212 of a clear pseudogap at $T_c$. The
pseudogap has basically the same amplitude as the superconducting gap. Both
appear to be essentially $T$-independent, seemingly filling up rather than
closing with increasing temperature \cite{Tao-1997} except for a slight tendency
of the pseudogap to increase upon approaching $T^*$ (see
Fig.~\ref{fig_Tdep-Bi2201}). We point out that, although it is well known that
in Nb the gap is closing, this is not obvious when looking at
Fig.~\ref{fig_Tdep-Nb-Bi2212UD}a. The superconducting gap of Nb is of the order
of 1.5~meV and $T_c\approx9$~K. Thermal smearing therefore prevents to follow
the coherence peaks in the DOS up to $T_c$. In contrast, in the case of UD
Bi2212 shown in Fig.~\ref{fig_Tdep-Nb-Bi2212UD}b, the gap magnitude is about
44~meV for $T_c=83$~K. Hence $\Delta_p \gg k_{\text{B}}T_c$ and thermal smearing
effects relative to $\Delta_p$ are much weaker. If the $T$-dependence of the gap
would be BCS-like, one should thus clearly see it in
Fig.~\ref{fig_Tdep-Nb-Bi2212UD}b. This is not the case.

\begin{figure}[tb]
\includegraphics[width=7cm]{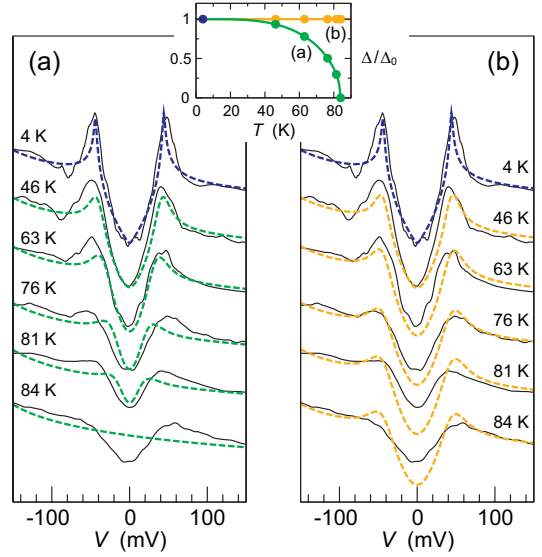}
\caption{\label{fig_Tdep-PGversusBCS}
$T$-dependence of the superconducting gap for $T\leqslant T_c$. The 4.2~K
experimental data (\citet{Renner-1998a}, see also
Fig.~\ref{fig_Tdep-Nb-Bi2212UD}b) is fitted by a $d$-wave function (blue), which
is then used to recalculate the spectra at each temperature corresponding to the
experimental data, assuming two different gap $T$-dependencies: (a) a BCS
$d$-wave $T$-dependence (green); (b) a constant gap magnitude (yellow).
}
\end{figure}

The observation that in Bi2212 the gap magnitude is to first approximation
$T$-independent rather than following a BCS-like dependence is highlighted in
Fig.~\ref{fig_Tdep-PGversusBCS}. Here, the 4.2~K data of
Fig.~\ref{fig_Tdep-Nb-Bi2212UD}b is fitted by a $d$-wave BCS DOS (blue curve).
The asymmetric background is generated by pushing the van Hove singularity of
the tight-binding dispersion down to $-200$~meV. Subsequently, the spectrum is
\emph{recalculated} at \emph{each} temperature corresponding to the experimental
data of Fig.~\ref{fig_Tdep-Nb-Bi2212UD}b up to $T=84~\text{K}\approx T_c$ by
assuming two different $T$-dependencies for the gap: (i) a BCS $d$-wave
$T$-dependence which simulates a closing gap (Fig.~\ref{fig_Tdep-PGversusBCS}a);
(ii) a $T$-independent gap (Fig.~\ref{fig_Tdep-PGversusBCS}b). This latter
simulation reproduces much better the $T$-behavior of the gap magnitude than the
BCS case. Looking more carefully to the data (Fig.~\ref{fig_Tdep-Nb-Bi2212UD}b)
reveals other key features which are generic for the $T$-dependence of the DOS
in Bi2212.

\begin{figure}[tb]
\includegraphics[width=8.5cm]{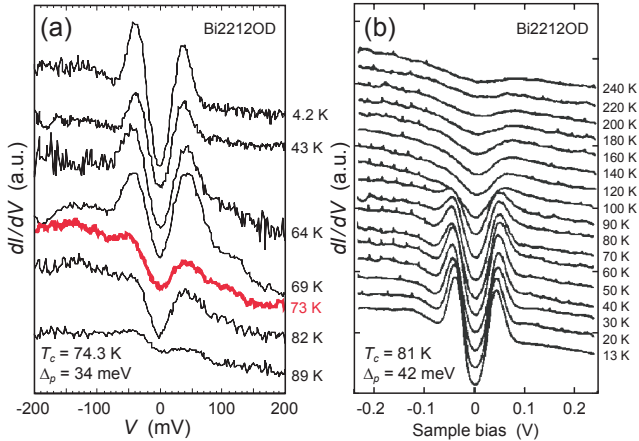}
\caption{\label{fig_Tdep-Bi2212OD}
Tunneling spectra versus temperature in OD Bi2212. (a) Strongly OD sample with
$T_c=74.3$~K and $\Delta_p=34$~meV; adapted from \citet{Renner-1998a}. (b)
Slightly OD sample with $T_c=81$~K and $\Delta_p=42$~meV; adapted from
\citet{Matsuda-1999b}. The tunnel resistances are of the order of 1~G$\Omega$.
}
\end{figure}

\begin{figure}[tb]
\includegraphics[width=7cm]{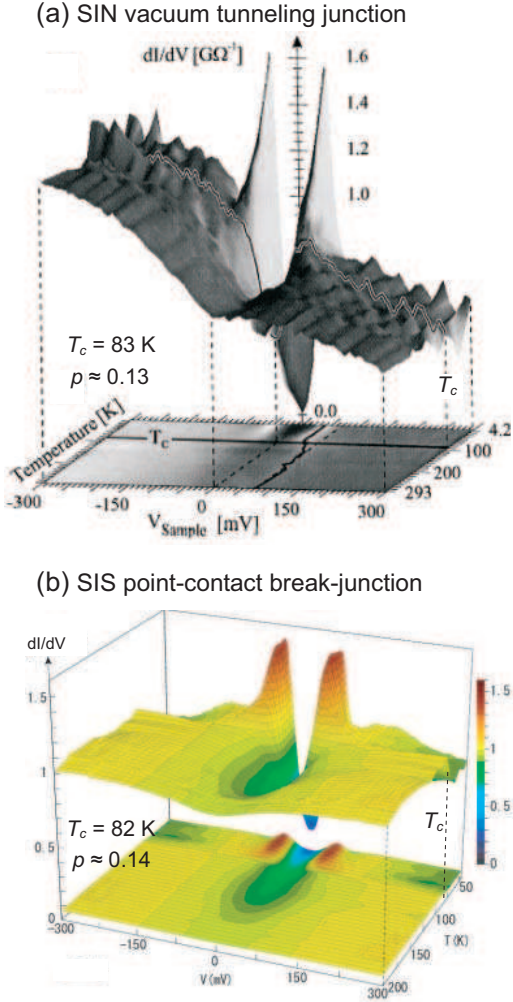}
\caption{\label{fig_Tdep-SIN-SIS}
$T$-dependence of the DOS of Bi2212 for two different junction types and
tunneling techniques. (a) SIN vacuum tunneling using STM; $T_c=83$~K,
$\Delta_p=44$~meV, $T^*\approx300$~K, and $R_t=0.1$--1~G$\Omega$; from
\citet{Renner-1998a}. (b) SIS junction using PC-BJT; $T_c=82$~K,
$\Delta_p\approx27$~meV, $T^*\approx180$~K, and $R_t=10$--100~k$\Omega$; from
\citet{Dipasupil-2002}.
}
\end{figure}

First, by rising the temperature, the coherence-peak intensity is rapidly
reduced at the bulk $T_c$. The coherence peak at negative bias, as well as the
dip-hump feature, are suppressed and a reduced peak shifting to a slightly
higher energy remains at positive bias. The suppression of the coherence peaks
at $T=T_c$ clearly observed in Figs.~\ref{fig_Tdep-Nb-Bi2212UD}b,
\ref{fig_Tdep-Bi2212OD}a, and \ref{fig_Tdep-SIN-SIS}a, is a strong evidence that
in Bi2212, $T_c$ at the surface is identical to the one of the
bulk.\footnote{This is true under the assumption that the coherence peaks are
indeed due to superconductivity. This fact is generally accepted and has
furthermore been demonstrated in detail by ARPES \cite{Feng-2000}.} We note
however that in some cases, and predominantly on underdoped samples which are
more prone to inhomogeneous oxygen distributions (see
Sec.~\ref{sect_gap_spectroscopy}), the observation of this typical PG signature
does not always coincide with the bulk $T_c$ \cite{Matsuda-1999a,Matsuda-1999b}.
As mentioned earlier, a possible explanation could be that, due to thermal drift
and sample inhomogeneity, the tip probes regions with different doping levels as
the temperature is changed. Conversely, a $T$-independent gap magnitude, as
observed in Fig.~\ref{fig_Tdep-Nb-Bi2212UD}b, proves a high level of homogeneity
and it demonstrates that the local doping level is not changing with
temperature. This point is reinforced by the finding that the asymmetric
background conductance is not changing with temperature
(Fig.~\ref{fig_Tdep-Nb-Bi2212UD}b). The argument is based on the observation
that the slope of the background is varying with doping \cite{Kugler-2006},
possibly as a consequence of band-structure effects \cite{Hoogenboom-2003b}
and/or strong-correlation effects \cite{Rantner-2000,Anderson-2006}. For a
discussion of the background conductance, see also
Sec.~\ref{sect_doping_dependence}.

Second, crossing $T_c$, the superconducting spectra evolve continuously into the
pseudogap with a gap magnitude which is to first approximation constant. It is
clear from Figs.~\ref{fig_Tdep-Nb-Bi2212UD}b and \ref{fig_Tdep-Bi2212OD} that
the pseudogap also does not change much with temperature except by a general
broadening of the gap edges due to thermal smearing at high temperatures. The
pseudogap is thus filling up---hence the zero-bias conductance increases
monotonically with temperature as reported first by \citet{Tao-1997}---and
smoothly vanishes at a crossover temperature $T^*$, which appears to be about
room temperature in the data shown in Fig.~\ref{fig_Tdep-Nb-Bi2212UD} on UD
Bi2212 (see Table~\ref{tab_T*-table}).

A third remarkable property is illustrated in Fig.~\ref{fig_Tdep-Bi2212OD}: The
pseudogap is not only bound to the underdoped regime, but also exists in
overdoped samples \cite{Renner-1998a,Matsuda-1999a,Matsuda-1999b}. The main
difference with the underdoped case is that the gap and pseudogap magnitudes are
smaller and that the pseudogap seems to vanish more quickly with increasing
temperature. Indeed, the PG temperature for UD Bi2212 is about room temperature
(Figs.~\ref{fig_Tdep-Nb-Bi2212UD}b and \ref{fig_Tdep-SIN-SIS}a). In overdoped
samples, with correspondingly smaller gaps, $T^*$ is smaller but well above
$T_c$ (Fig.~\ref{fig_Tdep-Bi2212OD}a). These results indicate that the PG
magnitude and the PG temperature are scaling with the superconducting gap
$\Delta_p$ in Bi2212. This key observation is discussed in
Sec.~\ref{sect_pseudogap_scaling}.

Recently \citet{Vershinin-2004,Vershinin-2004b} succeeded in studying the
spatial dependence of the tunneling conductance in the PG state above $T_c$ on
Bi2212. They observed spectra consistent with the generic PG signature described
above. Investigating conductance maps at low energies, they also found small
variations in the LDOS with an energy-independent incommensurate periodicity.
The origin of these modulations and their connection to the PG state are
discussed in Sec.~\ref{sect_modulations}.

STM junctions are generally of the SIN type, whereas techniques like
point-contact break-junction tunneling (PC-BJT) and intrinsic Josephson junction
tunneling (IJJT) provide SIS tunneling barriers. In the following we briefly
comment on the main findings of those techniques and compare them to STM vacuum
tunneling.

In the case of PC-BJT, a tip is driven into the sample surface. By slightly
reducing the pressure applied by the crashed tip it is possible to pick up a
cleaved crystal piece, thus providing a SIS vacuum tunnel junction with typical
tunnel resistances of 10--100~k$\Omega$ \cite{Ozyuzer-2000,Dipasupil-2002}. In
the case of IJJT, the Bi-based cuprate crystal is considered as a stack of
multiple Josephson junctions formed between the superconducting CuO$_2$ planes
separated by insulating BiO layers. Driving a current along the $c$-axis of
small crystal mesas provides intrinsic SIS junctions without any vacuum barrier
and with typical tunnel resistances below 10~$\Omega$ per junction
\cite{Kleiner-1992,Suzuki-1999,Yurgens-1999,Krasnov-2000b}.

In Fig.~\ref{fig_Tdep-SIN-SIS} we show results reported by
\citet{Dipasupil-2002} on Bi2212 SIS junctions using PC-BJT and compare them to
the STM data by \citet{Renner-1998a} on a Bi2212 SIN junction. The spectral
features at low temperatures, like the dip-hump and the characteristic $d$-wave
V-shaped DOS, as well as the overall $T$-dependence, including the observation
of the pseudogap on overdoped samples, and the scaling behavior between
$\Delta_p$ and $T^*$, are remarkably consistent. However, various studies by
other groups showed that strong discrepancies exist among data acquired with the
PC-BJT technique: data reported by \citet{Miyakawa-1998} did not show any
pseudogap on optimally-doped Bi2212 and results by \citet{Ozyuzer-2002} on
underdoped Bi2212 with $T_c=77$~K showed PG-like spectra that vanish already at
about 80~K. The origin of these diverging results is most likely due to the
ill-defined junction geometry and to the difficult junction control under
variable temperatures.

Over the past years IJJT in Bi2212 \cite{Suzuki-1999,Krasnov-2000b,Heim-2002}
and Bi2201 \cite{Yurgens-2003} mesas attracted considerable attention. These
studies revealed a double-gap structure in the DOS: a strongly peaked gap
developing on top of a much broader excitation gap, which were identified as the
superconducting gap and the pseudogap, respectively. The sharply peaked gap
shows a BCS-like $T$-dependence and closes at $T_c$, whereas the broad hump is
almost $T$-independent and visible even below $T_c$
\cite{Krasnov-2000b,Yurgens-2003}. The authors of these studies claimed to
distinguish and simultaneously observe both, the superconducting gap and the
pseudogap, and concluded that their origins are different. These results appear
to be in strong contradiction with STM. A possible explanation is that the
tunneling regime in IJJT is radically different from STM, since the tunneling
resistance $R_t$ in IJJT is typically below 10~$\Omega$, whereas for STM
$R_t\sim0.1$--1~G$\Omega$. Heating effects might thus play an important role in
IJJT. Indeed, very recently it was demonstrated that the spectral features
attributed to the pseudogap and to the superconducting gap are artifacts of
Joule heating in the mesas \cite{Zavaritsky-2004a,Zavaritsky-2004b,Yurgens-2004}.
Hence, the interpretation of IJJT experiments needs to be carefully
reconsidered.

\subsubsection{The case of Bi2201}
\label{sect_pseudogap_TdepBi2201}

The single CuO$_2$-layer Bi$_2$Sr$_2$CuO$_{6+\delta}$ for long stayed in the
shadow, having a much lower $T_c$ and being notoriously more difficult to grow
with reasonable homogeneity than Bi2212. However, with the need to explore how
far the behavior of Bi2212 is generic for the Bi-compound family and, in
extension, to HTS cuprates, Bi2201 recently regained a lot of interest.

\citet{Kugler-2001} reported the first STS study on Bi2201, which revealed
low-temperature spectra with well-developed coherence peaks at
$\Delta_p\approx\pm12$~meV. This is a remarkable result, since for a $T_c$ of
only 10~K the gap magnitude is extremely large. It is by a factor 7 larger than
the BCS $d$-wave prediction $\Delta_{\text{BCS}}=2.14k_{\text{B}}T_c=1.8$~meV.
This further leads to a very high reduced gap $2\Delta_p/k_{\text{B}}T_c\approx
28$, which is even more notable since the investigated sample was overdoped.
Based on the observations collected on Bi2212 and assuming that Bi2201 behaves
similarly, one would expect a PG state extending over an extremely wide
$T$-range above $T_c$.

\begin{figure}[tb]
\includegraphics [width=8.5cm]{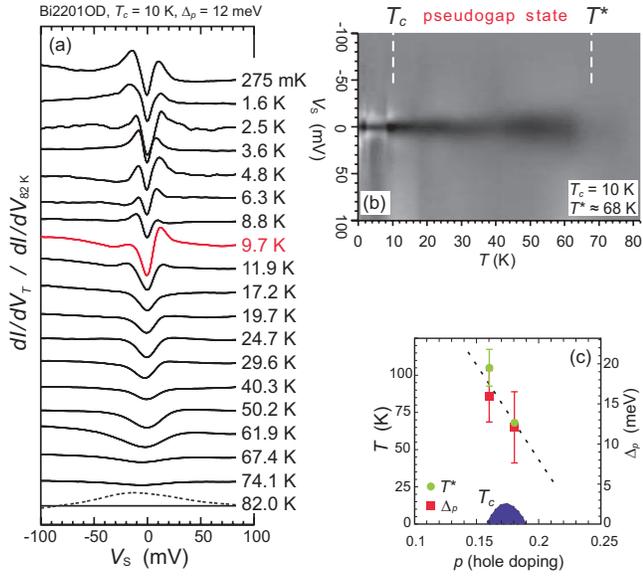}
\caption{\label{fig_Tdep-Bi2201}
(a) $T$-dependence of the DOS in Bi2201; (b) is the top-view representation of
(a). (c) Doping phase diagram showing the values of $T_c$, $\Delta_p$, and $T^*$
from Table~\ref{tab_T*-table}. The doping level $p$ was determined as described
by \citet{Ando-2000}. From \citet{Kugler-2001}.
}
\end{figure}

Figure~\ref{fig_Tdep-Bi2201} reproduces the $T$-dependence of the DOS obtained
by \citet{Kugler-2001} on a slightly overdoped sample. The spectral features are
less sharp than in Bi2212 and show a slight scattering, thus requiring the
averaging over a large number of spectra to extract the general behavior. In
spite of that, the data clearly shows that the pseudogap exists also on the
overdoped side, and that its magnitude is comparable to the superconducting gap.
Furthermore, it is obvious that the gap magnitude is to first approximation
$T$-independent. The overall DOS evolution is thus similar to what has been
observed in Bi2212 (Figs.~\ref{fig_Tdep-Nb-Bi2212UD}b, \ref{fig_Tdep-Bi2212OD},
and \ref{fig_Tdep-SIN-SIS}), although Bi2201 has a much lower $T_c$ and a
smaller superconducting gap.

The $T$-interval of the PG state is extremely large compared to the
superconducting one, as illustrated in Fig.~\ref{fig_Tdep-Bi2201}b. There is a
factor 7 between $T^*$ and $T_c$, the same as between $\Delta_p$ and
$\Delta_{\text{BCS}}$. $T^*$ thus scales with $\Delta_p$. This important
observation is confirmed in a strongly underdoped sample, although the
uncertainty on $T^*$ is much larger due to sample inhomogeneity (see
Table~\ref{tab_T*-table}).

Based on this data, \citet{Kugler-2001} draw a tentative doping phase diagram
for Bi2201 (Fig.~\ref{fig_Tdep-Bi2201}c), illustrating that the PG phase extends
significantly over the superconducting dome. This is particularly obvious in the
overdoped regime and highlights that in this compound $T_c$ is strongly
suppressed. For pure Bi2201 $T_c^{\text{max}}$ is about 13~K
\cite{Vedneev-1999,Gorina-1998}. The reasons for such a low $T_c$ relative to
the gap $\Delta_p$ are most probably disorder and strong phase fluctuations.
Whether this anomalous loss of phase coherence is a hallmark of single-layer
compounds remains an open issue. Preliminary results on La-doped Bi2201
\citep[slightly underdoped,][]{Kugler-2000} showed gap magnitudes $\Delta_p$
comparable to pure Bi2201, but for a $T_c$ as high as 29~K
\citep[$T_c=36$~K,][]{Ando-2000}. This suggests that La-doping does not change
the pairing energy but rather favors phase coherence. It yields a reduced gap of
the order of 10, which is consistent with Bi2212 and restores the
superconducting phase accordingly to the generic phase diagram shown in
Fig.~\ref{fig_fig5_phasediagram}, where the PG state is less dominant on the
overdoped side. These findings point out that in the case of HTS, the relation
between $\Delta_p$ and $T^*$ is much more robust than the relation between
$\Delta_p$ and $T_c$ (see Sec.~\ref{sect_pseudogap_scaling}).

\subsubsection{The case of Y123 and Nd123}
\label{sect_pseudogap_TdepY123}

\begin{figure}[tb]
\includegraphics [width=8.5cm]{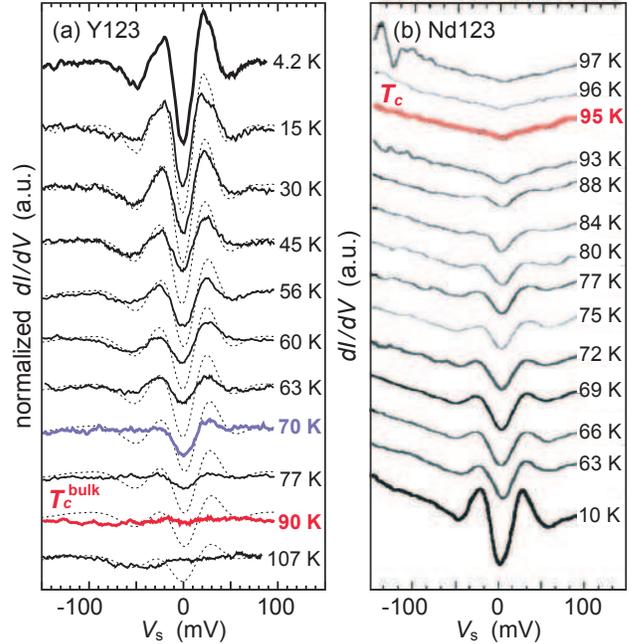}
\caption{\label{fig_Tdep-Y123-Nd123}
(a) $T$-dependence of optimally-doped Y123 ($T_c=92$~K, $\Delta_p=20$~meV). The
spectra (solid lines) are normalized to the background conductance taken at
$V>100$~mV and extrapolated to $V=0$. Dotted spectra correspond to the 4.2~K
spectrum thermally smeared to the different temperatures; from
\citet{Maggio-Aprile-2000}. (b) $T$-dependence of Nd123 ($T_c=95$~K,
$\Delta_p=30$~meV); adapted from \citet{Nishiyama-2002}.
}
\end{figure}

We finally turn to the $T$-dependence of compounds having both CuO$_2$ planes
and CuO chains, in particular Y123 and Nd123. Figure~\ref{fig_Tdep-Y123-Nd123}a
shows the results obtained by \citet{Maggio-Aprile-2000} on optimally-doped
Y123. The coherence peak position, hence the superconducting gap, remains
constant with rising temperature, which is consistent with the Bi-based
compounds. In contrast, the gap progressively fills up and eventually closes at
the bulk $T_c$, demonstrating that there is no signature of any pseudogap above
the bulk $T_c$ in optimally-doped Y123. Very similar results were reported on
Nd123 \cite{Nishiyama-2002} as shown in Fig.~\ref{fig_Tdep-Y123-Nd123}b. These
observations are consistent with results from other techniques \citep[see review
by][]{Timusk-1999}. Thus in spite of the fact that Y(Nd)123 and Bi2212 have the
same $T_c$, they display a very different behavior with respect to the
pseudogap. A possible key for understanding this difference is that Y(Nd)123 is
much less anisotropic and appears to be much closer to the BCS limit than Bi2212
\cite{Junod-1999}. Furthermore the reduced gap values for optimally-doped Y123
($\sim5$) and Nd123 ($\sim7$) are closer to the $d$-wave BCS ratio ($\sim4.3$).
The consequence is that Y(Nd)123 is less affected by superconducting
fluctuations than Bi2212, thus possibly explaining the absence of a pseudogap.
Another major difference comes from the CuO chains. How far these chains could
prevent the existence of a pseudogap remains to be elucidated.

Besides these differences, Fig.~\ref{fig_Tdep-Y123-Nd123}a reveals other
interesting aspects. A clear dip-hump feature is observed at about $\pm50$~meV.
This structure looks similar to what is seen in Bi2212 (see
Sec.~\ref{sect_dip_hump}) and it consistently disappears as $T$ approaches
$T_c$. \citet{Maggio-Aprile-2000} showed that this disappearance is not due to
thermal smearing, suggesting that the dip is related to the superconducting
state. In fact, it turns out that the dip disappears at 70~K where the spectral
line shape strongly resembles the Bi2212 spectrum at $T=T_c$
(Figs.~\ref{fig_Tdep-Nb-Bi2212UD} and \ref{fig_Tdep-Bi2212OD}). This could
indicate that at the surface of Y123 the coherent superconducting state
disappears at about 70~K and that above, a PG state exists up to 90~K. The
explanation would be that the surface $T_c$ is different from the bulk $T_c$ in
Y123 due to a higher effective anisotropy at the surface \cite{Meingast-2001}.
However, this interpretation remains speculative since the gap signal above 70~K
is too weak to make a decisive statement. A discussion of the doping phase
diagram of Y123 is presented in Sec.~\ref{sect_pseudogap_DPD}.

\subsubsection{The case of electron-doped cuprates}
\label{sect_pseudogap_e-doped}

\begin{figure}[tb]
\includegraphics[width=8.5cm]{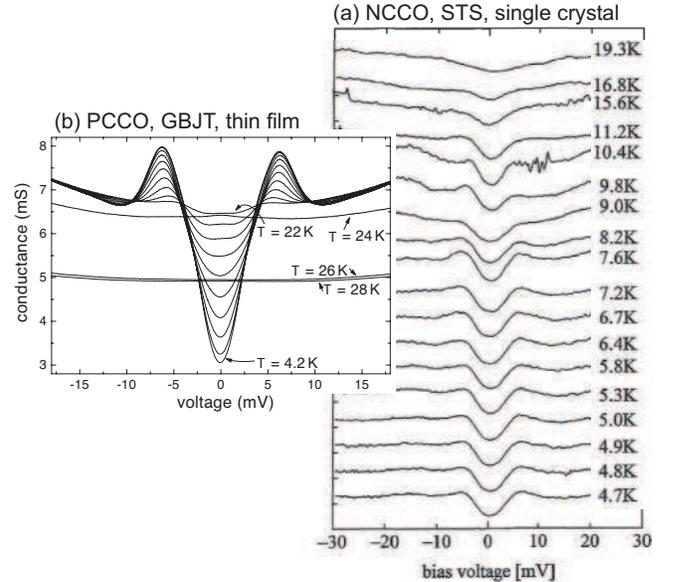}
\caption{\label{Fig_Tdep_e-doped}
(a) $T$-dependence by STS of a Nd$_{1.85}$Ce$_{0.15}$CuO$_4$ single crystal
(optimally doped, $T_c=20.5$~K, $\Delta_p\approx 4.5$~meV). The tip is parallel
to the [110] direction; from \citet{Hayashi-1998b}. (b) $T$-dependence by
grain-boundary junction tunneling (GBJT) on a Pr$_{1.85}$Ce$_{0.15}$CuO$_4$ thin
film (optimally doped, $T_c\approx 24$~K, $\Delta_p=3.1$~meV. Note that these
junctions are of the SIS type and in-plane; from \citet{Kleefisch-2001}.
}
\end{figure}

A crucial question for understanding the microscopic mechanism of HTS is to what
extent electron-doped cuprates R$_{2-x}$Ce$_x$CuO$_{4-y}$ (RCCO,
$\text{R}=\text{rare earth}$) are similar to hole-doped HTS. The specific issue
of the existence of the pseudogap has been investigated using various tunneling
techniques, although only very few reports exist. In
Fig.~\ref{Fig_Tdep_e-doped}a we reproduce STS data obtained on optimally-doped
NCCO single crystals by \citet{Hayashi-1998b}. These results were confirmed by
grain-boundary junction tunneling (GBJT) on optimally-doped NCCO and PCCO thin
films by \citet{Kleefisch-2001} (Fig.~\ref{Fig_Tdep_e-doped}b). The general
behavior is very similar to Y123 or Nd123, with the main conclusion that no
pseudogap is observed at optimal doping.

In contrast, when driving the superconductor into the normal state by applying a
magnetic field, a clear PG spectrum appears which has the same magnitude as the
superconducting gap at zero field \cite{Kleefisch-2001, Biswas-2001}. This is
consistent with what is observed in Bi2212 vortex cores
(Sec.~\ref{sect_pseudogap_PGinVC}). \citet{Alff-2003} investigated how this
pseudogap evolves with doping and determined $T^*$. They observed that $T^*$ has
a similar doping dependence to that of the hole-doped copper oxides, however
with the important difference that $T^*<T_c$ in the investigated doping range.
The authors concluded that in electron-doped cuprates the PG phase coexists and
competes with superconductivity and mentioned the possible existence of a
quantum critical point.

\subsubsection{Scaling properties: pseudogap versus superconducting gap}
\label{sect_pseudogap_scaling}

An important step towards understanding the PG origin is to search for relations
between the gap and pseudogap magnitude, as well as between $T_c$ and $T^*$. The
superconductors described in this section are an ideal stage for such an
analysis. Bi2212 acts as a template, whereas Bi2201 and Y123 allow to test if
the scaling behaviors are generic. Bi2201 presents the advantage to have a
crystal structure similar to Bi2212 but radically different superconducting
parameters $T_c$ and $\Delta_p$, whereas Y123 and Bi2212 have comparable $T_c$'s
but different crystal structures.

\begin{figure}[tb]
\includegraphics[width=8.5cm]{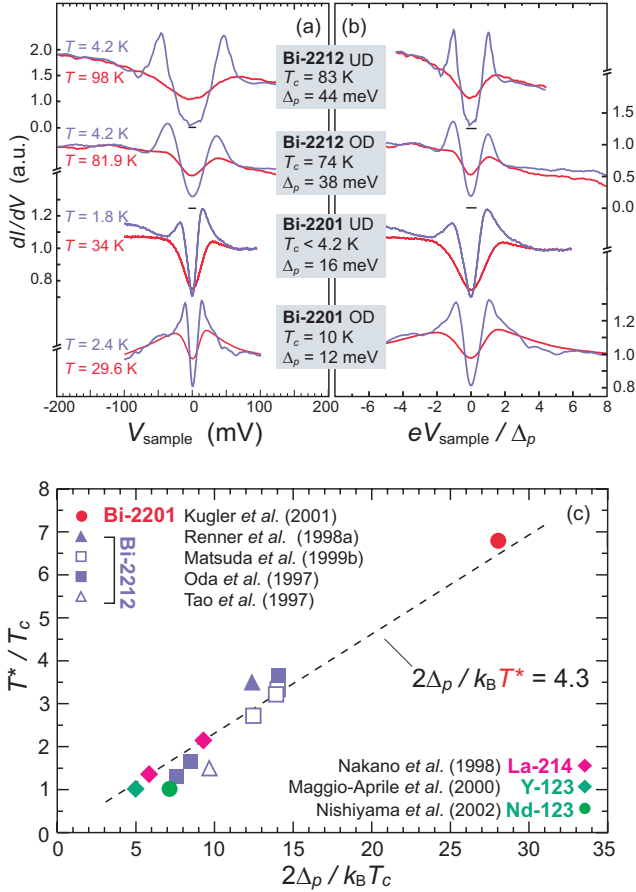}
\hphantom{
\cite{Kugler-2001, Renner-1998a, Matsuda-1999a, Oda-1997, Tao-1997, Nakano-1998,
Maggio-Aprile-2000, Nishiyama-2002}
}
\vspace*{-2em}
\caption{\label{fig_Scaling}
Scaling relations \citep[adapted from][]{Kugler-2001}. (a) Comparison of the
pseudogap and the superconducting gap for Bi2212 and Bi2201. (b) Same data with
energy rescaled by $\Delta_p$. (c) $T^*/T_c$ as a function of
$2\Delta_p/k_{\text{B}}T_c$ for various cuprates investigated by tunneling
spectroscopy. The dashed line corresponds to the mean-field $d$-wave relation
$2\Delta_p/k_{\text{B}}T^*=4.3$, where $T_c$ is replaced by $T^*$. See
Table~\ref{tab_T*-table} for detailed superconducting parameters.
}
\end{figure}

Figure~\ref{fig_Scaling}a compares the low-$T$ superconducting DOS and the PG
spectrum measured slightly above $T_c$ in Bi2212 and Bi2201. The superconducting
gap clearly increases with underdoping (see also Fig.~\ref{fig_fig5_doping}) and
the pseudogap behaves exactly the same in both compounds. Moreover, the relative
amplitude of these two gaps is nearly the same in both compounds independent of
doping as shown in Fig.~\ref{fig_Scaling}b \cite{Kugler-2001}. This consistency
is remarkable since the energy scale given by $\Delta_p$ varies by about 400\%.
Whether the scaling between the pseudogap and the superconducting gap observed
in Bi-cuprates also applies to Y123 remains to be seen. This question may be
answered by investigating the DOS $T$-dependence of underdoped samples, where
other techniques have suggested that a pseudogap exists \citep[see review
by][]{Timusk-1999}.

The existence of a scaling relation between $\Delta_p$ and $T^*$ was first
noticed by \citet{Oda-1997} in Bi2212 and by \citet{Nakano-1998} in
La214.\footnote{In both cases the $T^*$ values where determined by in-plane
resistivity measurements.} As shown in Fig.~\ref{fig_Scaling}c, \emph{the same}
scaling law also holds for Bi2201 \cite{Kugler-2001}, although
$2\Delta_p/k_{\text{B}}T_c$ is 2--3 times larger. Optimally-doped Y123, where
$T^*\lesssim T_c$ \cite{Maggio-Aprile-2000}, also fits into the picture. This
consistency among various compounds is strong evidence that the scaling is
robust. Moreover, the graph reveals that $2\Delta_p/k_{\text{B}}T^*\approx4.3$,
which corresponds to the BCS $d$-wave relation with $T^*$ playing the role of
$T_c$. This finding clearly supports the idea that $T^*$ is the mean-field BCS
temperature and reflects the interaction energy leading to superconductivity.

\def\fnlt#1#2{\multicolumn{8}{l}{$~^#1$\footnotesize #2}\\ [-0.1em]}
\begin{table*}[tb]
\caption{\label{tab_T*-table}
Low-$T$ gap amplitude $\Delta_p$ and PG temperature $T^*$ as obtained by
tunneling experiments. The references are classified from underdoped to
overdoped for each compound.$^a$ \\ [-2em]
}
\begin{tabular*}{\textwidth}[t]{l@{\extracolsep{\fill}}r@{\hspace{-1cm}}c@{\extracolsep{\fill}}
                                c@{\extracolsep{\fill}}c@{\extracolsep{\fill}}
                                d@{\extracolsep{\fill}}d@{\extracolsep{\fill}}l}
\hline\hline \\ [-1em]
Compound & \multicolumn{2}{c}{$T_c$ (K)} & $\Delta_p$ (meV) & $T^*$ (K) &
\multicolumn{1}{c}{$2\Delta_p/k_{\text{B}}T_c$} &
\multicolumn{1}{c}{$2\Delta_p/k_{\text{B}}T^*$} & Reference\\
\hline \\ [-0.5em]
Bi2212$^b$ & 60       & UD & $36\pm2$ & $217\pm9$  & 13.9 & 3.8 & \citet{Oda-1997}\\
Bi2212$^c$ & 77       & UD & $40$     & $200\pm8$  & 12.1 & 4.6 & \citet{Dipasupil-2002}\\
Bi2212$^c$ & 82       & UD & $36$     & $180\pm8$  & 10.2 & 4.6 & \citet{Dipasupil-2002}\\
Bi2212     & 83       & UD & $44\pm2$ & $\sim290$  & 12.3 & 3.5 & \citet{Renner-1998a}\\
Bi2212     & 90       & OP & $54\pm3$ & $\sim300$  & 13.9 & 4.2 & \citet{Matsuda-1999a}\\
Bi2212$^b$ & 88       & OP & $32\pm2$ & $142\pm9$  &  8.4 & 5.2 & \citet{Oda-1997}\\
Bi2212$^d$ & 87       & ?  & $36\pm1$ & $130\pm20$ &  9.6 & 6.4 & \citet{Tao-1997}\\
Bi2212     & 84       & OD & $50\pm4$ & $\sim270$  & 13.8 & 4.3 & \citet{Matsuda-1999a}\\
Bi2212$^c$ & 82       & OD & $25$     & $120\pm8$  &  7.1 & 4.8 & \citet{Dipasupil-2002}\\
Bi2212     & 81       & OD & $43\pm2$ & $\sim220$  & 12.3 & 4.5 & \citet{Matsuda-1999a}\\
Bi2212$^b$ & 81       & OD & $27\pm2$ & $104\pm9$  &  7.7 & 6.0 & \citet{Oda-1997}\\
\hline \\ [-0.5em]
Bi2201     & $<4$     & UD & $16\pm3$ & $105\pm15$ &      & 3.5 & M. Kugler (unpublished)\\
Bi2201     & 10       & OD & $12\pm3$ & $68\pm2$   & 28   & 4.1 & \citet{Kugler-2001}\\
\hline \\ [-0.5em]
Y123       & 92       & OP & $20\pm1$ & $92$       &  5   & 5   & \citet{Maggio-Aprile-2000}\\
Nd123      & 95       & OP & $30$     & $95$       &  7.3 & 7.3 & \citet{Nishiyama-2002}\\
\hline \\ [-0.5em]
La214$^b$  & $\sim40$ & UD & $\sim16$ & $82\pm15$  &  9.3 & 4.5 & \citet{Nakano-1998}\\
La214$^b$  & $\sim40$ & OD & $\sim10$ & $53\pm10$  &  5.8 & 4.4 & \citet{Nakano-1998}\\ [0.3em]
\hline\hline \\ [-0.6em]
\fnlt{a}{The identification as underdoped or overdoped is the one given by the
authors.}
\fnlt{b}{$T^*$ is determined by in-plane resistivity measurements.}
\fnlt{c}{SIS point-contact break-junction.}
\fnlt{d}{Pb/Bi2212 planar junction.}
\end{tabular*}
\end{table*}

\subsection{The pseudogap at low temperatures}
\label{sect_pseudogap_LTPG}

Various STS studies demonstrated that the characteristic PG signature is not
only observed above $T_c$, but also occurs at low temperatures inside vortex
cores \cite{Renner-1998b} and on strongly disordered surfaces
\cite{Cren-2000,McElroy-2004a}.

\subsubsection{The pseudogap inside vortex cores}
\label{sect_pseudogap_PGinVC}

The spatial dependence of the LDOS at low temperatures and in magnetic fields
has been investigated by different groups on Bi2212
\cite{Renner-1998b,Pan-2000b,Hoogenboom-2000a} and Y123
\cite{Maggio-Aprile-1995,Maggio-Aprile-1997,Shibata-2003b} and revealed the
unconventional spectral signature of the vortex cores in these compounds, as
discussed in detail in Sec.~\ref{sect_vortices}. Here we focus on the relation
between vortex-core spectra observed on Bi2212 and the pseudogap above $T_c$.

\begin{figure}[tb]
\includegraphics[width=6cm]{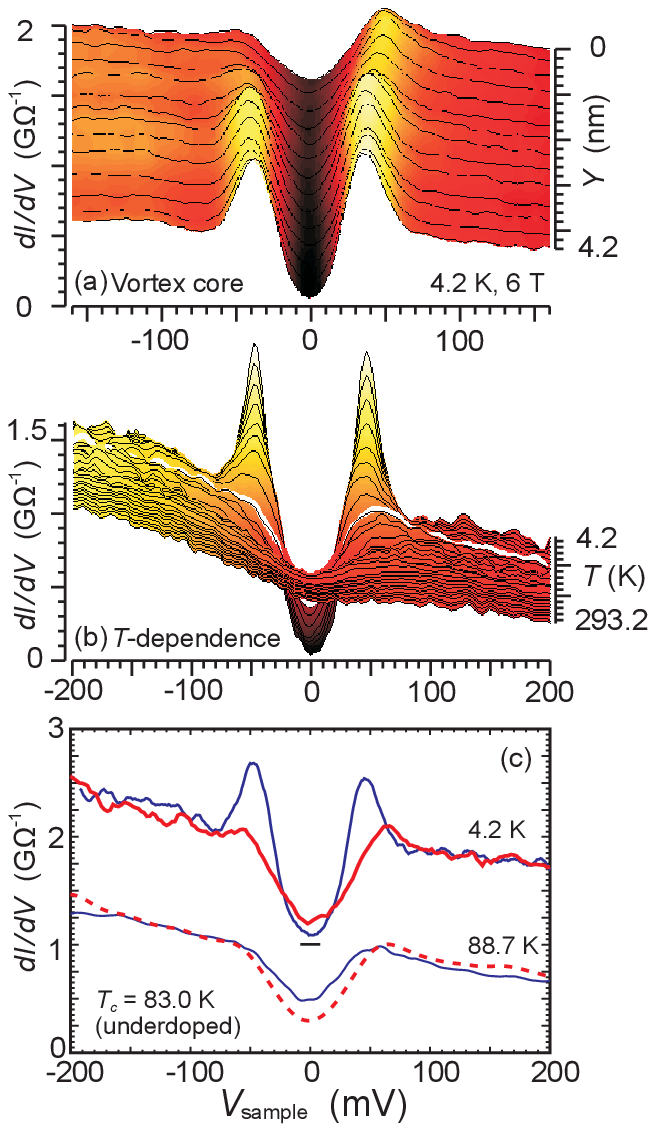}
\caption{\label{fig_PG-VC-Tdep2}
Comparison of the pseudogap measured above $T_c$ and the LDOS inside the vortex
core of Bi2212. (a) Spatial evolution of the LDOS when moving the tip from the
center ($Y=0$) to the periphery of the vortex core ($Y=4.2$~nm) at 4.2~K and 6~T
\citep[Bi2212OD,][]{Renner-1998b}. (b) $T$-dependence of the DOS. The pseudogap
measured at $T\approx T_c=83$~K is highlighted in white \citep[Bi2212UD, adapted
from][]{Renner-1998a}. (c) Top: superconducting gap (blue) measured in between
vortices at 4.2~K and 6~T compared to a vortex-core spectrum (red). Bottom:
pseudogap measured on the same sample in zero field at $T=88.7~\text{K}>T_c$
(solid line), compared to the vortex-core spectrum numerically smeared to 88.7~K
(dashed red) \citep[Bi2212UD, adapted from][]{Renner-1998b}.
}
\end{figure}

\citet{Renner-1998b} measured the evolution of the spectra when the tip is moved
across a vortex core (VC) on a Bi2212 surface (Fig.~\ref{fig_PG-VC-Tdep2}a).
Outside the VC the spectra do not depend on tip position and present
well-defined coherence peaks and the characteristic dip-hump. Moving into the VC
the coherence peak at negative bias is suppressed over a typical distance of
10~\AA, whereas the peak at $+\Delta_p$ is reduced and shifts to slightly higher
energy. This behavior is strikingly similar to the $T$-dependence of the DOS
when crossing $T_c$, as illustrated in Fig.~\ref{fig_PG-VC-Tdep2}. In
Fig.~\ref{fig_PG-VC-Tdep2}c the VC spectrum is numerically smeared to the
temperature where the pseudogap is measured, showing again the similarity of
both line shapes. From this \citet{Renner-1998b} conclude that the LDOS measured
at the vortex center is the normal-state PG signature, and that the spectrum
measured above $T_c$ is essentially the thermally-smeared low-$T$ LDOS. This
implies that the PG magnitude, like the superconducting gap
(Fig.~\ref{fig_Tdep-PGversusBCS}), is $T$-independent. Furthermore, as shown in
Fig.~\ref{fig_PG-VC-Tdep2}c, the VC gap magnitude is to first approximation
equal to the superconducting gap measured just outside the core.
\citet{Renner-1998b} also showed that these findings are doping independent and
therefore consistently obey the scaling properties derived for the pseudogap
measured above $T_c$ (Fig.~\ref{fig_Scaling}).

The correspondence between the normal-state pseudogap and the vortex-core
spectra, as well as the doping and $T$-dependence of the spectra reviewed
respectively in Sec.~\ref{sect_gap_spectroscopy} and \ref{sect_pseudogap_Tdep},
have important consequences and raise crucial questions with respect to the
origin of the pseudogap.

Does the superconducting state develop on top of a gapped conductance background
originating from an independent PG state, similar to the charge-density wave gap
observed in dichalcogenides \citep[see \textit{e.g.}][]{Coleman-1991,Wang-1990}?
If the pseudogap would have an origin different from superconductivity, then the
normal-state gap would not necessarily need to have the same amplitude as the
superconducting gap. This scenario could explain why the pseudogap is observed
in the vortex cores, but is difficult to reconcile with the fact that the two
gaps scale with each other as the doping is changed.

An important question related to the correspondence between the pseudogap and
the vortex core is whether the PG state between $T^*$ and $T_c$ is due to
fluctuating vortices. Thermal transport experiments by \citet{Wang-2001b}
support this idea. They found an anomalous Nernst signal above $T_c$ in La214
and Bi2201, which they attributed to vortex-like excitations. They pointed out
that this regime penetrates deep into the PG phase and appears predominantly in
underdoped samples. They interpreted these results in terms of strong
fluctuations between the PG state and the $d$-wave superconducting phase
\cite{Emery-1995, Norman-1998, Franz-2001b, Lee-2001}. These vortex-like
excitations could be electronic excitations specific to the PG state or, as
suggested by tunneling data \citep[][Fig.~\ref{fig_PG-VC-Tdep2}]{Renner-1998b},
they could be Abrikosov vortices of the superconducting condensate formed in the
normal state by phase fluctuations. In the latter scenario the pseudogap would
have the same origin as superconductivity, raising the interesting question of
its magnetic field dependence.

The field dependence of the pseudogap above $T_c$ has not yet been investigated
systematically by STS.\footnote{\citet{Ng-2000} investigated Bi2212 using
PC-BJT. No field dependence of the pseudogap was observed up to 1.1~T;
\citet{Krasnov-2001b} investigated Bi2212 mesas by IJJT. The spectral signature
they attributed to the pseudogap is invariant at least up to 14~T. Note however
that this structure has been attributed to an artifact of Joule heating in the
mesas (see Sec.~\ref{sect_pseudogap_TdepBi2212}). \citet{Anagawa-2003} performed
similar investigations up to 9~T. They observe that both the pseudogap structure
and the superconducting gap are essentially field independent.} However,
\citet{Hoogenboom-2000a} showed on Bi2212 at 4.2~K that the energy of the core
states and the superconducting gap in between cores scale and are field
independent, at least at moderate fields up to 6~T (Sec.~\ref{sect_vortices}).
The PG state under fields as high as 60~T has been investigated by $c$-axis
transport measurements in Bi2212 by \citet{Krusin-Elbaum-2003}. These
experiments showed that above $\sim20$~T the $c$-axis resistivity, which is
believed to be proportional to the Fermi-level DOS, increases notably,
indicating that the pseudogap is filling up with increasing field.
\citeauthor{Krusin-Elbaum-2003} also determined the pseudogap closing field
$H_{\text{pg}}$, which for an overdoped sample extrapolates to about 90~T at
$T_c=67$~K. Remarkably, $H_{\text{pg}}$ scales with $T^*$ as
$g\mu_{\text{B}}H_{\text{pg}} \approx k_{\text{B}}T^*$, hence with $\Delta_p$
(Sec.~\ref{sect_pseudogap_scaling}). These results fit well into a picture where
the PG state is formed of incoherent pairs, that eventually break at the
depairing field $H_{\text{pg}}$.

In this context, we mention grain-boundary junction tunneling experiments
performed on the electron-doped cuprates NCCO and PCCO \cite{Kleefisch-2001,
Biswas-2001}. The suppression of superconductivity by a strong magnetic field
results in a PG-like spectrum at $H_{c2}$ (See also
Sec.~\ref{sect_pseudogap_e-doped}). \citet{Kleefisch-2001} also demonstrated
that the pseudogap closes at a field corresponding to the Clogston paramagnetic
limit \cite{Clogston-1962, Yang-1998}.

\subsubsection{The pseudogap on disordered surfaces}
\label{sect_pseudogap_PGinDisorder}

Over the past years many studies focused on the spatial variations of the LDOS
at the nanometer scale under the influence of disorder. Here we discuss the
local observation of the pseudogap at low temperatures and zero magnetic field
in Bi2212 thin films \cite{Cren-2000} which have a high degree of structural
disorder, in UD Bi2212 single crystals \cite{Howald-2001} where disorder has
been introduced intentionally by local damage, and in inhomogeneous samples
where disorder is presumably due to a local variation in oxygen concentration
\cite{McElroy-2004a}.

\begin{figure}[tb]
\includegraphics[width=8.5cm]{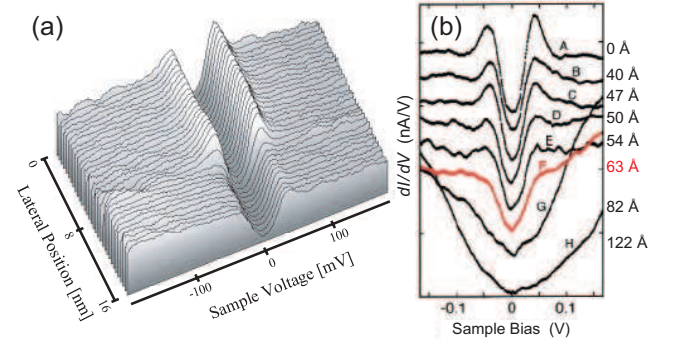}
\caption{\label{fig_PG-disorder}
(a) LDOS as a function of tip position on a Bi2212 300~\AA-thick film at 4.2~K.
Spectra were acquired every 5~\AA; from \citet{Cren-2000}. (b) LDOS evolution
through a locally damaged Bi2212 region. The values on the right indicate the
distance of each spectrum relative to spectrum A; adapted from
\citet{Howald-2001}.
}
\end{figure}

Figure~\ref{fig_PG-disorder}a shows the evolution of the LDOS as the tip is
moved out of a superconducting region on a Bi2212 thin film \cite{Cren-2000}: a
gap with well-defined coherence peaks abruptly changes into a spectrum with the
same gap magnitude, and which is identical to the one observed at $T>T_c$ and in
the vortex cores. Moving further away from this pseudogapped zone the spectra
finally become semiconductor-like (not shown in Fig.~\ref{fig_PG-disorder}a),
indicating that strong disorder is indeed present. The gap-pseudogap transition
occurs over roughly 10~\AA, consistently with the spectral evolution across a
vortex core (Fig.~\ref{fig_PG-VC-Tdep2}a and Sec.~\ref{sect_vortices}). This
characteristic length is of the order of $\xi$ and sets the scale for the loss
of superconducting phase coherence. Similar results were obtained on Pb-doped
Bi2212 single crystals \cite{Cren-2001} and on an intentionally ``damaged''
Bi2212 single crystal \cite{Howald-2001}, as shown in
Fig.~\ref{fig_PG-disorder}b (spectrum F). These experiments show that in
strongly disordered samples superconducting and pseudogapped regions can locally
coexist.

\begin{figure}
\includegraphics[width=8.5cm]{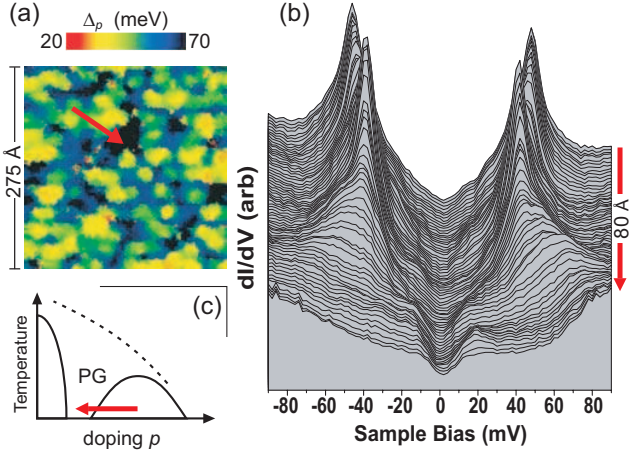}
\caption{\label{fig_McElroy-trace}
(a) Gap map at 4.2~K and zero field acquired on a Bi2212 sample which on average
is underdoped. The red arrow indicates the location of the 80~\AA\ trace shown
in (b); adapted from \citet{McElroy-2004a}. (c) Hypothesis on the local doping
variation along the spectral trace.
}
\end{figure}

Many other experiments on native and generally UD Bi2212 surfaces also
demonstrated patches with variable superconducting characteristics
\cite{Pan-2001, Howald-2001, Lang-2002, Matsuda-2003, Kinoda-2003,
McElroy-2004a}. However, as illustrated in Fig.~\ref{fig_McElroy-trace}b, these
results show a \textit{central} difference: instead of presenting a constant gap
magnitude when crossing the limits of a patch, the coherence peaks shift to
considerably higher energies, they gradually broaden and decrease in intensity
to eventually end up in a PG-like DOS. This behavior is characteristic for local
variations of the oxygen content, as demonstrated by \citet{Pan-2001}
(Fig.~\ref{fig_fig5_inhomogene}d). In terms of the doping phase diagram, these
experiments show, on a local scale, how the DOS evolves when reducing the doping
level (Fig.~\ref{fig_McElroy-trace}c). This is in clear contrast with the
observations by \citet{Cren-2000}, who found that the gap magnitude remains
constant. These authors interpreted their results using a theoretical model by
\citet{Huscroft-1998}. It shows that in the presence of strong disorder, the
phase coherence is lost, while the pairing amplitude which manifests itself as
the pseudogap, is not necessarily suppressed.

\begin{figure}[tb]
\includegraphics[width=8.5cm]{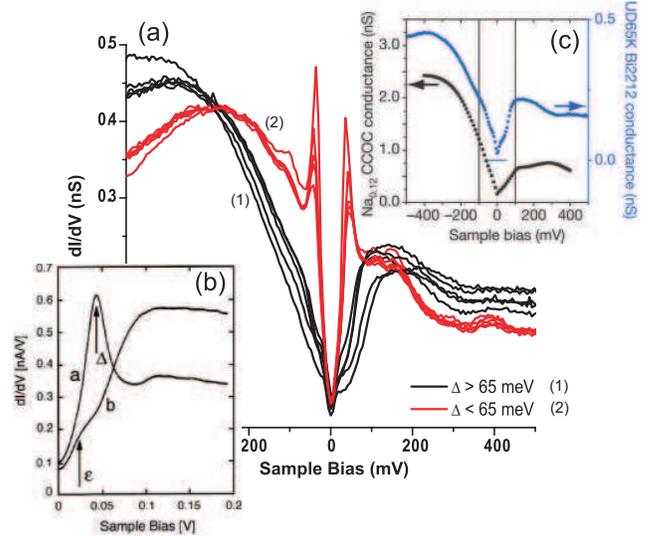}
\caption{\label{fig_ZTPG}
(a) Representative spectra from nanoscale Bi2212 regions exhibiting coherence
peaks (red) and from regions exhibiting the ZTPG (black) acquired at 4.2~K; from
\citet{McElroy-2005a}. (b) Comparison of the positive-bias LDOS of a ZTPG and a
superconducting region in Bi2212; from \citet{Howald-2001}. (c) Comparison
between spatially averaged tunneling conductance spectra on heavily UD
Ca$_{2-x}$Na$_x$CuO$_2$Cl$_2$ (black) and equivalently UD Bi2212 (blue); from
\citet{Hanaguri-2004}.
}
\end{figure}

The systematic study of microscopic inhomogeneities in the LDOS at different
dopings on Bi2212 by \citet{McElroy-2004a} revealed that the total area showing
regions with large gap values increases with underdoping. For strongly
underdoped samples, regions appear where the coherence peaks are absent. The
authors claim that in those regions all spectra are identical to each other, and
that no changes are observed on further underdoping. Instead, the total area
covered by such regions increases. \citet{McElroy-2004a} therefore conclude that
they observed a limiting class of spectra and identified it as the
zero-temperature pseudogap (ZTPG). Similar spectra were observed by
\citet{Hanaguri-2004} on heavily underdoped and even non-superconducting
Ca$_{2-x}$Na$_x$CuO$_2$Cl$_2$. As illustrated in Fig.~\ref{fig_ZTPG}c, these
spectra have a V-shaped gap with a broad peak at positive bias, and resemble the
PG spectrum described earlier, however for an extremely underdoped region, since
the peak at positive bias is around 150~mV. It is suggested that these spectra
characterize the DOS between the superconducting dome and the antiferromagnetic
phase \cite{Hanaguri-2004}. In contrast to the $T$-dependence of the DOS
(Sec.~\ref{sect_pseudogap_scaling}) where the pseudogap and superconducting gap
have the same magnitude, we emphasize that in the case described in
Fig.~\ref{fig_ZTPG}a the ZTPG does not scale with the superconducting spectrum
since they do not correspond to the same doping level: the superconducting
spectra are typical of optimally or slightly underdoped regions, whereas the
ZTPG spectra correspond to extremely underdoped regions. This aspect is clearly
seen in Fig.~\ref{fig_McElroy-trace}.

These observations suggest that there exist two types of disorder which have
distinct spectral signatures: (i) disorder which does not change the local
doping level, hence the gap magnitudes are the same in both the superconducting
and pseudogap regions; (ii) disorder due to a locally varying oxygen
concentration, hence the observed pseudogap will correspond to an extremely
underdoped sample.

As first pointed out by \citet{Howald-2001} and later by \citet{McElroy-2005a},
the low-energy features of the spectra shown in Fig.~\ref{fig_ZTPG}a and b
coincide up to roughly 20~meV, above which a kink occurs in the ZTPG. This might
indicate that the so-called ZTPG has two components: a first component
originating from the nodal points with coherent $d$-wave quasiparticles at
energies below the kink, and a second one at higher energies with incoherent
quasiparticles predominantly from the anti-nodal points, yielding the
pseudogapped envelope due to the reduced quasiparticle lifetime. Other
interpretations in terms of phase separation between ``good'' and ``bad''
superconducting regions have also been proposed \cite{Howald-2001,Lang-2002}.
The issue of a two-component behavior is discussed in more detail in
Sec.~\ref{sect_modulations} in relation with the spatial modulations of the
LDOS.

\subsection{The doping phase diagram}
\label{sect_pseudogap_DPD}

\begin{figure}[tb]
\includegraphics[width=8.5cm]{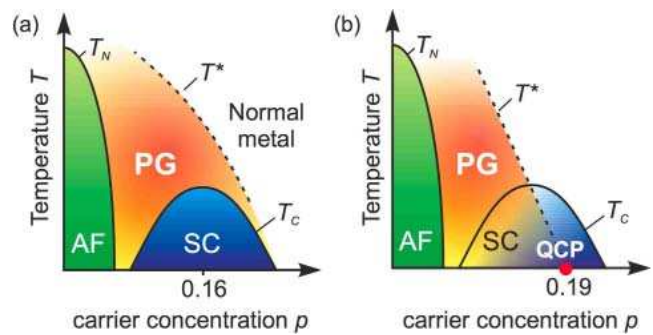}
\caption{\label{fig_DPD-QCP}
Two scenarios for the hole-doped HTS phase diagram. (a) $T^*$ merges with $T_c$
on the overdoped side. (b) $T^*$ crosses the superconducting dome (SC) and falls
to zero at a quantum critical point (QCP). $T_N$ is the N\'{e}el temperature for
the antiferromagnetic (AF) state.
}
\end{figure}

Conventional BCS superconductors are characterized by their $T_c$, which is
closely related to the pairing energy $\Delta$ via the coupling constant
$2\Delta/k_{\text{B}}T_c\approx3.5$. The phase diagram of HTS is much more
complex (Fig.~\ref{fig_fig5_phasediagram}). The superconducting gap seems to be
unrelated to $T_c$ and a new state---the PG state---appears above $T_c$ as
discussed in the previous paragraphs. Although it is generally accepted that the
pseudogap is intimately related to the occurrence of high $T_c$, there is
currently no consensus as to its nature \citep[see reviews
by][]{Timusk-1999,Phillips-2003}. Two alternative phase diagrams, depicted in
Fig.~\ref{fig_DPD-QCP}, are generally proposed. Either $T^*$ merges with $T_c$
on the overdoped side, or $T^*$ crosses the superconducting dome and possibly
falls to zero at a quantum critical point (QCP).

Different scenarios can be associated with different phase diagrams. For
instance, if the pseudogap would be the signature of a phase-disordered $d$-wave
superconductor, a phase diagram like the one shown in Fig.~\ref{fig_DPD-QCP}a
would naturally result, although in this scenario the $T^*$ line could also
cross the superconducting dome. In the scenario that the pseudogap would be a
phenomenon different and largely independent of superconductivity, like a charge
density wave, it has been proposed that the dome results from the competition of
the two phenomena, the pseudogap winning out at low doping and superconductivity
at higher doping. In such a scenario it is expected that $T^*$ would fall below
$T_c$ roughly at optimal doping and end in a QCP hidden below the dome as shown
in Fig.~\ref{fig_DPD-QCP}b.  If this would be the correct explanation, the phase
diagram shown in Fig.~\ref{fig_DPD-QCP}b should be a universal phase diagram. It
was in fact for a long time assumed that the pseudogap was associated with the
underdoped region of the phase diagram, since this seems to be the case for
Y123. However, STS has clearly shown that for Bi2212 \cite{Renner-1998a} and
Bi2201 \cite{Kugler-2001} the pseudogap exists also in overdoped samples and
that the phase diagram for these compounds is of the type shown in
Fig.~\ref{fig_DPD-QCP}a. Therefore STS has ruled out the phase diagram in
Fig.~\ref{fig_DPD-QCP}b as a universal phase diagram for HTS materials.

\begin{figure}[tb]
\includegraphics[width=6cm]{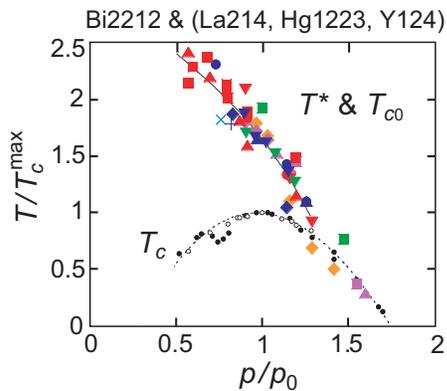}
\caption{\label{fig_DPD-Nakano}
Doping phase diagrams of various hole-doped HTS cuprates. From
\citet{Nakano-1998}.
}
\end{figure}

Although the phase diagram has not yet been investigated systematically by
tunneling spectroscopy, results obtained so far and discussed in this section
already allow to draw some conclusions for Bi2212 and Bi2201: (i) the
superconducting gap and the pseudogap are $T$-independent, (ii) they have the
same amplitude, (iii) the PG temperature $T^*$ increases monotonically when
underdoping and scales with $\Delta_p$, and (iv) the pseudogap clearly exists in
the overdoped region. These results obtained by STS---a method which probes
directly the quasiparticle DOS---give strong support to the scenario depicted in
Fig.~\ref{fig_DPD-QCP}a. This picture is corroborated by ARPES \citep[see review
by][p.~520]{Damascelli-2003} and by various bulk techniques for several other
cuprates (La214, Y124, Hg1223, Tl2201), as shown in Fig.~\ref{fig_DPD-Nakano}.
There are compounds, however, which deviate from the diagram of
Fig.~\ref{fig_DPD-QCP}a. In Bi2201 the phase fluctuations seem to suppress $T_c$
even in the strongly overdoped region (Fig.~\ref{fig_Tdep-Bi2201}c). In the case
of Y123 (Sec.~\ref{sect_pseudogap_TdepY123}) the superconducting gap scales with
$T_c$ \cite{Yeh-2001}. Furthermore, STS data at optimal doping does not reveal
any clear pseudogap \cite{Maggio-Aprile-2000}, suggesting that the $T^*$ line is
shifted towards the underdoped side. This is consistent with the early NMR
experiments \cite{Warren-1989,Alloul-1989} and with a recent high-resolution
dilatometry study by \citet{Meingast-2001}. These authors showed that phase
fluctuations dramatically suppress $T_c$ from its mean-field value
$T_c^{\text{MF}}$ in UD Y123, whereas for overdoped samples $T_c \approx
T_c^{\text{MF}}$. Thus $T_c^{\text{MF}}$ exhibits a similar doping dependence as
$T^*$, suggesting that the pseudogap in Y123 is due to phase-incoherent Cooper
pairs. Particularly interesting in this regard are the recent thermal
conductivity measurements of \citet{Sutherland-2005} which indicate the presence
of $d$-wave like nodal fermions even \emph{outside} the superconducting dome of
Fig.~\ref{fig_DPD-QCP}a. The presence of such nodal excitations provides strong
support for the picture of the pseudogap as a superconducting ground state
disordered by strong quantum \emph{phase} fluctuations which nonetheless
maintains a robust $d$-wave pairing \emph{amplitude} \cite{Franz-2001b}. In
contrast, results obtained using other bulk techniques like transport, heat
capacity, and infrared spectroscopy tend to favor the QCP scenario \citep[see
\textit{e.g.}][]{Tallon-2001}. No consensus is yet reached on this debate, which
involves the non-trivial question of what each experimental technique is
effectively attributing to the pseudogap and how these characteristics are
related to each other.

From the point of view of tunneling, which gives the most direct access to the
pseudogap, the results discussed here for hole-doped cuprates give strong
support to the idea that the pseudogap is a precursor of the superconducting
gap, reflecting fluctuating or non-coherent pairs. However, in the case of
electron-doped HTSs more investigations are needed to reach a general view. The
data available at present have been interpreted in terms of a QCP scenario
\cite{Alff-2003} as briefly mentioned in Sec.~\ref{sect_pseudogap_e-doped}.

\subsection{Summary}
\label{sect_pseudogap_summary}

Scanning tunneling spectroscopy allowed to directly visualize the existence of a
pseudogap in the high-temperature DOS of Bi2212 and Bi2201. As the temperature
is raised from $T<T_c$ the superconducting gap $\Delta_p$ remains constant and
smoothly evolves into the pseudogap across $T_c$. The coherence peaks are
abruptly reduced at $T_c$ giving rise to a generic PG spectrum. Above $T_c$, the
pseudogap is gradually filling up and remains essentially constant, with an
apparent tendency to increase at higher temperature before vanishing at the
crossover temperature $T^*$.

The generic PG spectrum measured at $T_c$ in Bi2212 presents the following
hallmarks: (i) the pseudogap and the superconducting gap have a comparable
magnitude and scale with each other as a function of doping; (ii) the coherence
peak at negative bias is completely suppressed; (iii) the coherence peak at
positive bias is strongly reduced and shifts to slightly higher energies; (iv)
the zero-bias conductance is increased. Furthermore (v) this PG spectrum is
observed on both underdoped and overdoped samples, and (vi) the PG temperature
$T^*$ scales with $\Delta_p$ for all doping levels. These characteristics are
common to Bi2212 and Bi2201 in spite of their radically different
superconducting parameters. Furthermore, the PG signature is also observed at
low temperatures in vortex cores and on disordered samples, where phase
coherence is reduced or suppressed. Finally, it appears that for several HTS
cuprates the values of $T^*$ and $\Delta_p$ measured by STS roughly obey the
$d$-wave BCS scaling law, $2\Delta_p/k_{\text{B}}T^*=4.3$, consistent with other
experimental techniques.

\section{Imaging of vortex matter and core structure by STS}
\label{sect_vortices}

The vortex state study provides a very direct access to the fundamental
properties of superconductors. The electronic structure of the vortex cores, as
well as the interaction between the flux lines, are intimately connected with
the nature and the behavior of the charge carriers. STS offers unique
capabilities to investigate these properties. STS provides a way to detect the
individual flux lines and visualize the vortex distribution in real space, as
well as the structure of the cores. In addition, the probe can access the
electronic excitations within the cores, revealing important and often
unexpected properties of the superconducting pairing states.

Ten years after the theoretical prediction of type-II superconductivity by
\citet{Abrikosov-1957}, real-space imaging of vortices was achieved using Bitter
decoration in lead \cite{Essmann-1967}. Various techniques of vortex imaging in
real space have been developed since then, all relying on the magnetic field
generated by the flux lines, and hence restricted to relatively low fields. In
the case of extreme type-II superconductors like HTS, where the penetration
depth $\lambda$ amounts to hundreds of nanometers, the magnetic imaging contrast
starts to be damped at fields not exceeding very much $H_{c1}$.

A breakthrough occurred when \citet{Hess-1989} used STS to map the vortex
lattice of NbSe$_2$, by detecting systematic spatial variations of the tunneling
conductance. Since the latter occur over distances of the order of the coherence
length $\xi$ around the vortex center, this technique can in principle be used
up to fields where the cores start to overlap, \textit{i.e.} $H_{c2}$. In
practice, because of the small size of the cores, the relatively slow sampling
and the limited field of view of the instrument, vortex imaging by STS should be
considered as a complementary tool to other vortex mapping techniques,
especially at very low fields or in the case of vortex dynamics studies.

This section will start with a brief overview of vortex matter studies by STS.
The focus will then be set on the specific electronic signature of the vortex
cores. One of the most important observation is that the vortex cores in Bi2212
exhibit pseudogap-like spectra, revealing the non-BCS character of
superconductivity in this compound. The existence of low-energy discrete states
in both Y123 and Bi2212 is another surprising property of the cores. The
properties of the vortices in conventional BCS superconductors will also be
presented, in order to establish comparisons with the striking electronic
signatures found in HTS.

\subsection{Vortex matter imaging by STS}
\label{sect_vortex_matter}

STS imaging of the vortices relies on the fact that their cores affect the
quasiparticle excitation spectra. Images are usually obtained by mapping the
tunneling conductance in real space at a particular bias voltage (see
Sec.~\ref{sect_technique}). The imaging contrast depends strongly upon the
chosen bias voltage, since the spatial variations of the tunneling conductance
close to the vortex are usually energy dependent. Consequently, the visual
aspect of the vortex cores in the images depends on the mapping energy. The
performances of the technique are also inherently related to the quality of the
sample surfaces, since the spectroscopic signatures of superconductivity are
easily affected by a non adequate surface layer
(Sec.~\ref{sect_characterization}). Inhomogeneities like structural defects,
chemical adsorbates, impurities, or electronic inhomogeneities will generally
perturb the detection of individual vortices.

Vortex imaging by STS finds an obvious interest in the study of the spatial
distribution of the flux lines. Most conventional superconductors have an
isotropic BCS order parameter, and when pinning is not effective, vortices
arrange in a regular lattice. When pinning is more relevant, vortex lattices
become disordered. The STS observation of the flux lines on BCS compounds
confirms these expectations. In addition, STS imaging provides information on
the shape of the cores. Both informations give a direct access to the anisotropy
of the superconducting state.

Because obtaining surfaces with homogeneous electronic characteristics is
difficult, vortex imaging by STS was achieved on only a few materials,
essentially conventional superconductors. For HTS, degradable surfaces and the
very small coherence lengths imply that the superconducting spectral
characteristics may be affected by the nature of the topmost layer. The
technological developments allowing the synthesis of large and ultra pure single
crystals largely contributed to the recent breakthroughs in spectroscopic
studies and vortex imaging of some HTS.

\subsubsection{NbSe$_2$ and other conventional superconductors}

\begin{figure}[tb]
\includegraphics[width=8.6cm]{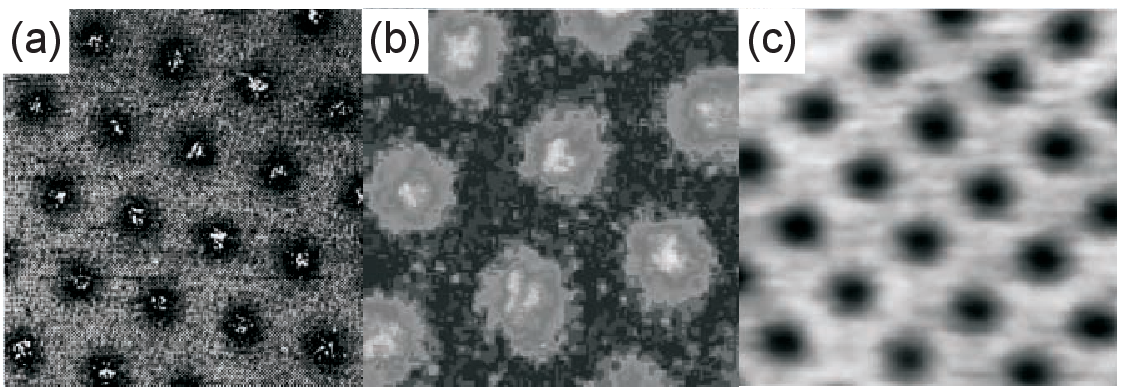}
\caption{\label{VorticesSTSordered}
STS images of ordered vortex lattices in conventional superconductors. (a)
$400\times400$~nm$^2$ hexagonal lattice in NbSe$_2$ ($T=1.3$~K, $H=0.3$~T) from
\citet{Renner-1991}. (b) $250\times250$~nm$^2$ hexagonal lattice in MgB$_2$
($T=2$~K, $H=0.2$~T) from \citet{Eskildsen-2002}. (c) $170\times170$~nm$^2$
square lattice in LuNi$_2$B$_2$C ($T=4.2$~K, $H=1.5$~T) from
\citet{DeWilde-1997a}. The mapping energy was $\Delta_p$ for (a) and (c), and 0
for (b), explaining the inverted contrast.
}
\end{figure}

With easily cleavable planes and a reasonably high $T_c$ (7.2~K), NbSe$_2$ is an
ideal candidate for vortex studies using STS. The first experiments of
\citeauthor{Hess-1989} were followed by many other studies \cite{Renner-1991,
Behler-1994}.

The standard way to detect vortices in NbSe$_2$ is to map the conductance
measured at the energy of the coherence peaks. An example is shown in
Fig.~\ref{VorticesSTSordered}a revealing vortices arranged into a perfect
hexagonal lattice with a lattice parameter of $\sim90$~nm. The expected
inter-vortex spacing is
    $
        a_{\triangle}=(2\Phi_0/\sqrt{3}H)^{\frac{1}{2}}
        =48.9/\sqrt{H~\text{[T]}}~\text{nm},
    $
yielding $a_{\triangle}=89$~nm for $H=0.3$~T. Since the magnetic profile of a
single flux line in Fig.~\ref{VorticesSTSordered}a extends over about one
quarter of the total image, techniques sensitive to the magnetic field would be,
in the same conditions, incapable to distinguish individual vortices. The vortex
cores look isotropic with an apparent radius of 15 to 20~nm, larger than the
coherence length of NbSe$_2$ estimated to be 7.7~nm from critical field
anisotropy measurements \cite{Trey-1973}.

The relatively long acquisition times do not preclude vortex dynamics studies by
STS, provided that the flux line motion is sufficiently slow.
\citet{Renner-1993} observed vortices with an elongated shape due to a slow
motion of the lattice during the line by line scanning of the tip. After some
time, this motion stopped and the vortex cores appeared perfectly isotropic.
More recently, much faster vortex imaging was achieved by controlling the tip
position, without interrupting the feedback loop, using a voltage close to the
gap energy, rather than the commonly-used biases in the CITS technique
\cite{Troyanovski-1999,Troyanovski-2002}.

Disordered lattices in irradiated NbSe$_2$ samples illustrated the pinning of
vortices by columnar defects \cite{Behler-1994}. This observation was helped by
the fact that the coherence length in NbSe$_2$ is much larger than the size of
the defects. When the cores and the defects have similar sizes, the
spectroscopic signatures of the cores may be masked by the defects, and thus
difficult to distinguish \cite[see \textit{e.g.} Zn-doped Bi2212
by][]{Pan-2000b}.

A hexagonal vortex lattice was also seen in MgB$_2$ single crystals, as shown
in Fig.~\ref{VorticesSTSordered}b \cite{Eskildsen-2002}. The expected vortex
spacing of 109~nm is readily verified in the image. More surprisingly,
\citet{DeWilde-1997b} found a square vortex lattice in the borocarbide
LuNi$_2$B$_2$C (Fig.~\ref{VorticesSTSordered}c) at a field of 1.5~T. The
expected spacing for a square lattice,
    $
        a_{\square}=(\Phi_0/H)^{\frac{1}{2}}
        =45.5/\sqrt{H~\text{[T]}}~\text{nm},
    $
is again respected (37~nm at 1.5~T). The square symmetry was explained in a
Ginzburg-Landau theoretical approach, considering a fourfold perturbation term
in the free energy arising from the underlying tetragonal structure of the
crystal \cite{DeWilde-1997a}. A transition from a distorted hexagonal lattice at
low fields to a square lattice at high fields was predicted, and experimentally
confirmed by small angle neutron scattering \cite{Eskildsen-1997}. Later, this
transition was directly observed by STS in another borocarbide compound,
YNi$_2$B$_2$C \cite{Sakata-2000}, and in the cubic A15 compound V$_3$Si
\cite{Sosolik-2003}.

\subsubsection{High-temperature superconductors: Y123 and Bi2212}

The behavior of vortices in HTS cuprates differs radically from what is observed
in BCS superconductors. At first sight, the high $T_c$'s and the large
superconducting gaps make them ideal compounds for STS, since for most of them,
the electronic characteristics can be probed at 4.2~K or above, and the
characteristic DOS features are situated at relatively high energies. But
because of difficulties in preparing surfaces with sufficient homogeneity, the
study of vortices by STS remained a considerable challenge. Up to now, imaging
of vortices has been possible only in Y123 and Bi2212. The very small coherence
length in these materials have two direct observable implications. First, the
vortex cores are tiny in comparison to conventional superconductors, and thus
difficult to localize. Second, the vortices get easily pinned by defects. As a
consequence, at the relatively high fields which are suitable for STS, flux
lines lattices show no perfect order in HTS.

\begin{figure}[tb]
\includegraphics[width=7cm]{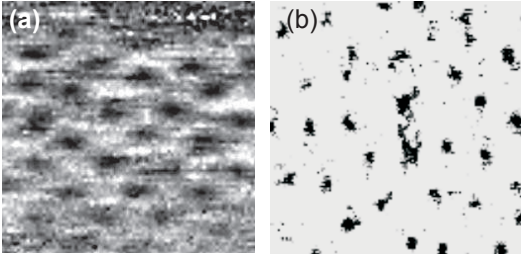}
\caption{\label{VorticesSTSDisordered}
$100\times100$~nm$^2$ STS images of (a) a slightly disordered oblique lattice in
Y123 \cite[from][]{Maggio-Aprile-1995} and (b) a disordered vortex distribution
in Bi2212 \cite[from][]{Hoogenboom-2001a}. Both images were acquired at
$T=4.2$~K and $H=6$~T. The contrast is defined by the conductance measured at
the gap energy (20~meV for Y123 and 30~meV for Bi2212) normalized by the
zero-bias conductance.
}
\end{figure}

Examples of vortices imaged with STS in a 6~T field in both Y123 and Bi2212 are
presented in Fig.~\ref{VorticesSTSDisordered}. In Y123 the vortices arrange in a
slightly disordered oblique lattice (Fig.~\ref{VorticesSTSDisordered}a). In
Bi2212 they are more disordered (Fig.~\ref{VorticesSTSDisordered}b). Note that
the entire scan range of Fig.~\ref{VorticesSTSDisordered} fits into a single
vortex core of Fig.~\ref{VorticesSTSordered}b, graphically illustrating the
small size of the cores in HTS.

\paragraph{Y123}
\label{sect_vortex_matter_Y123}

\begin{figure}[b]
\includegraphics[width=8.6cm]{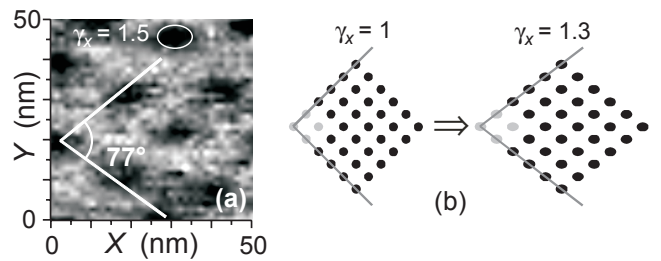}
\caption{\label{VorticesY123}
(a) $50\times50$~nm$^2$ STS image of vortices in Y123. The primitive vectors of
the oblique lattice form an angle of 77$^{\circ}$, and the cores present an
elliptic distortion of about 1.5 ratio. (b) Schematic diagram showing the
formation of an oblique lattice, starting from a square lattice distorted by an
anisotropy factor of 1.3. Adapted from \citet{Renner-1996}.
}
\end{figure}

Y123 was the first non-conventional superconductor whose vortex lattice could be
investigated by STS, and this was achieved by \citet{Maggio-Aprile-1995} on
as-grown surfaces. Since then, two other groups reported the STS observation of
flux lines in Y123, either on as-grown surfaces \cite{Hubler-1998}, or on
chemically etched surfaces \cite{Shibata-2003b}. Fig.~\ref{VorticesY123}a shows
an image acquired in optimally-doped Y123 ($T_c=92$~K). The observed vortex
density matches the 6~T applied field. The lattice shows a local oblique
symmetry, with an angle of about 77$^{\circ}$. Vortex cores reveal an apparent
radius of about 5~nm, but systematically present an elliptical shape with an
axis ratio of about 1.5 (see Fig.~\ref{VorticesY123}a).
\citet{Maggio-Aprile-1995} noticed that this distortion is independent of
scanning direction and not related to any flux line motion, and they ascribed it
to the $ab$-plane anisotropy of Y123.

In the framework of an anisotropic Ginzburg-Landau model, the observed oblique
lattice in Y123 can be interpreted as a square lattice distorted by an
anisotropy factor of $\sim 1.3$, or as a hexagonal lattice distorted by an
anisotropy factor of $\sim 2.2$. Since it is expected that the anisotropy
controlling the distortion of the lattice is also responsible for the intrinsic
distortion of the cores, the observed value of $\sim 1.5$ for the latter led
\citet{Renner-1996} to conclude that the lattice corresponds to a deformed
square lattice as illustrated in Fig.~\ref{VorticesY123}b. Similar anisotropy
factors have been derived from penetration depth measurements in Y123
\cite{Zhang-1994}. At analog magnetic fields, neutron scattering measurements
showed similar distortions \cite{Keimer-1994}. More recent neutron scattering
investigations evidenced a continuous transition from a distorted hexagonal to a
nearly perfect square lattice at high fields \cite{Brown-2004}. Square vortex
lattices can be expected in a number of situations where a four-fold
perturbation field acts. In particular, a square lattice was predicted for
$d$-wave superconductors \cite{Won-1995, Affleck-1997}. It is not yet clear
whether the tendency to order into a square lattice at high field is due to the
$d$-wave symmetry of the order parameter, or to other factors. On the other
hand, Bitter decoration measurements at low fields revealed a regular
\cite{Gammel-1987, Vinnikov-1988, Dolan-1989a} or slightly distorted
\cite{Dolan-1989b} hexagonal lattice, as expected at low vortex density.

\begin{figure}[b]
\includegraphics[width=8.6cm]{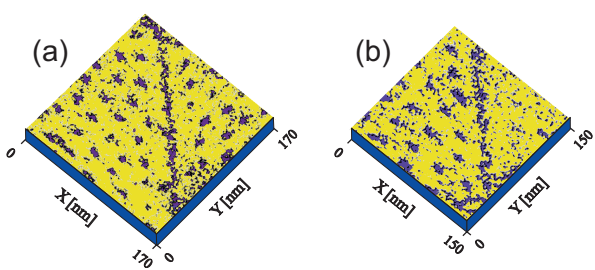}
\caption{\label{VorticesTwinY123}
STS observation of a twin boundary in Y123. (a) $170\times170$~nm$^2$ image at
$H=3$~T; the flux lines are distributed on both sides of the boundary with an
equal density. (b) After reducing the field to 1.5~T, the flux lines pile up on
the left domain and reach a density corresponding to about 2.5~T. From
\citet{Maggio-Aprile-1997}.
}
\end{figure}

The disorder observed in the vortex lattice underlines the influence of strong
pinning in compounds with small coherence lengths. In Y123, a substantial vortex
creep is observed after switching on the magnetic field, with flux lines
continuously moving for a couple of days. The influence of twin boundaries on
pinning in Y123 single crystals has stimulated various debates and
investigations \cite{Gammel-1992, Vlasko-Vlasov-1994, Welp-1994, Crabtree-1996,
Herbsommer-2000}. STS measurements demonstrate that the twin boundaries act as
very efficient barriers, blocking the movements of the flux lines perpendicular
to the twin plane. This phenomenon is illustrated in
Fig.~\ref{VorticesTwinY123}a, which shows vortices mapped out in a field of 3~T
on both sides of a twin boundary \cite{Maggio-Aprile-1997}, seen as the dark
line running along the $[110]$ direction. Vortices are found on both sides with
the same average density. The anisotropic shape of the cores is rotated by
90$^\circ$ between both domains, demonstrating that the dark line separates two
regions with inverted $a$ and $b$ crystallographic axes. The field was then
reduced to 1.5~T, and a piling up of flux lines on the left side of the twin
boundary was observed, corresponding to a field of 2.5~T, substantially higher
than the applied field. The right side showed a vortex free region extending
over more than hundred nanometers (Fig.~\ref{VorticesTwinY123}b). The gradients
of the magnetic field could be directly measured from the flux line
distribution, and allowed the authors to estimate a current density close to the
depairing current limit for Y123 ($\sim3\times10^8~$A/cm$^2$). The dark line on
the twin boundary was attributed to trapped vortices, too densely packed or
mobile to be resolved.

\paragraph{Bi2212}

\begin{figure}[tb]
\includegraphics[width=8.6cm]{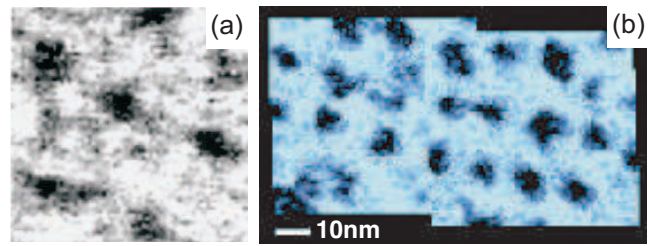}
\caption{\label{VorticesBi2212}
STS images of vortices in Bi2212. (a) $40\times40$~nm$^2$ image of vortex cores
in overdoped Bi2212 at $H=6$~T; from \citet{Renner-1998b}. (b) Ordered phase at
$H=8$~T, revealing a nearly square symmetry; from \citet{Matsuba-2003a}.
}
\end{figure}

In Bi2212, the quasiparticle spectra taken outside and inside the vortex cores
show rather small differences (see Fig.~\ref{fig_PG-VC-Tdep2}).
\citet{Renner-1998b} imaged for the first time vortices in overdoped Bi2212
using negative energies. As seen in Fig.~\ref{VorticesBi2212}a the cores in
Bi2212 are tiny, consistent with a coherence length of the order of a few atomic
unit cells. They usually exhibit irregular shapes (see
Sec.~\ref{sect_vortex_shape_Bi2212}).

In the STS images, the vortices are distributed in a totally disordered manner,
but with an average density matching the 6~T field \cite{Renner-1998b}. This can
be understood since Bi2212, unlike Y123, has a nearly two-dimensional electronic
structure and the vortices are therefore constituted by a stack of 2D elements
(pancakes), weakly coupled between adjacent layers, and easily pinned by any
kind of inhomogeneity. The observed disordered lattice fits into the commonly
accepted vortex phase diagram for Bi2212, where the low-temperature high-field
region is associated with a disordered vortex solid. At very low fields takes
place an ordered Bragg glass phase, difficult to probe with STS because of the
large inter-vortex distances. At 8~T, however, \citet{Matsuba-2003a} observed a
short-range ordered phase in Bi2212, presenting a nearly square symmetry almost
aligned with the crystallographic $ab$ directions (Fig.~\ref{VorticesBi2212}b).

\subsection{Electronic structure of the cores}
\label{sect_vortex_core}

\subsubsection{BCS superconductors}

In \citeyear{Caroli-1964} \citeauthor*{Caroli-1964} provided an approximate
analytical solution of the Bogoliubov-de Gennes equations for an isolated vortex
in a clean $s$-wave superconductor. They predicted the existence of electronic
states bound to the vortex at energies below the superconducting gap. These
bound states form a discrete spectrum with a typical inter-level spacing
$\Delta^2/E_{\text{F}}$. For most superconductors, this spacing is so small that
the detection of individual bound states is hampered by the finite temperature
and by impurity scattering. In NbSe$_2$, $\Delta^2/E_{\text{F}}$ is typically
$\sim0.05-0.5$~K, and so far, it was not possible to resolve an ideal discrete
spectrum. Instead, a strong zero-bias conductance peak (ZBCP) is observed,
understood as the continuous envelope of a set of low-energy states.

\begin{figure}[tb]
\includegraphics[width=7cm]{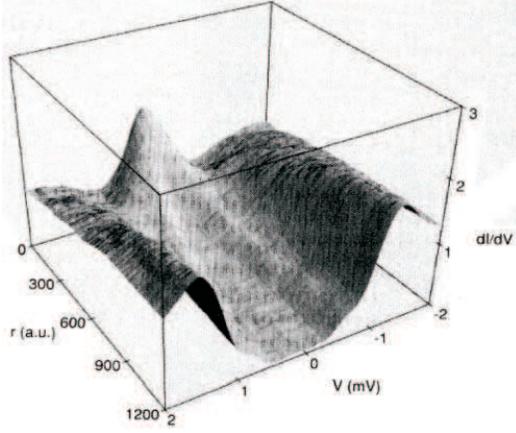}
\caption{\label{BoundStatesNbSe2}
Conductance spectra acquired at $T=0.3$~K, $H=0.05$~T along a 50~nm path leading
from the center ($r=0$) to the outside ($r=1200$) of a vortex core in NbSe$_2$
\cite{Hess-1990}. The splitting of the zero-bias peak into one electron and one
hole branch shows the presence of many bound states in the core (see text). From
\citet{Gygi-1991}.
}
\end{figure}

\citet{Hess-1989} studied vortex cores by STS in pure NbSe$_2$ single crystals
at $T=1.85$~K, and observed features in agreement with theoretical expectations
\cite{Shore-1989}. The progressive splitting of the ZBCP at a distance $r$ from
the center provided another verification of the BCS predictions for vortex
cores in $s$-wave superconductors (Fig.~\ref{BoundStatesNbSe2}). This splitting
reflects the presence of many core states with different angular momenta $\mu$.
The lowest-lying state ($\mu=\frac{1}{2}$) has energy
$E_{\frac{1}{2}}=\frac{1}{2}\Delta^2/E_{\text{F}}\approx0$ and maximum
probability at $r=0$, giving rise to the ZBCP at the vortex center. The energy
of higher-angular momentum vortex states departs from zero as
$E_{\mu}=\mu\Delta^2/E_{\text{F}}$ at low $\mu$ \cite{Caroli-1964}, and these
states have maximum amplitude at a distance $r=r_{\mu}\approx|\mu|/k_{\text{F}}$
from the core center. In the LDOS, these latter states show up as one electron
branch ($E_{\mu}>0$) and one hole branch ($E_{\mu}<0$) at energy $E_{\mu}$ and
position $r_{\mu}$. With increasing $|\mu|$, the electron and hole branches
approach the gap edges at $\pm\Delta$, and merge asymptotically with the
continuum of scattering states with energies $|E|>\Delta$. Thus the
superconducting spectrum is eventually recovered outside the core, at
$r=5$--$10\xi$. These observations were successfully described by a complete
numerical solution of the Bogoliubov-de Gennes equations for an isolated
$s$-wave vortex \cite{Gygi-1990a, Gygi-1991}.

\begin{figure}[tb]
\includegraphics[width=7cm]{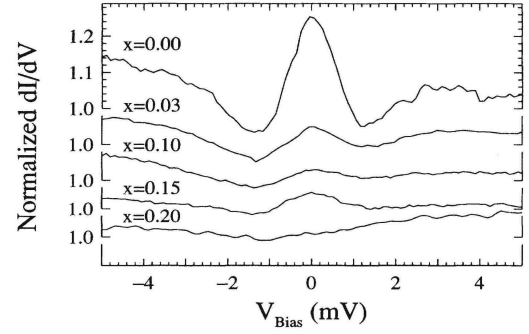}
\caption{\label{DirtyCoresNbSe2}
Tunneling conductance spectra acquired at the vortex-core center of
Nb$_{1-x}$Ta$_x$Se$_2$ at $T=1.3$~K and $H=0.3$~T for different impurity dopings
$x$. For $x=0$ (clean limit), the bound states are observed as a broad ZBCP,
which progressively vanishes for larger amounts of impurities
\cite{Renner-1991}.
}
\end{figure}

In dirty superconductors, quasiparticle scattering mixes states with different
angular momenta, and the zero-bias peak transforms into a quasi-normal (flat)
DOS spectrum when the quasiparticle mean free path is smaller than the coherence
length. This effect was observed by STS in Ta-doped NbSe$_2$ \cite{Renner-1991},
as shown in Fig.~\ref{DirtyCoresNbSe2}. The spectra were acquired in different
Nb$_{1-x}$Ta$_x$Se$_2$ samples, each doped with a specific amount of tantalum.
As seen in the figure, the low-energy spectra taken at the core center
progressively lose the pronounced zero-bias feature, and for dopings $x>0.15$
show a nearly constant conductance reminiscent of the normal-state DOS.

\begin{figure}[tb]
\includegraphics[width=8.6cm]{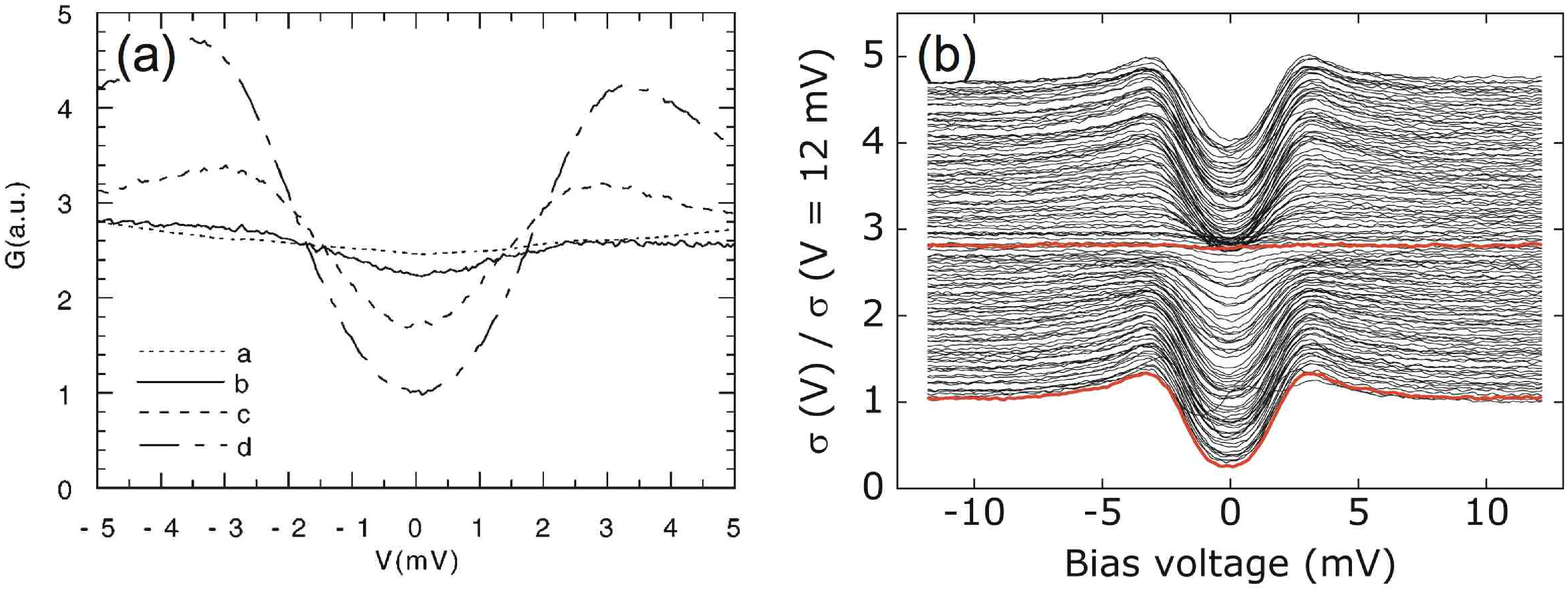}
\caption{\label{DirtyCoresBCS}
Examples of missing zero-bias peak in the vortex cores of conventional BCS
superconductors. (a) Spectra acquired at 47~nm (dashed line) to the center of
the core (dotted line) in LuNi$_2$B$_2$C at $T=4.2$~K and $H=1.5$~T; from
\citet{DeWilde-1997b}. (b) 250~nm long spectroscopic trace crossing a single
vortex core in MgB$_2$ at $T=2$~K and $H=0.2$~T; from \citet{Eskildsen-2002}.
}
\end{figure}

\citet{DeWilde-1997b} reported similar vortex-core spectra for LuNi$_2$B$_2$C
(Fig.~\ref{DirtyCoresBCS}a), with a nearly constant tunneling conductance.
Recently \citet{Nishimori-2004} reported the observation of a zero-bias peak in
pure samples of this compound. In the two-band superconductor MgB$_2$, the
vortex-core spectra acquired on the (001) surface show also a totally flat
conductance (\citet{Eskildsen-2002}, Fig.~\ref{DirtyCoresBCS}b). This
observation is more surprising, since the mean free path in MgB$_2$ is believed
to be much larger than the superconducting coherence length. The absence of a
zero-bias peak was attributed to the fact that the $\pi$-band is predominantly
probed in this geometry. Since this band becomes superconducting through
coupling with the $\sigma$-band, the presence of a flux line, suppressing the
superconducting character of the $\sigma$-band, also affects the $\pi$-band,
which becomes metallic, with an energy independent DOS \cite{Eskildsen-2002}.

\subsubsection{High-temperature superconductors}
\label{sect_CoreHTS}

\paragraph{Y123}

\begin{figure}[tb]
\includegraphics[width=8.6cm]{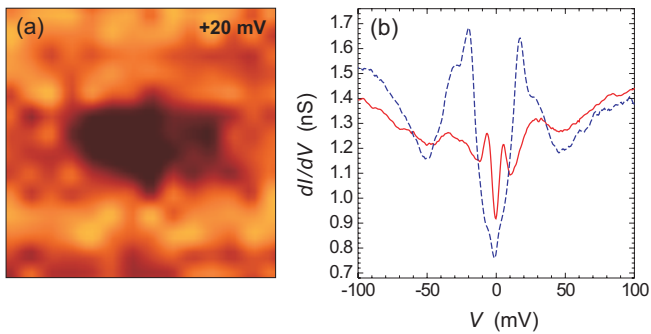}
\caption{\label{CoreStatesY123}
Core states in Y123. (a) $20\times20$~nm$^2$ conductance map acquired at
$+20$~mV, showing a detailed view of a vortex core ($T=4.2$~K, $H=6$~T). (b)
Conductance spectra acquired at the center (red) and at about 20~nm from the
center of the vortex core (dashed blue). The low-energy core states are located
at about $\pm5.5$~meV. Adapted from \citet{Maggio-Aprile-1995}.
}
\end{figure}

The predictions and experimental confirmations of core bound states in BCS
$s$-wave superconductors raised important questions when the exotic properties
of HTS were uncovered. Among these, the question of the electronic signature of
the vortex cores in HTS was at the time completely open. The first answers came
from measurements on optimally-doped Y123 single crystals by
\citet{Maggio-Aprile-1995}. Fig.~\ref{CoreStatesY123}a shows a tunneling
conductance map of a vortex recorded at an energy close to the positive
coherence peak, and illustrates the suppression of the latter in the core
region. The most striking observation comes from the conductance spectrum at the
core center (Fig.~\ref{CoreStatesY123}b). Contrary to the case of conventional
superconductors, the core spectra in Y123 neither reveal any broad zero-bias
feature nor a flat conductance. Instead, a pair of low-energy peaks clearly
emerging above the conductance background is observed at an energy of about
$\pm5.5$~meV. While both coherence peaks near $\pm20$~meV disappear inside the
core, a broad feature of reduced intensity persists at slightly higher energies
(30--40~meV).

\begin{figure}[b]
\includegraphics[width=7cm]{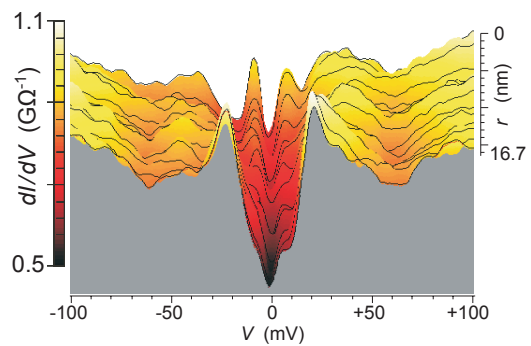}
\caption{\label{CoreStatesTraceY123}
Conductance spectra taken along a 16.7~nm path into a vortex core in Y123 at
$T=4.2$~K and $H=6$~T. Together with the vanishing of the 20~mV coherence peaks,
a pair of low-energy peaks emerges when the tip approaches the center of the
vortex core ($r=0$). From \citet{Renner-1998b}.
}
\end{figure}

The observation of isolated core states at finite energy in Y123 appeared
contradictory in various respects. On one hand, attempts to interpret the
observed bound states like classical BCS $s$-wave localized states failed for
two reasons:~first, it would imply an unusually small Fermi energy
($E_{\text{F}}\sim36$~meV); second, there is a complete absence of energy
dispersion at different positions within the core
(Fig.~\ref{CoreStatesTraceY123}). When the tip starts entering the core
($r$=16.7~nm), the 20~meV coherence peaks progressively vanish, and the
low-energy core states develop while the tip approaches the core center ($r$=0).
The energy at which the core states are observed does not vary, indicating that
only one pair of localized states exists below the gap. Note that a weak
low-energy structure persists in the spectra outside the vortex cores (see
spectrum at $r=16.7$~nm in Fig.~\ref{CoreStatesTraceY123} and the dashed
spectrum in Fig.~\ref{CoreStatesY123}b). A similar structure also exists in the
absence of magnetic field (see Sec.~\ref{sect_gap_spectroscopy}). Although the
energy of this structure observed outside the cores is approximately the same as
the core state energy, the question of its origin and the relation it has with
the core state is still open.

On the other hand, this observation does not fit with the $d$-wave character of
the cuprate compounds. Following the observation of anomalous core spectra in
Y123, an important effort was dedicated to calculating the quasiparticle spectra
in the vortex cores of $d$-wave superconductors, where the presence of nodes in
the order parameter changes the properties of the low-energy excitations. Based
on the Bogoliubov-de Gennes \cite{Soininen-1994, Wang-1995, Morita-1997a,
Franz-1998b, Yasui-1999} or Eilenberger \cite{Ichioka-1996, Ichioka-1999}
equations, it was shown that the vortex-core energy spectrum is continuous
rather than discrete, and that the quasiparticles in the core are not
exponentially localized as in $s$-wave superconductors. The theoretical LDOS at
the center of the core shows a broad peak at zero energy.

Therefore the observation of discrete states at finite energy is in
contradiction with the $d$-wave BCS picture. A possible magnetic origin of these
states has been ruled out because (i) the Zeeman splitting can be estimated to
be about ten times smaller than the measured energy of the states, and (ii) the
peak energy position was found to be independent of the applied magnetic field.
Finite-energy core states in Y123 were confirmed in more recent STS studies by
\citet{Shibata-2003b}. Infrared absorption measurements \cite{Karrai-1992} also
clearly identified the existence of low-energy quasiparticle excitations, in
quantitative agreement with all STS observations.

\paragraph{Bi2212}

\begin{figure}[b]
\includegraphics[width=7cm]{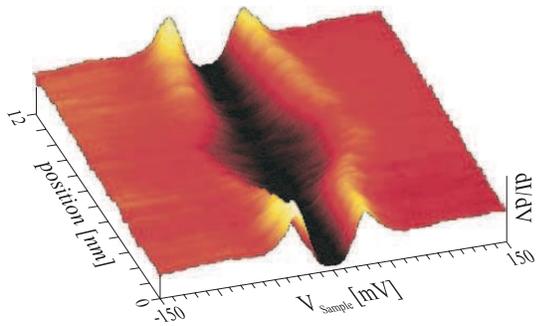}
\caption{\label{CoreStatesBi2212PG}
Conductance spectra taken along a 12~nm path across a vortex core in overdoped
Bi2212 ($T_c=77$~K) at $T=2.6$~K and $H=6$~T. The spectra acquired at the center
of the core reveal a pseudogap-like shape similar to the one observed above
$T_c$. From \citet{Kugler-2000}.
}
\end{figure}

Vortex core spectroscopy in Bi2212 provided new insights into the debate
concerning the origin of the pseudogap, starting with the STS measurements of
\citet{Renner-1998b} on both underdoped and overdoped Bi2212 single crystals.
The evolution of the spectra across a vortex core in overdoped Bi2212 is shown
in Fig.~\ref{CoreStatesBi2212PG}. When entering the core, the spectra evolve in
the same way as when the temperature is raised above $T_c$:~the coherence peak
at negative energy vanishes over a very small distance, the coherence peak at
positive energy is reduced and shifts to slightly higher energies, and the
dip/hump features disappear. The observation of such pseudogap-like spectra in
the cores of Bi2212 is discussed in more details in Sec.~\ref{sect_pseudogap}.

\begin{figure}[tb]
\includegraphics[width=8.6cm]{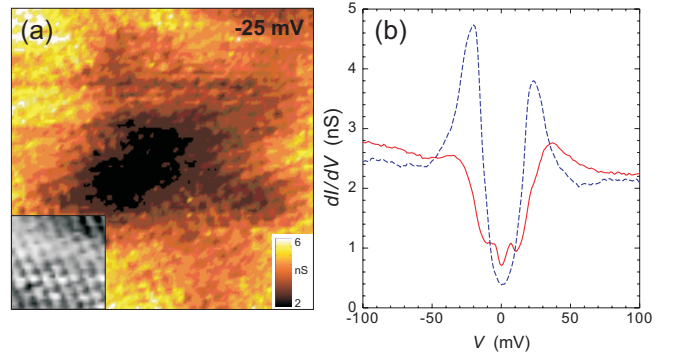}
\caption{\label{CoreStatesBi2212}
(a) $8.7\times8.7$~nm$^2$ conductance map of a vortex core in Bi2212, acquired
at $-25$~mV and 6~T. The inset shows the simultaneously acquired topography at
the same scale. (b) Spectra taken close to the core center with two peaks at
$\pm6$~meV corresponding to a pair of core states (red) and outside the vortex
core (blue). Adapted from \citet{Levy-2005}.
}
\end{figure}

While the first vortex core investigations in Bi2212 focused on pseudogap
aspects, \citet{Pan-2000b} and \citet{Hoogenboom-2000a} showed that weak
low-energy structures were also present in the vortex core spectra.
Fig.~\ref{CoreStatesBi2212}a shows a spectroscopic image, mapped out at
$-25$~meV, of a vortex core in slightly overdoped Bi2212. The full conductance
spectra taken inside and outside the core are shown in
Fig.~\ref{CoreStatesBi2212}b. Contrary to Y123, where the peaks clearly emerge
in the core center, the core states appear in Bi2212 as weak peaks in an overall
pseudogap-like spectrum. For slightly overdoped Bi2212, the energy of these
states is of the order of $\pm6$~meV. A slight increase of the zero-bias
conductance is also systematically measured in the core center.

\begin{figure}[b]
\includegraphics[width=8.6cm]{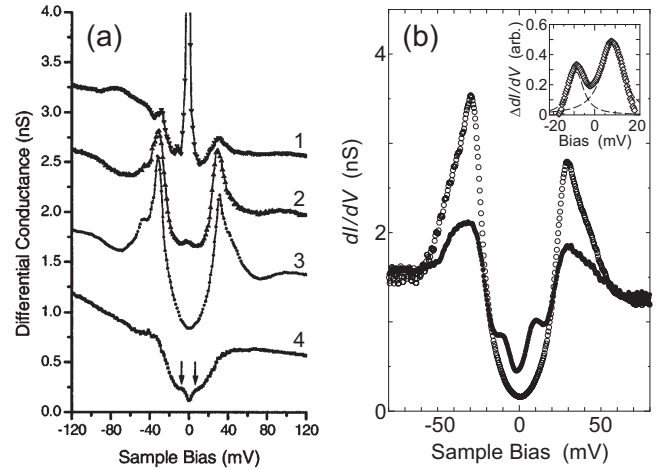}
\caption{\label{CoreStatesBi2212others}
STS measurements of vortex core states in Bi2212. (a) Tunneling spectra acquired
in various sites of Zn-doped Bi2212 samples, reflecting strong (1) and weak (2)
impurities scattering resonances, defect-free superconducting (3) and vortex
core (4) spectra. In the latter case, the arrows indicate the position of the
low-energy core states; adapted from \citet{Pan-2000b}. (b) Tunneling spectra
taken in a vortex core showing localized states at $\pm9$~meV (filled circles)
and outside a vortex core (open circles); from \citet{Matsuba-2003a}.
}
\end{figure}

Various studies reported the observation of such states in Bi2212.
\citet{Pan-2000b} observed that vortices sitting away from impurity sites show
core spectra with the low-energy gapped structure, as indicated with arrows in
the bottom curve of Fig.~\ref{CoreStatesBi2212others}a. These spectra are
compared with the ones acquired on strong or weak impurities shown in the top
two curves of Fig.~\ref{CoreStatesBi2212others}a, where vortices are likely to
be pinned. As a consequence, they deduce that the pair of localized states
reflects an intrinsic character of the vortex cores. The systematic presence of
core states in Bi2212 at different oxygen dopings was also reported by
\citet{Matsuba-2003a}, shown in Fig.~\ref{CoreStatesBi2212others}b. They
observed an asymmetry of the peak heights, which they attributed to
positively-charged vortices.

\begin{figure}[tb]
\includegraphics[width=6cm]{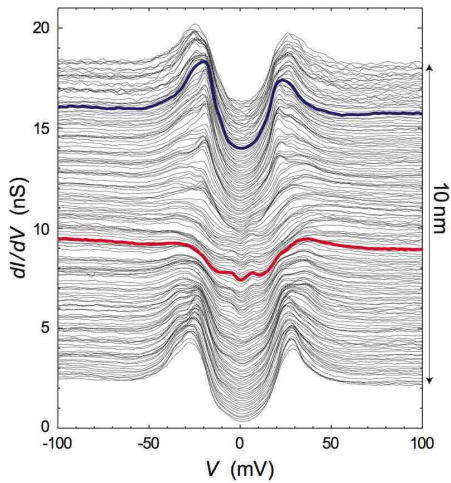}
\caption{\label{CoreStatesBi2212Trace}
10~nm long spectroscopic trace across a vortex in overdoped Bi2212, at $T=1.8$~K
and $H=6$~T. The core-states peaks lie at $\pm6$~meV, and their position does
not change across the vortex (G. Levy, unpublished).
}
\end{figure}

\citet{Hoogenboom-2000a} investigated various samples where a large number of
vortices showed the low-energy structures, measured at an energy independent of
the position within the core, as shown in Fig.~\ref{CoreStatesBi2212Trace}. We
stress that later investigations \cite{Levy-2005} showed that the intensity of
the core-states peaks depends on the precise measurement position within the
core (see Sec.~\ref{sect_modulations_vortex}).

\begin{figure}[tb]
\includegraphics[width=6cm]{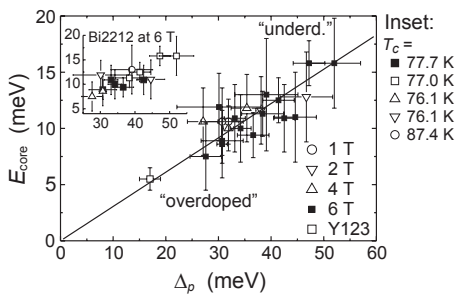}
\caption{\label{CoreVsGap}
Energy of vortex core states $E_{\text{core}}$ as a function of the
superconducting gap $\Delta_p$. The plot shows Bi2212 single crystals of various
dopings, and optimally-doped Y123, at different magnetic fields. The solid line,
with a slope of 0.3, is a linear fit to all Bi2212 data points. The inset shows
data for Bi2212 samples with different $T_c$, at a fixed field of 6~T
\cite{Hoogenboom-2000a}.
}
\end{figure}

The low-energy core states observed in both Y123 and Bi2212 raise two important
questions. Are the core states observed in Y123 and Bi2212 of the same origin?
And how can these core states be understood within the framework of a $d$-wave
symmetry of the order parameter? In order to answer the first point, detailed
core spectra studies were carried out in both Y123 and Bi2212. One first notes
that the spectra acquired in both Y123 and Bi2212 cores present a similar shape.
In addition, although they are much more pronounced in Y123, the low-energy
structures behave similarly in both types of compounds: they do not disperse
when measured at various distances from the center of the core
(Figs.~\ref{CoreStatesTraceY123} and \ref{CoreStatesBi2212Trace}) and they are
magnetic field independent \cite{Maggio-Aprile-1995, Hoogenboom-2001a}. The most
striking fact comes from the scaling between the energy of the core state
$E_{\text{core}}$ and the superconducting gap amplitude $\Delta_p$, as shown in
Fig.~\ref{CoreVsGap}. The correlation found between $E_{\text{core}}$ and
$\Delta_p$ for Y123 (optimal doping) and Bi2212 (various dopings) strongly
suggests that the core states in the two compounds have a common origin.
Moreover, since this correlation is linear (with a slope of about 0.3), the
interpretation of these bound states as \citeauthor*{Caroli-1964} states of an
$s$-wave superconductor with a large gap to Fermi energy ratio can be ruled out,
as one would expect a $\Delta_p^2$ dependence.

Many theoretical studies were undertaken in order to explain the observed
electronic signatures of the vortex cores in HTS. The absence of a large ZBCP at
the vortex center in Y123 and Bi2212 has been tentatively attributed to an
anisotropic tunneling matrix element \cite{Wu-2000}. The general belief,
however, is that the qualitative difference between the vortex-core spectra
measured by STM in HTS and the spectra expected for a $d$-wave BCS
superconductor \cite{Soininen-1994, Wang-1995, Franz-1998b} reflects an
intrinsic property of the superconducting ground state, which distinguishes this
state from a pure BCS ground state, and is presumably related to the anomalous
normal (pseudogap) phase. \citet{Franz-1998b} showed that the admixture of a
small magnetic-field induced complex component with $d_{xy}$ symmetry
\cite{Laughlin-1998} leads to a splitting of the ZBCP due to the suppression of
the $d$-wave gap nodes; as noted by the authors, this interpretation would imply
a strong field dependence of the vortex core spectra, in contradiction with the
observation in Bi2212 (see Fig.~\ref{CoreVsGap}). The effect of an
antiferromagnetic order nucleated at the vortex core was investigated within the
SO(5) theory \cite{Arovas-1997, Andersen-2000} and the Bogoliubov-de Gennes
mean-field framework \cite{Zhu-2001a}. Both approaches lead to a suppression of
the ZBCP at the core center; in the latter model, two symmetric core-state peaks
are found when the on-site repulsion is sufficiently large. Some calculations
based on the $t$-$J$ model have predicted a splitting of the ZBCP at low doping
due to the formation of an $s$-wave component in the order parameter
\cite{Himeda-1997, Han-2000a, Tsuchiura-2003}. A similar conclusion was reached
by \citet{Chen-2002b} who investigated the formation of spin- and charge-density
waves around the vortex and their effect on the spectrum in the core. The
structure of the vortex was also studied within the staggered-flux
\cite{Kishine-2001} or $d$-density wave \cite{Maska-2003} scenarios of the
pseudogap. This model is able to explain the absence of ZBCP in the core, but
does not seem to account for the formation of the core states. A good
qualitative agreement with the experimental spectra was obtained by
\citet{Berthod-2001b} using a model where short-range incoherent pair
correlations coexist with long-range superconductivity in the vortex state. This
model correctly reproduced the measured exponential decay of the LDOS at the
energy of the core states \cite{Pan-2000b}, as well as the increase of the
core-state energy with increasing gap $\Delta_p$.

\subsection{The shape of the vortex cores}
\label{sect_vortex_shape}

The size and shape of the vortex cores mapped out by STS depend directly on how
the mapping contrast is defined in the images. Since the superconducting
coherence peaks at $\Delta_p$ disappear in the core, the bias $eV=\Delta_p$ is
routinely selected as the mapping energy. In $s$-wave superconductors, the decay
length of localized states increases to several $\xi$ as their energy approaches
the gap edge. As a consequence, vortex cores imaged at $eV=\Delta_p$ always
appear larger than expected from the coherence length \cite[][see
Fig.~\ref{BoundStatesNbSe2}]{Gygi-1991}. This is clearly the case in all the
spectroscopic images of flux lines presented in Sec.~\ref{sect_vortex_matter}.
\citet{Volodin-1997} studied the core size in NbSe$_2$ as a function of the bias
imaging voltage, and indeed found that the apparent core size is reduced when
the selected bias voltage decreases.

\subsubsection{Intrinsic shape of vortices}

\begin{figure}[tb]
\includegraphics[width=8.6cm]{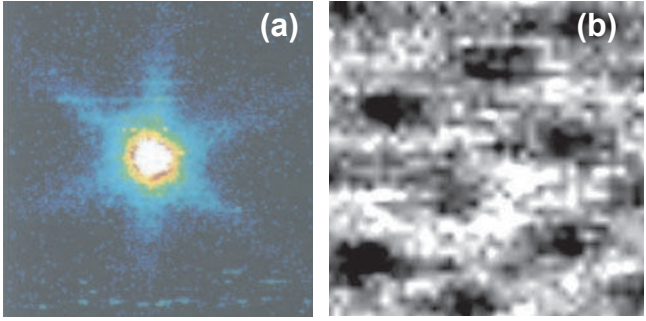}
\caption{\label{IntrinsicShape}
(a) 150$\times$150~nm$^2$ zero-bias conductance map taken at $H=500$~Gauss and
$T=0.3$~K, revealing the sixfold shape of a vortex core in NbSe$_2$; from
\citet{Hess-1990}. (b) 50$\times$50~nm$^2$ spectroscopic image of flux lines in
Y123, at $H=6$~T and $T=4.2$~K; from \citet{Maggio-Aprile-1996}.
}
\end{figure}

In NbSe$_2$, detailed measurements performed at 0.3~K allowed \citet{Hess-1990}
to resolve fine structures in the imaged cores, revealing star-shaped patterns
with sixfold symmetry (Fig.~\ref{IntrinsicShape}a). Theoretical studies based on
the Bogoliubov-de Gennes formalism and using a sixfold perturbation term were
able to explain the shape of the objects \cite{Gygi-1990b}. The fact that the
pattern orientation strikingly depends on the mapping energy was elucidated
later by \citet{Hayashi-1996}, considering an anisotropic $s$-wave pairing with
a hexagonal symmetry.

In $d$-wave BCS superconductors the anisotropic order parameter leads to a
fourfold anisotropy of the low-energy LDOS around vortices. The LDOS is larger
in the direction of the nodes where the spatial decay of the core states is
slower \cite{Morita-1997b, Franz-1998b}. In the microscopic Bogoliubov-de Gennes
calculations this LDOS anisotropy is weak, but can be as large as $\sim 10\%$
close to the core center \cite{Morita-1997b, Berthod-2001b}, and should be
accessible experimentally in clean samples. The Eilenberger formalism leads to a
more complex and somewhat stronger fourfold anisotropy \cite{Schopohl-1995,
Ichioka-1996}. Both semiclassical \cite{Franz-1999} and microscopic
\cite{Berthod-2001b} calculations showed that self-energy effects can reduce the
anisotropy considerably. Although the scope and conclusions of these studies
differ, a common prediction is that a weak but characteristic fourfold LDOS
anisotropy around vortices should be observed in the conductance maps of
$d$-wave superconductors. Up to now, such an observation has never been firmly
reported by STS in high-$T_c$ cuprates. There could be several reasons for this.
The effect is expectedly very weak, the fourfold features being only a slight
deviation from a completely isotropic signal. Then, in case of mixed-symmetry
models (\textit{e.g.} \citet{Franz-1998b}), the additional components may damp
the fourfold character of the spectroscopic signatures. Added to this comes the
fact that the coherence length in HTS is extremely short: in Y123 the spectra
turn back to be superconducting over a distance of 6--8~nm
(Fig.~\ref{CoreStatesY123}a), while in Bi2212 this distance drops to less than
2~nm (Fig.~\ref{CoreStatesBi2212}a), suggesting that $\xi$ for this material is
3 to 4 times smaller than in Y123. \citet{Kugler-2000} performed a quantitative
fit of the zero-bias conductance profile across a vortex core, within the
framework of a $d$-wave symmetry model \cite{Franz-1999}. The fitting parameters
lead to a value for $\xi$ of the order of 1~nm. These short characteristic
lengths imply stringent experimental requirements on spatial resolution, and
make the electronic signatures very sensitive to inhomogeneities. Accurate
analysis of the conductance images have been performed for both Y123 and Bi2212.
Although certain vortices mapped by STS appear with a square shape, the
observations report mostly irregular patterns \cite{Hoogenboom-2000b}. For
Bi2212, the cores present no apparent specific symmetry. For Y123, the cores
show an overall elliptical shape (Fig.~\ref{IntrinsicShape}b) attributed to the
$ab$-plane anisotropy of the carrier effective mass
(Sec.~\ref{sect_vortex_matter_Y123}). The observation of the weak LDOS
anisotropy around the vortex cores remains a considerable challenge for future
STS studies. It largely depends on the availability of samples with an extremely
high degree of homogeneity.

\subsubsection{Irregular shapes of vortices in Bi2212}
\label{sect_vortex_shape_Bi2212}

\begin{figure}[tb]
\includegraphics[width=8.6cm]{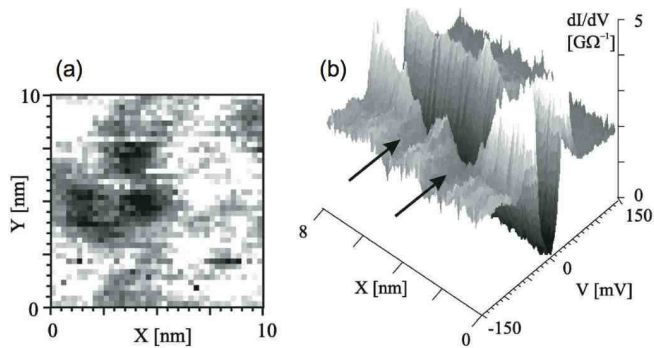}
\caption{\label{Bi2212cores}
(a) 10$\times$10~nm$^2$ zero-bias conductance map acquired in Bi2212 ($T=4.2$~K,
$H=6$~T). The whole core pattern is formed by split subcomponents belonging to
one single flux line. (b) 8~nm trace through such a core, illustrating how the
spectra recover the superconducting characteristics between two elements, whose
positions are marked by arrows. From \citet{Hoogenboom-2000b}.
}
\end{figure}

In Bi2212, images reveal that the vortices often present very irregular shapes,
and are in some cases even split into several subcomponents, as shown in
Fig.~\ref{Bi2212cores}a \cite{Hoogenboom-2000b}. The full pattern (in black in
the zero-bias conductance map) extends over an area of about 25~nm$^2$, and in
this particular case three subcomponents separated by about 3~nm each can be
distinguished. As seen in the spectroscopic trace shown in
Fig.~\ref{Bi2212cores}b, in between these core elements, the spectra turn back to
the full superconducting characteristics.

The possibility of vortices occupying neighboring sites may be ruled out for two
reasons. Since the expected distance between the vortices at 6~T is of the order
of 20~nm, a considerable energy would have to be spent to place the flux lines
only 3~nm apart. Moreover, images showing a large number of flux lines ensure
that a pattern like the one observed in Fig.~\ref{Bi2212cores}a is accounting
for only one vortex, and that the small subcomponents must belong to the same
flux line. The irregular shape of the vortex cores certainly results from doping
inhomogeneities. The latter are currently the subject of an intense debate
concerning the role they play in the fine electronic structure of HTS (see
Sec.~\ref{sect_gap_spectroscopy}). Since $\xi$ is so small in Bi2212 the
superconductor adapts to these inhomogeneities even at short length scales.

The results reported by \citet{Hoogenboom-2000b} were interpreted as follows.
Fig.~\ref{Bi2212cores}a would correspond to a situation where a vortex hops back
and forth between three pinning sites. This could be either thermally activated
motion or quantum tunneling of vortices. Estimates of the tunneling barriers
calculated from the positions of the sub-components combined with measurements
at 4.2~K and 2.5~K gave evidence for the quantum tunneling scenario.
Measurements at lower temperatures have to be carried out in order to reach
definitive conclusions. \citeauthor{Hoogenboom-2000b} did also observe a slow
shift of weight between the different substructures as a function of time, and
concluded that the vortex creep proceeds by this mode rather than by a slow
motion of the vortices themselves.

It should be pointed out that the vortex core subcomponents presented in this
section are different and should be distinguished from the electronic
modulations seen by \citet{Hoffman-2002} and \citet{Levy-2005} in the vortex
core. In the case of the core subcomponents presented above, the sizes of the
structures and the distances separating them are larger. Moreover, the
conductance spectra acquired in between these components present the full
superconducting characteristics. The electronic modulations are spatial periodic
variations in the low-energy part of the conductance spectra, all being observed
within the pseudogap spectra characteristic of one single vortex core. That
point will be discussed in more details in Sec.~\ref{sect_modulations_vortex}.

\subsection{Field dependence of the spectra}
\label{sect_Dopplershift}

\begin{figure}[tb]
\includegraphics[width=6cm]{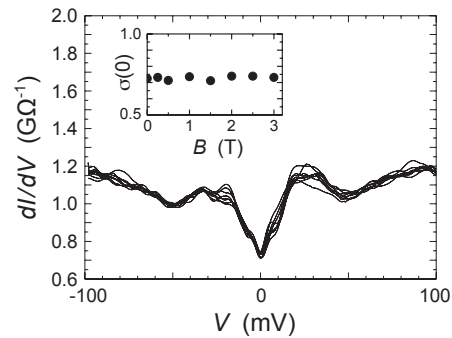}
\caption{\label{SpecFieldY123}
Conductance spectra acquired between vortices in Y123 at $T=4.2$~K under fields
ranging from 0 to 3~T. The inset shows the zero-bias tunneling conductance as a
function of the applied field \cite{Maggio-Aprile-1995}.
}
\end{figure}

Although STS is not sensitive to the phase of the order parameter, it can reveal
electronic features which are a direct consequence of the order parameter
symmetry. In the case of the $d_{x^2-y^2}$-wave symmetry, the presence of nodes
in the gap function gives rise to low-energy nodal quasiparticles with a linear
DOS at the Fermi energy (see a critical discussion in \citet{Hussey-2002}). With
a magnetic field applied parallel to the $[001]$ direction, the superfluid
screening currents flow within the $ab$ planes around the vortices with a
velocity $\bm{v}_s(\bm{r})\propto1/r$. \citet{Volovik-1993} suggested that the
superfluid velocity would locally shift the quasiparticle energies by an amount
$\delta E_{\vec{k}}=\bm{v}_s(\bm{r})\cdot\bm{k}$, the so-called Doppler shift.
He showed that this leads to a $1/r$ behavior of the LDOS outside the core, and
to an increase of the zero-energy \emph{total} DOS proportional to $\sqrt{H}$.
This behavior of the total DOS was also found in the microscopic calculations
\cite{Wang-1995}. Up to now, the only known possible experimental confirmation
of this prediction was provided by specific heat measurements on optimally-doped
Y123, by \citet{Moler-1994}, who found that the magnetic field dependence
followed quite precisely the $\sqrt{H}$ law. In principle, this effect should
also be directly measurable by tunneling spectroscopy, through integration of
the zero-bias LDOS over the unit cell of the vortex lattice. In Bi2212, the
conductance spectra acquired at zero field are similar to the ones acquired at
6~T in between the vortex cores, with no noticeable evolution of the zero-bias
conductance. The zero-bias conductance has also been locally measured in
optimally-doped Y123 by \citet{Maggio-Aprile-1995}, and appeared to be field
independent up to 3~T (Fig.~\ref{SpecFieldY123}). However, since Y123 presents
spectra with a large zero-bias conductance, it is not clear yet whether this
observation is reflecting an intrinsic property, or whether other competing
factors mask the predicted LDOS behavior, like for instance lifetime effects
which may wash out the $1/r$ behavior responsible for the Doppler effect.
\citet{Franz-1999} also argued that an anisotropic matrix element could lead to
a measured tunneling conductance $\propto1/r^3$, thus hiding the $1/r$ LDOS
behavior. All these factors may explain why the observation of the Volovik
effect by STS is challenging.

\subsection{Summary}

The possibility to detect vortices with nanometer-scale resolution in HTS is a
considerable success, allowing to uncover properties inaccessible to other
techniques. Overcoming the problems due to the sample surface quality and
homogeneity, individual vortices were probed in both Y123 and Bi2212. For Y123,
at high fields and low temperatures, vortices are arranged in an oblique
lattice, whose anisotropy is consistent with the observed intrinsic anisotropy
of the cores. The direct observation of a twin boundary evidenced its ability to
block the perpendicular flux line motion.

For Bi2212, the tiny size of the cores reveals the very small coherence length
of this material. The observation of a mostly disordered lattice underlines the
strong pinning effect due to defects or inhomogeneities. The latter are also
believed to be at the origin of the irregular shapes of the cores. Evidence for
quantum tunneling of vortices between close pinning sites has been reported.

For both compounds, the vortex cores revealed striking electronic signatures,
which can not be understood within the framework of a conventional BCS $s$-wave
or $d$-wave theory. In both underdoped and overdoped Bi2212, the core reveal
pseudogap-like spectra, similar to those measured above $T_c$. A pair of states
at finite energy in the vortex cores is systematically observed for both Y123
and Bi2212. Furthermore, the energy of these states scales linearly with the
superconducting gap, strongly suggesting that the states for both materials have
a common origin.

\section{Local electronic modulations observed by STM}
\label{sect_modulations}

The unusual properties of high-temperature superconductors and the existence of
a pseudogap has led to numerous theoretical predictions for the superconducting
state as well as the pseudogap state. Some of these imply the existence of
spatial charge modulations, either static or dynamic, like stripes, charge
density waves, etc. The STM can in principle resolve such charge density
variations if they are static or possibly slowly fluctuating, and this has led
to a search for such spatial structures in STM topographs or in STS surface maps
on Bi$_2$Sr$_2$CaCu$_2$O$_{8+\delta}$. Two different types of spatial variations
have been seen by STS: (i) large but irregular spatial variations of the gap,
with typical lengths scales of the order of 3--10~nm, in samples which are not
specially treated for homogeneity, and (ii) weaker but spatially periodic LDOS
modulations with a wavelength of about 1.6--2~nm. The first type of spatial
variations are discussed in Sec.~\ref{sect_gap_spectroscopy}, whereas the
periodic modulations appear as a very different phenomenon and are the topic of
this section.

The first indication of the presence of such periodic spatial modulations was
the observation that around the center of a vortex there is a modulation of the
LDOS with a periodicity of about $4a_0$ \cite{Hoffman-2002}. Subsequently,
\citet{Howald-2003} found that charge modulations were also present in the
absence of a magnetic field. They reported that the structure appeared at an
energy around 25~meV, and that the super-period did not disperse with energy.
\citet{Hoffman-2002b} reported similar zero-field electronic modulations but in
contrast to \citet{Howald-2003}, they found that these modulations disperse with
energy. They successfully interpreted their findings in terms of quasiparticle
interference due to scattering from impurities and other inhomogeneities. More
recently \citet{Vershinin-2004} observed electronic modulations in the pseudogap
phase above $T_c$. Contrary to the case of quasiparticle interference patterns,
where a large number of $\vec{q}$ vectors were observed, here a single square
pattern was detected with a $\vec{q}$ vector that does not disperse with energy.
This finding has to be contrasted with the results obtained by
\citet{McElroy-2005a}. They observed at low temperature, in regions of strong
underdoping where the coherence peaks in the STS spectra have disappeared, that
a non-dispersing square pattern is found at high energy with a period similar to
the one seen in the pseudogap phase. A non-dispersing square lattice was
subsequently found at low temperature by \citet{Hanaguri-2004} in strongly
underdoped Ca$_{1-x}$Na$_x$CuO$_2$Cl$_2$ (NCCOC). Finally, \citet{Levy-2005}
have recently studied the electronic modulations inside the vortex core in
detail. They confirmed the initial observation by \citet{Hoffman-2002}, but
showed that this modulation does not disperse with energy and thus appears to be
similar to the one found in the pseudogap phase. They were also able to link the
local square modulation to the localized state structure observed in the vortex
core spectra \cite{Maggio-Aprile-1995, Pan-2000b, Hoogenboom-2000a}. They thus
demonstrated the existence of a direct relation between the superconducting
state and the vortex-core electronic modulations.

In the following sections we shall first review the electronic modulations
resulting from the quasiparticle interference. This will be followed by a
discussion of the square pattern in the pseudogap state and of the electronic
modulations observed in the vortex cores. Finally we shall discuss the relations
between these different findings.

\subsection{Quasiparticle interference oscillations in the superconducting state}
\label{sect_modulations_interference}

\begin{figure}[tb]
\includegraphics[width=8.6cm]{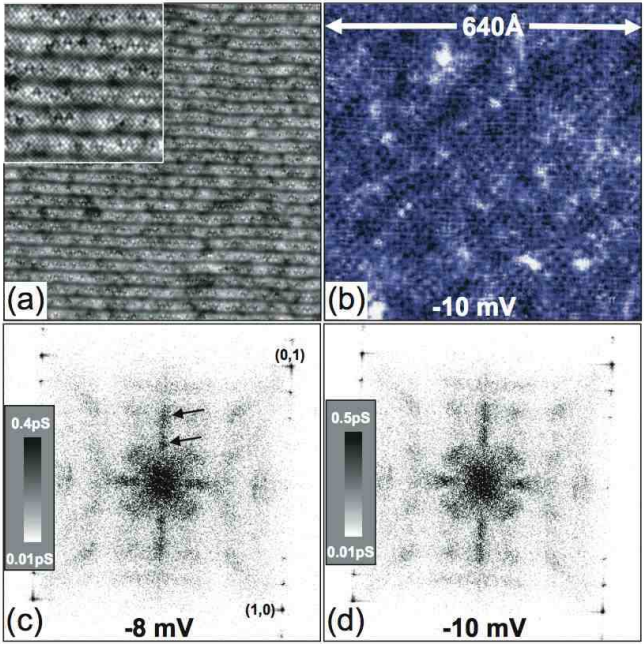}
\caption{\label{fig_McElroy-2003-1}
(a) Atomic resolution image of a $64\times64$~nm$^2$ region on near-optimal
doped Bi2212; the inset shows a $16\times16$~nm$^2$ magnification. (b)
Conductance map of the same field of view acquired at $V=-10$~mV. (c--d)
Fourier-space images of conductance maps at $-8$~mV and $-10$~mV, respectively.
The square of intense points near the corners of each panel corresponds to the
atomic lattice, and the arrows in (c) indicate the $\vec{q}$-vectors of the
supermodulation. These two sets of spots do not disperse with energy. The maps
clearly show an additional pattern with four-fold symmetry. The corresponding
wave-vectors disperse in energy. From \citet{McElroy-2003}.
}
\end{figure}

The detection of electronic modulations in the superconducting state is made
difficult by the strong tendency of these materials to show inhomogeneities
masking the weak periodic modulations. In order to overcome this difficulty
\citet{Hoffman-2002b} and \citet{Howald-2003} made large maps of the conductance
at a given energy. These maps were then Fourier transformed to search for other
periodicities than the atomic lattice and the well-known supermodulation in the
$b$ direction. In Fig.~\ref{fig_McElroy-2003-1} we display such a map and the
Fourier transform for different energies. Fig.~\ref{fig_McElroy-2003-1}a shows
the topography and Fig.~\ref{fig_McElroy-2003-1}b the conductance map of the
same area at $-10$~mV. Figs.~\ref{fig_McElroy-2003-1}c and d show the Fourier
transform of such conductance maps for two different energies.

\begin{figure}[t]
\includegraphics[width=7cm]{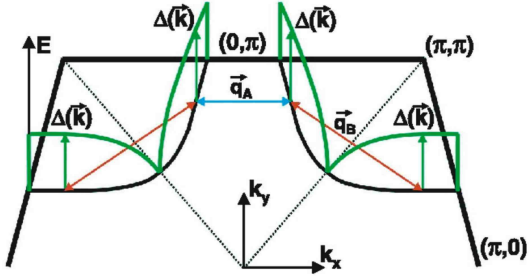}
\caption{\label{fig_Hoffman-2002b}
Representation of the quasiparticle energy along the Fermi surface. $\vec{q}_A$
and $\vec{q}_B$ are two possible vectors connecting quasiparticle states with
identical energies, giving rise to interference patterns. From
\citet{Hoffman-2002b}.
}
\end{figure}

As can easily be seen, in addition to the fine points at high $q$ values
reflecting the atomic lattice and the super modulation in the $b$ direction, one
observes a number of more diffuse spots at lower $q$-values reflecting periodic
modulations of the local density of states. These spots evolve with energy in a
characteristic way. Whereas the $q$-values for the spots in the $(\pi,\,0)$ and
$(0,\,\pi)$ directions decrease with increasing energy, the $q$-values for the
spots in the $(\pm\pi,\,\pm\pi)$ directions increase with increasing energy.
This can be understood in the framework of quasiparticle interference of a
$d$-wave superconductor \cite{Hoffman-2002b, Wang-2003, Zhang-2003,
McElroy-2003}. Fig.~\ref{fig_Hoffman-2002b} illustrates the gap variation at the
Fermi surface of a $d$-wave superconductor. At a given energy $\Delta(\vec{k})$
above or below the Fermi energy, interference through potential scattering can
occur between points at the Fermi surface linked by the vectors $\vec{q}_A$ or
$\vec{q}_B$. This will give rise to oscillations in the LDOS with wave vectors
$\vec{q}_A$ and $\vec{q}_B$ corresponding to the eight spots discussed above.
One sees easily that the length of $\vec{q}_A$ will decrease with increasing
energy (from the Fermi level) and the vector $\vec{q}_B$ will increase with
increasing energy exactly as observed in the experiment.

\begin{figure}[b]
\includegraphics[width=8.6cm]{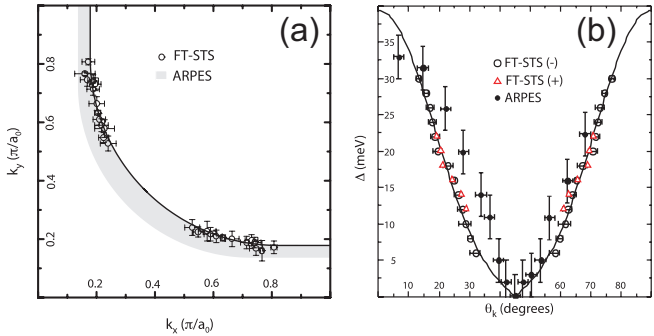}
\caption{\label{fig_McElroy-2003-2}
Analysis of the data shown in Fig.~\ref{fig_McElroy-2003-1}. (a) Fermi surface;
the solid line is a fit to the data, assuming the Fermi surface is the
combination of a circular arc joined with two straight lines. The grey band
represents the Fermi surface determined by ARPES \cite{Ding-1996}. (b)
Superconducting energy gap; the solid line is a $d$-wave fit to the data, and
the filled circles represent ARPES results \cite{Ding-1996}. From
\citet{McElroy-2003}.
}
\end{figure}

A more detailed analysis of these results was carried out by
\citet{McElroy-2003}. By using the simple relations between the vectors
$\vec{q}$ and the wave vectors $\vec{k}$ one can calculate the locus
$(k_x,\,k_y)$ of the Fermi surface. The result reported by \citet{McElroy-2003}
is shown in Fig.~\ref{fig_McElroy-2003-2}a with a comparison of the ARPES
results for the Fermi surface. A striking correspondence is found reinforcing
the interpretation that these LDOS oscillations result from quasiparticle
interference. Thus, although STM on homogeneous samples does not have $k$-space
resolution, this can be obtained when local scattering potentials are
introduced, producing quasiparticle interference. The relation $\Delta(\vec{k})$
can also be deduced from these results as illustrated in
Fig.~\ref{fig_McElroy-2003-2}b, and compares well to the results reported from
ARPES measurements. Such quasiparticle interference oscillations in the LDOS at
low temperature were also reported by \citet{Vershinin-2004} on a Bi2212 sample
containing 0.6\% of Zn impurities. It is important to stress that this
quasiparticle interference interpretation of the results implies that there is a
sufficient amount of scattering centers in the sample, \textit{i.e.} that the
sample either contains impurities, put in on purpose, or other imperfections
resulting from the sample preparation. Thus the intensity of the reflection
spots in the Fourier transform is necessarily sample dependent. This may explain
why \citet{Levy-2005} did not observe such quasiparticle interference patterns
in relatively homogeneous samples although they observed the square modulations
in the vortex cores.

The initial observations of \citet{Howald-2003} contrast with those of
\citet{Hoffman-2002b} and \citet{Vershinin-2004} in that the dominant features
in the Fourier transform are four spots in the Cu-O bond direction. These do not
disperse with energy and correspond to an oscillation with a period of about
$4a_0$. \citet{Howald-2003} conclude that these spots are there in addition to
the spots reflecting the quasiparticle interference. Using a filtered inverse
Fourier transform of their images they claim to see evidence for one dimensional
stripe-like patterns.

\subsection{Electronic modulations in the pseudogap state}
\label{sect_modulations_pseudogap}

\begin{figure}[tb]
\includegraphics [width=7cm] {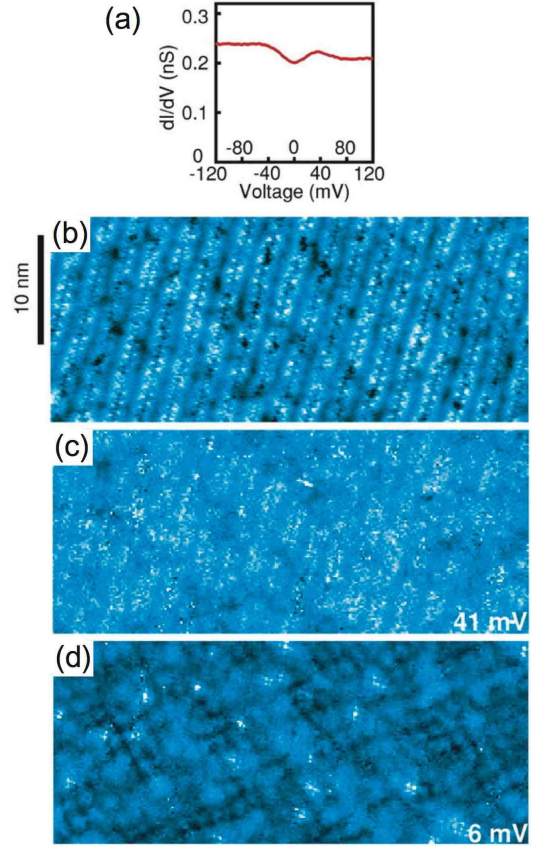}
\caption{\label{fig_Vershinin-2004-1}
Energy and spatial dependence of the DOS at 100~K on Bi2212. (a) A typical
spectrum showing the pseudogap. (b) A typical $45\times19.5$~nm$^2$ topograph
showing atomic corrugation and the incommensurate supermodulation along the $b$
axis. (c--d) Real-space conductance maps recorded simultaneously at 41 and 6~mV,
respectively. From \citet{Vershinin-2004}.
}
\end{figure}

\begin{figure}[tb]
\includegraphics [width=8.6cm] {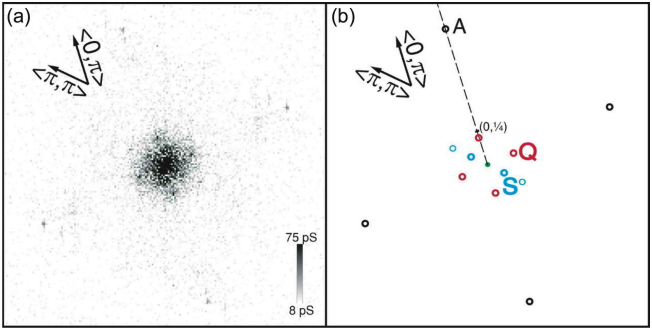}
\caption{\label{fig_Vershinin-2004-2}
Fourier analysis of DOS modulations observed at 100~K (see also
Fig.~\ref{fig_Vershinin-2004-1}). (a) Fourier-space image of a conductance map
($38\times38$~nm$^2$) acquired at 15~mV. (b) The Fourier-space image shows peaks
corresponding to atomic sites (in black, labeled A), primary (at $2\pi/6.8a_0$)
and secondary peaks corresponding to the $b$-axis supermodulation (in cyan,
labeled S), and peaks at $\sim2\pi/4.7a_0$ along the $(\pm\pi,\,0)$ and
$(0,\,\pm\pi)$ directions (red, labeled Q). From \citet{Vershinin-2004}.
}
\end{figure}

As discussed in Sec.~\ref{sect_pseudogap} the pseudogap state is still not
understood. In a recent experiment on a slightly underdoped sample
\citet{Vershinin-2004} succeeded to obtain detailed conductance maps at
temperatures above $T_c$ and thus to explore the pseudogap state
(Fig.~\ref{fig_Vershinin-2004-1}). The striking result of this investigation is
the observation of a square superstructure with $q$-values equal to
$2\pi/4.7a_0$. As the energy is lowered below the pseudogap one sees a square
like disordered pattern appearing. The Fourier transform of one such map is
shown in Fig.~\ref{fig_Vershinin-2004-2}a and illustrated in
Fig.~\ref{fig_Vershinin-2004-2}b. Four spots in the $(\pm\pi,\,0)$ and
$(0,\,\pm\pi)$ directions are seen in addition to the spots reflecting the
atomic lattice. The intensity of these peaks is energy dependent and they are
best seen at low energy around and below 20~meV as illustrated in
Fig.~\ref{fig_Vershinin-2004-1}c,d. On the other hand the positions of these
spots in $q$-space do not vary with energy, in contrast to the quasiparticle
interference seen at lower temperature in the superconducting state at the same
energies. In a more detailed analysis of these results \citet{Misra-2004}
demonstrated that the results of \citet{Vershinin-2004} cannot be understood in
terms of a quasiparticle interference scheme.

\begin{figure}[tb]
\includegraphics [width=5cm] {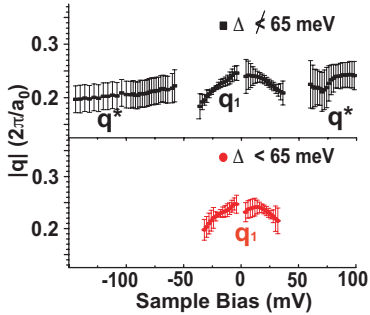}
\caption{\label{fig_McElroy-2005a-2}
Dispersion of the $\vec{q}$-vectors observed in regions with SC
coherence-peaked spectra (red circles) and in regions with
pseudogap-like spectra (black squares). The vector $\vec{q}^*$ has a
length $\pi/4.5a_0\pm15$\% \cite{McElroy-2005a}.
}
\end{figure}

Another situation related to the pseudogap was investigated by
\citet{McElroy-2005a}. They studied samples with a strong inhomogeneity in the
gap distribution at low temperature, presumably resulting from a local variation
in the doping (see Sec.~\ref{sect_spatial_homogeneity}). The larger the gap,
\textit{i.e.} the lower the doping, the smaller are the coherence peaks. Typical
spectra from such a sample are shown in Fig.~\ref{fig_fig5_inhomogene}c.
Selecting strongly underdoped (in average) samples they find large regions where
the coherence peaks are completely suppressed. The spectra seen in such regions
are illustrated in Fig.~\ref{fig_ZTPG}a together with spectra in higher doping
regions with smaller gaps and well developed coherence peaks. They assigned the
regions without coherence peaks to a pseudogap behavior and they studied the
spatial variations of the local density of states in these regions by
suppressing the other regions from the conductance maps. The result is that at
low energy they find the same quasiparticle interference patterns as in the
regions where a coherence peak is seen. However, in contrast to the latter
regions they find at higher energy, above 65~meV, a square pattern similar to
the one seen by \citet{Vershinin-2004} in the pseudogap state. This is
illustrated in Fig.~\ref{fig_McElroy-2005a-2}. Their conclusion is that this
square pattern, seen at high energy, is a different phenomenon than the
quasiparticle oscillations seen at lower energy. Comparing the two reports
discussed here \cite{Vershinin-2004, McElroy-2005a}, we note the following: the
non-dispersing square pattern observed in the pseudogap phase above $T_c$ in a
nearly optimally-doped sample is strongest at low energies. On the other hand,
in the strongly underdoped samples investigated at low temperature an apparently
different behavior is found. Here a non-dispersing pattern is found at high
energy while a dispersing quasiparticle interference pattern is found at low
energy. We are going to discuss this question further below in
Sec.~\ref{sect_modulations_summary}.

\begin{figure}[b]
\includegraphics [width=8.6cm] {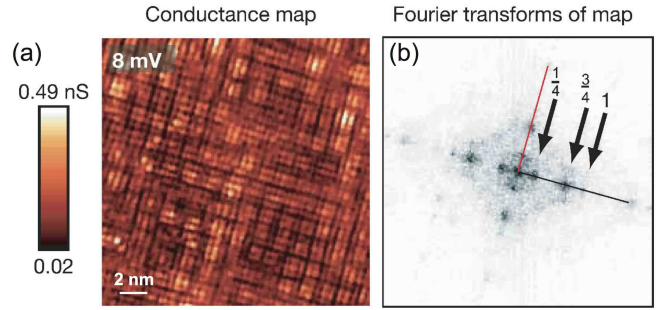}
\caption{\label{fig_Hanaguri-2004}
Conductance map on Ca$_{1-x}$Na$_x$CuO$_2$Cl$_2$ at 8~mV and the corresponding
Fourier transform. From \citet{Hanaguri-2004}.
}
\end{figure}

In order to study the pseudogap state in strongly underdoped samples
\citet{Hanaguri-2004} investigated Ca$_{1-x}$Na$_x$CuO$_2$Cl$_2$ (NCCOC). The
parent compound CaCuO$_2$Cl$_2$ is a Mott insulator and doping with Na makes it
a superconductor for $x>0.1$. The three compositions studied in this paper all
showed a pseudogap-like tunneling spectrum. Conductivity maps at various
energies showed a striking electronic order with a dominant $4a_0\times 4a_0$
superstructure similar to the one seen at high temperature in the pseudogap
state in Bi2212. Fig.~\ref{fig_Hanaguri-2004} shows these modulations and their
Fourier transform at 8~meV. The characteristic periods are energy independent,
as found by \citet{Vershinin-2004} in the pseudogap state above $T_c$ in Bi2212.
Another result implicit in the study on NCCOC is that the order appears to be
independent of doping. However, since the spectra show no signature of
superconductivity, no information about the precise doping level at the surface
was given. Another striking result of this study was the observation of
additional periodicities inside the $4a_0\times 4a_0$ unit cell. A
$\frac{4}{3}a_0$ structure was also found, suggesting a non commensurate
electronic modulation and thus a more complex structure than observed by
\citet{Vershinin-2004}. Very recently, \citet{Momono-2005} studied two different
surfaces of Bi2212 at low temperature, one (A) showing pseudogap tunneling
spectra and the other (B) superconducting spectra with coherence peaks. Surface
A showed a well-developed $4a_0\times4a_0$ type non-dispersive modulation, whereas
surface B did not show this modulation except for a small area.

\subsection{Electronic modulations in the vortex core}
\label{sect_modulations_vortex}

\begin{figure}[tb]
\includegraphics [width=5.5cm] {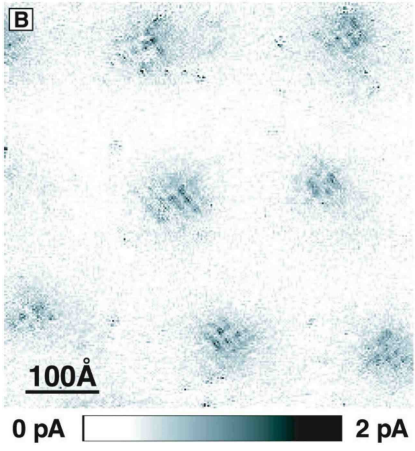}
\caption{\label{fig_Hoffman-2002}
Local ordering in the vortex cores of Bi2212 seen in the difference map using
the differential conductivity integrated from 0 to 12~meV. The difference map
was obtained by subtracting the conductivity map obtained in zero field from the
one obtained in the presence of a field of 3~T. From \citet{Hoffman-2002}.
}
\end{figure}

\citet{Hoffman-2002} analyzed conductance maps of Bi2212 taken in the presence
of a magnetic field. These maps were obtained by plotting the differential
conductance integrated between 0 and 12~meV. Because of the inhomogeneity of
their samples they measured also the maps in zero field and obtained a
difference map by subtracting the latter from the former. This is a delicate and
difficult operation, but has the advantage that much of the perturbations due to
inhomogeneities are eliminated. In this way they were able to show that the
local density of states has a $4a_0\times4a_0$ spatial modulation in the vortex
core. In Fig.~\ref{fig_Hoffman-2002} we display such a difference map showing a
few ``lattice points'' per vortex due to this local order. Note that the
variation in the conductance is of the order of 10\% of the zero-field signal.

More recently \citet{Levy-2005} have carried out a detailed study of the
behavior of an individual vortex in Bi2212. Fig.~\ref{CoreStatesBi2212}a shows a
vortex imaged at $-25$~meV as well as the spectra taken just outside the vortex
and close to the vortex center. The salient features of the spectrum close to
the center is that one sees a large gap-like structure reflecting the pseudogap
as discussed in Secs.~\ref{sect_pseudogap} and \ref{sect_vortices}, and two
peaks placed at $\sim\pm0.3\Delta_p$ suggesting the presence of one pair of
localized states. Two questions motivated the investigations: (i) How does the
square pattern observed in the vortex cores relate to the square pattern
observed in the pseudogap state above $T_c$? (ii) How does this square pattern
in the vortex core relate to the localized state?

\begin{figure}[tb]
\includegraphics [width=8.6cm] {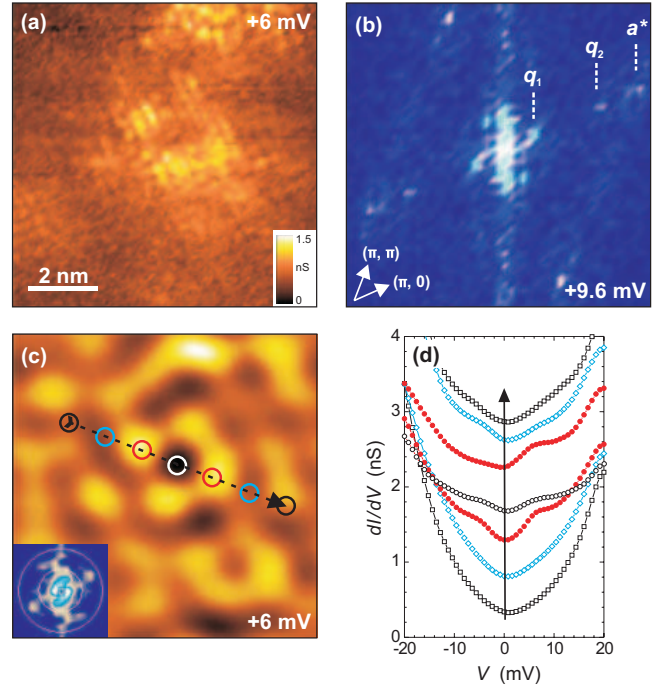}
\caption{\label{fig_Levy-2004-2}
(a) Conductance map at $+6$~mV of the area shown in
Fig.~\ref{CoreStatesBi2212}a, revealing a square pattern around the vortex
center. (b) Fourier transform at $+9.6$~mV with $a^*=2\pi/a_0$,
$q_1\approx\frac{1}{4}a^*$, and $q_2\approx\frac{3}{4}a^*$. (c) Filtered inverse
FT image. The inset shows the filter applied to the image acquired at $+6$~mV
and which selects the region around the four $q_1$ peaks. (d) Spectra averaged
in the 7 circles along the arrow in (c). The curves are offset for clarity.
Adapted from \citet{Levy-2005}.
}
\end{figure}

When this vortex was imaged at the energy of the core state
(Fig.~\ref{fig_Levy-2004-2}a) a square pattern oriented in the Cu-O bond
direction was obtained at the center of the vortex. The Fourier transform of
this pattern (Fig.~\ref{fig_Levy-2004-2}b) showed, in addition to the atomic
lattice, four spots with a period close to $4a_0$. A filtered inverse Fourier
transform, selecting an area including the four spots, illustrates the
correspondence between the latter and the pattern seen in the 6~meV image
(Fig.~\ref{fig_Levy-2004-2}c). The dominant square pattern in
Fig.~\ref{fig_Levy-2004-2}a is well reproduced in Fig.~\ref{fig_Levy-2004-2}c.
Note the difference in scale between Fig.~\ref{fig_Levy-2004-2}a and
Fig.~\ref{fig_Hoffman-2002}. The individual spots inside each vortex of the
latter correspond to the large spots in Figs.~\ref{fig_Levy-2004-2}a. The fine
structures in Fig.~\ref{fig_Levy-2004-2}a show the atomic lattice on which is
superposed the $\frac{4}{3}a_0$ period in one spatial direction.

Since it has been found that the vortex-core spectra show the pseudogap
(Sec.~\ref{sect_pseudogap_PGinVC}) one might expect that the spatially ordered
structure in the vortex cores has the same origin as the one observed in the
pseudogap state. One important property of the spatial modulations in the
pseudogap, as discussed in Sec.~\ref{sect_modulations_pseudogap}, is the absence
of dispersion of the $\vec{q}$ vectors, contrary to the modulations observed in
the low-temperature superconducting state. Thus one criterion in comparing the
vortex cores with the pseudogap is to check whether the observed order disperses
or not. This was also studied by \citet{Levy-2005}, who plotted the intensity of
the Fourier transform along the $(\pi,\,0)$ direction for different energies.
The position of the local order was indeed found to be independent of energy,
exactly as observed for the pseudogap. The period in this overdoped sample was
$(4.3\pm0.3)a_0$, somewhat less than observed in the underdoped sample by
\citet{Vershinin-2004}. This difference could be due either to a difference in
doping or to the difference in temperature. Note that in the $(\pi,\,0)$
direction \citet{Levy-2005} also observed the $\frac{4}{3}a_0$ modulation seen
in NCCOC by \citet{Hanaguri-2004}.

The relation to the localized states was also studied by \citet{Levy-2005} who
demonstrated that the intensity of the peaks corresponding to these states is
maximal at the four spots reflecting the local order, whereas in between and
outside these four spots the peaks at $\pm0.3\Delta_p$ are absent
(Fig.~\ref{fig_Levy-2004-2}d). Since the spatial dependence of the peaks (no
dispersion with position) suggests that these two peaks reflect only \emph{one}
pair of localized states (see Sec.~\ref{sect_vortex_core}), the STS measures the
square of the wave function of this state. Thus the four spot pattern in the
center of the vortex may be understood as a plot of the wave function of the
localized state. Note that this pattern is very different from what one would
obtain for a classical $s$-wave localized state as predicted by
\citet{Caroli-1964}.

\subsection{Discussion}
\label{sect_modulations_summary}

In this section we have discussed some of the striking recent scanning tunneling
spectroscopy observations on high-temperature superconductors. A detailed
analysis of the tunneling spectra taken with high spatial resolution over large
areas has revealed periodic variations with a spatial wavelength of the order of
4--5$a_0$. These variations correspond to small, sometimes very small, spatial
changes of the tunneling spectra. This has to be distinguished from the large
gap inhomogeneities often found in these crystals as discussed in
Sec.~\ref{sect_gap_spectroscopy}.

The observed phenomena can be divided into two classes: (i) the modulations with
dispersing wave lengths and (ii) the ones which do not disperse or have a much
weaker dispersion than the former. The first category is observed in the
superconducting state by two groups. These modulations are characterized by a
large number of points in the Fourier transform and the position of these points
disperse in a characteristic way with energy. In a series of papers,
\citet{Wang-2003, Zhang-2003, McElroy-2003, Hoffman-2002b} interpreted
convincingly these structures as modulations due to quasiparticle interference
resulting from scattering on impurities and other imperfections. Note that with
the observed behavior this explanation can only work if we have a $d$-wave
superconductor. An $s$-wave superconductor would not give the same pattern.
Also, the intensity of such charge modulations must depend on the amount of
scattering centers in the sample and thus depends on the quality of the sample.
This may explain why in some cases this type of density of states modulations
have not been observed \cite{Levy-2005}. From these results one can determine
the band structure around the Fermi surface as well as the angular dependence of
the gap. These data, which agree largely with the results from ARPES,
demonstrate beautifully how the STM, in addition to its spatial resolution also
can have $k$-space resolution.

The second category of spatial modulations have so far been observed in the
pseudogap state either at high temperature, at low doping, or in the vortex
cores. In one case \cite{Howald-2003} such modulations were reported in a Bi2212
sample which was inhomogeneous but on the average superconducting. In another
case \cite{Momono-2005} weak signs of modulations were observed on a
superconducting surface, but modulations appeared strongly on a surface showing
pseudogap behavior. In yet another case this was reported on NCCOC samples which
were nominally superconducting but whose surfaces showed dominantly
pseudogap-like spectra. Comparing all these data, there is considerable evidence
that the non-dispersing order appears when $d$-wave superconductivity is
suppressed, thus that this order is a characteristic behavior of the pseudogap
state. In addition to the absence of dispersion, all these square lattices have
similar periodicities and the same orientation with respect to the atomic
lattice. The reported periods are $4.7a_0$ in the pseudogap above $T_c$ in an
underdoped Bi2212 sample \cite{Vershinin-2004}, $4.3a_0$ in the vortex cores at
low temperature in a slightly overdoped Bi2212 sample \cite{Levy-2005} and
$4.0a_0$ at low temperature in a strongly underdoped oxychloride
\cite{Hanaguri-2004}. A possible common interpretation of these numbers is that
the square local ordering occurs at $4a_0$ at low doping and that the period
increases with doping and temperature.

The observation of \citet{McElroy-2005a} that the square pattern is seen at high
energies in strongly underdoped samples and that in the same samples the
dispersing quasiparticle interference pattern is seen at low energies seem at
first to be contrary to the above conclusions. However, in this experiment the
authors studied strongly underdoped regions in an inhomogeneous sample. In these
regions superconducting coherence seems to be lost at high energy, \textit{i.e.}
for states around $(\pi,\,0)$ but present at low energy, \textit{i.e.} around
$(\pi/2,\,\pi/2)$. In this sense the sample is partly in the superconducting
state and partly in a pseudogap state. Thus depending on the energy one observes
quasiparticle interference, characteristic of a superconducting state, or a
square pattern characteristic of a pseudogap.

The experiments on the vortex cores by \citet{Levy-2005} bring a new dimension
into this picture. The square pattern observed in the vortex core represents the
spatial distribution of the localized state. On the other hand this pattern is
very similar to the order found in the pseudogap state: both do not disperse and
both have a maximum amplitude at low energy below 20~meV. Furthermore, the
overall spectra of the pseudogap state and the vortex core have been shown to be
the same. This raises the question about the precise relation between the
localized state and the square pattern and in particular the question if the
so-called localized state is also a characteristics of the pseudogap phase. So
far no signature of a state at $\sim0.3\Delta_p$ has been observed in the
pseudogap above $T_c$, but such a structure would most probably be thermally
smeared so that one would not see it at temperatures of the order of 100~K. Thus
it remains a challenge to determine whether or not there is such a state related
to the zero-field pseudogap. Another scenario is that the square pattern is the
characteristic feature of the pseudogap, and that the localized state develops
in the vortex core thanks to the presence of the square pattern.

The recent STM observations reviewed here, either in the heavily underdoped
NCCOC at low temperature, in the slightly underdoped Bi2212 above $T_c$, or in
the vortex cores of slightly overdoped Bi2212, consistently reveal a tendency
towards local electronic ordering when the $d$-wave superconductivity is
suppressed and the pseudogap is formed. The relation between the superconducting
state and the pseudogap state, and the appearance of spatial ordering, have been
the origin of numerous papers and discussions.\footnote{See \textit{e.g.}
\citet{Zaanen-1989, Zhang-1997, Emery-1999, Vojta-1999, Wen-1996,
Chakravarty-2001}. Several theories have predicted that the pseudogap
corresponds to an ordered state. One-dimensional charge modulations, stripes,
could be one candidate. \citet{Howald-2003} argue that their observations show
evidence for one-dimensional stripes. However, subsequent observations all show
a clear two-dimensional pattern. One possible way out within this scenario would
be that the two-dimensional pattern result from fluctuating one-dimensional
domains, or that the two-dimensional order result from the observation of two
CuO$_2$ planes with crossed one-dimensional orders as put forward by
\citet{Fine-2004}. However, there is at present, with the exception of
\citet{Howald-2003}, no experimental STM evidence in favor of any of those
scenarios. Another possibility would be that the local ordering is due to a
spin-density wave (SDW) pinned by vortices. The SDW has been attributed to the
proximity of a quantum phase transition \cite{Demler-2001, Zhang-2002} or to
strong quantum phase fluctuations \cite{Franz-2002}. Neutron scattering
experiments have given ample evidence for the presence of such fluctuations
\cite{Cheong-1991, Lake-1999}. If such fluctuations are pinned by imperfections
a modulation in the LDOS may result. It has also been argued that there is a
local anti-ferromagnetic order in the vortex cores \cite{Zhang-1997,
Arovas-1997, Zhu-2002, Takigawa-2004} which may be reflected in the local
ordering observed in STM. Several other possibilities remain as to the nature of
this electronic order. Conventional charge-density waves \cite{Vojta-2002} seem
inconsistent with the data: this type of order, as well as charged stripes
discussed above, would induce LDOS modulations which are approximately odd with
respect to the bias polarity \cite{Podolsky-2003, Chen-2004a}. In these
scenarios one would also not understand why the gaps in the LDOS are always
pinned at the Fermi energy \cite{Franz-2004}. Other types of order involve the
doped holes themselves. Using a variational approach to study a generalized
Hubbard model, \citet{Fu-2004} found an insulating ground state at doping
$x=1/16$, made of square-ordered solitons on top of an anti-ferromagnetic
background. They argued that additional holes would make this system metallic
and eventually superconducting at higher doping levels. The existence of this
state was questioned by \citet{White-2004}, who found that this type of
structures are unstable in the $t$-$J$ model; instead, they suggested that
pinned CDW along the one-dimensional stripes would lead to patterns similar to
the ones observed experimentally.}

One important experimental finding brought about by the STM investigations is a
strong evidence that the superconducting state, the pseudogap state and the
local square order have a common origin. \citet{Emery-1997} proposed that the
pseudogap reflects fluctuating superconductivity with pair correlations
persisting well above $T_c$. A possible scenario within this scheme would be
that in the pseudogap state pairs localize and form a disordered static lattice
which can be seen in STM only below the pairforming energy, \textit{i.e.} the
pseudogap. Several authors have proposed that the LDOS modulations reflect the
spatial ordering of hole pairs rather than single holes \cite{Chen-2002,
Anderson-2004b, Tesanovic-2004, Melikyan-2005}.

The new findings by STM reported in this section clearly shine new light on the
quest for an understanding of the microscopic nature of both the superconducting
state and the pseudogap state. Further investigations along these lines will
certainly be necessary to unravel the detailed mechanisms behind the relation
between the two phenomena.

\section{Conclusion}
\label{sect_conclusion}

Over the last decade scanning tunneling spectroscopy has evolved into a major
experimental tool in the quest for an understanding of the high-temperature
superconductors. The first main achievement of STS on HTS was to bring tunneling
experiments into focus again after the first tunneling experiments on
high-temperature superconductors using various methods failed to give
reproducible results. In fact, studying the spatial dependence it was possible
to demonstrate reproducible and homogeneous tunneling on high-quality Bi2212
samples \cite{Renner-1994, Renner-1995}. It thus became clear that the tunneling
spectra were consistent with $d$-wave symmetry, but with an unusual high value
of the gap. Later on, other groups stressed the fact that samples often are
inhomogeneous \cite{Howald-2001, Pan-2001} even if other measurements tend to
indicate higher homogeneity. Thus, STS with its very high spatial resolution
revealed an inhomogeneity that was not fully recognized by other methods.

STS has led to a number of important discoveries. In 1995 the vortex cores could
be imaged on Y123 and anomalous vortex-core spectra were observed
\cite{Maggio-Aprile-1995}, very different from expectations of the BCS theory
($d$-wave or $s$-wave). In 1998 the pseudogap was observed in Bi2212 by STM both
in underdoped and overdoped samples \cite{Renner-1998a}. Prior to that the
pseudogap was believed to be only present in underdoped samples. STS has also
uncovered the striking doping dependence of the density of states, including the
doping dependence of the gap and the pseudogap and their scaling relations. It
was recognized that the pseudogap phase is present in the vortex cores at low
temperature \cite{Renner-1998b}. \citet{Pan-2000b} and \citet{Hoogenboom-2000a}
found later that weak signals from localized states are also present in the
vortex cores in Bi2212. \citet{Hoogenboom-2001a} showed that the localized
states in Bi2212 and Y123 both follow a universal linear scaling with the
superconducting gap and thus that the vortex-core structures in Y123 and Bi2212
have the same origin. Two groups reported spectroscopic studies of impurities in
Bi2212 and showed that a zero-bias resonance appears at the site of a
non-magnetic impurity, as expected for a $d$-wave superconductor
\cite{Yazdani-1999, Hudson-1999}. The introduction of Fourier transformed
conductance maps opened new possibilities \cite{Hoffman-2002b, Howald-2003}.
This led to the discovery of periodic modulations of the local density of states
in the superconducting phase \cite{Hoffman-2002b, Howald-2003} as well as in the
pseudogap phase \cite{Vershinin-2004}, and in strongly underdoped NCCOC at low
temperature \cite{Hanaguri-2004}. Whereas the modulations in the superconducting
state can be understood in terms of quasiparticle interference, the simple
square pattern observed in the pseudogap may turn out to be a characteristic
signature of order in this phase. The observation of such patterns in the vortex
cores by \citet{Hoffman-2002} was recently confirmed by \citet{Levy-2005} who
demonstrated the close relations between the localized states and the square
pattern. These and other results reported in this review have uncovered a large
number of remarkable and unexpected behaviors of the high-temperature
superconductors. The STM/STS studies have thus made essential contributions to
our understanding of high-temperature superconductivity.

\section*{Acknowledgments}

We thank the current and former members of the Geneva STM group for their
contributions to experiments, to discussions, and for their comments:
P.-E. Bisson,
J.-G. Bosch,
J. Cancino,
C. Dubois,
M. R. Eskildsen,
B. W. Hoogenboom,
A. D. Kent,
G. Levy,
A. A. Manuel,
Ph. Niedermann,
L. Stark.
Over the years, we have benefited from numerous scientific discussions with
P. Aebi,
G. Aeppli,
H. Alloul,
O. K. Andersen,
P. W. Anderson,
Y. Ando,
A. V. Balatsky,
B. Barbiellini,
M. R. Beasley,
H. Beck,
L. Benfatto,
G. Blatter,
J. C. Campuzano,
T. Claeson,
P. Coleman,
J. C. Davis,
D. Dessau,
G. Deutscher,
A. Erb,
T. Feher,
L. Forro,
M. Franz,
H. Fukuyama,
T. H. Geballe,
T. Giamarchi,
E. Giannini,
B. Giovannini,
W. Hanke,
T. Jarlborg,
A. Junod,
K. Kadowaki,
A. Kapitulnik,
H. Keller,
J. W. Loram,
K. Maki,
D. van der Marel,
P. Martinoli,
J. Mesot,
V. Mikheev,
A. J. Millis,
N. Momono,
C. de Morais Smith,
K. A. M\"uller,
M. R. Norman,
S. H. Pan,
M. Peter,
M. Randeria,
T. M. Rice,
D. Roditchev,
C. Rossel,
S. Sachdev,
D. J. Scalapino,
T. Schneider,
S. G. Sharapov,
Z.-X. Shen,
M. Sigrist,
J. L. Tallon,
Y. Tanaka,
Z. Te\v{s}anovi\'{c},
J.-M. Triscone,
C. M. Varma,
A. Yazdani,
A. Yurgens,
J. Zaanen,
S.-C. Zhang.

We also thank the Swiss National Science Foundation for support through its
Divisions II and IV, and through the National Center of Competence in Research
MaNEP.

\appendix

\section{The tunneling theory in real space}
\label{app_tunneling_theory_r}

In this appendix we reproduce the detailed calculation of the single-particle
current within the tunneling-Hamiltonian formalism. The derivation follows the
procedure given by \citet{Mahan-1990}; however, we use the real-space
formulation of the problem rather than the momentum-space formulation, since our
focus is on local probes like the STM.

\subsection{Calculation of the tunneling current}
\label{app_tunneling_Hamiltonian}

The tunnel junction is described by the Hamiltonian
	\begin{equation}\label{eq_tunneling_Hamiltonian}
		\mathcal{H}=\mathcal{H}_L+\mathcal{H}_R+\mathcal{H}_T
		\equiv\mathcal{H}_0+\mathcal{H}_T.
	\end{equation}
$\mathcal{H}_L$ and $\mathcal{H}_R$ are the Hamiltonians of the left and right
materials, respectively (Fig.~\ref{fig_junction}), which contain
all surface and geometric effects such as surface
states and defects, surface curvature (\textit{e.g.} in an STM tip), etc. The
central assumption of the model is that the two materials are independent,
meaning that the
operators on one side of the junction all commute with
the operators on the other side, in particular
$\mathcal{H}_L$ commutes with $\mathcal{H}_R$. The last term
in Eq.~(\ref{eq_tunneling_Hamiltonian}) is the tunneling
Hamiltonian, Eq.~(\ref{eq_transfer_Hamiltonian}), which is responsible for the
transfer of electrons across the junction. It will prove useful to write
$\mathcal{H}_T=\mathcal{X}+\mathcal{X}^{\dagger}$ with
	\begin{equation}\label{eq_X_operator}
		\mathcal{X}=\int d\vec{l}d\vec{r}\,T(\vec{l},\,\vec{r})
		\psi^{\dagger}(\vec{r})\psi(\vec{l}).
	\end{equation}
$\psi^{\dagger}(\vec{r})$ and $\psi(\vec{l})$ are the electron creation and
destruction operators in the right and left materials, respectively, and
$T(\vec{l},\vec{r})$ gives the probability for an electron to tunnel from a
point $\vec{l}$ in the left system to a point $\vec{r}$ in the right system.
$T(\vec{l},\vec{r})$ relates to the the matrix element $T_{\lambda\rho}$ and the
single-particle states $\varphi_{\lambda,\,\rho}$ of Sec.~\ref{sect_theory}
through
	\begin{equation}\label{eq_matrix_element_r}
		T(\vec{l},\vec{r})=\sum_{\lambda\rho}\varphi_{\lambda}^*(\vec{l})
		T_{\lambda\rho}^{Ê}\varphi^{Ê}_{\rho}(\vec{r}).
	\end{equation}
The tunneling current is simply given by the rate of change of the particle
number on the right side \cite{Cohen-1962}:
	\begin{equation}\label{eq_tunnel_current1}
		I=e\langle\dot{\mathcal{N}}_R\rangle.
	\end{equation}
Our convention is that $e=|e|$ and the current is taken positive when the
electrons flow from left to right. Since $\mathcal{N}_R$ commutes with
$\mathcal{H}_L$ and $\mathcal{H}_R$ we have
$i\dot{\mathcal{N}}_R=[\mathcal{N}_R,\,\mathcal{H}]=
[\mathcal{N}_R,\,\mathcal{H}_T]=\mathcal{X}-\mathcal{X}^{\dagger}$, in units
such that $\hbar=1$. We now calculate $\langle\dot{\mathcal{N}}_R\rangle$ to
first order in $\mathcal{H}_T$. The linear-response theory gives
	\begin{equation}\label{eq_linear_response}
		\langle\dot{\mathcal{N}}_R\rangle_t = -i\int_{-\infty}^{+\infty}dt'\,
		\theta(t-t')\langle[\dot{\mathcal{N}}_R(t),\,\mathcal{H}_T(t')]\rangle.
	\end{equation}
In the r.h.s. of Eq.~(\ref{eq_linear_response}) the time evolution of the
operators is governed by $\mathcal{H}_0$, and the angular brackets represent a
thermal average with respect to $\mathcal{H}_0$. Since $\mathcal{H}_L$ and
$\mathcal{H}_R$ commute this average factorizes into left and right components.
We can therefore assume that the left and right systems are in independent
thermodynamic equilibria, and are characterized by two different chemical
potentials $\mu_L$ and $\mu_R$. Introducing
$\mathcal{K}=\mathcal{H}_0-\mu_L\mathcal{N}_L-\mu_R\mathcal{N}_R$ we rewrite
the time evolution of the operator $\mathcal{X}$ as
	\begin{multline}\label{eq_evolution_X}
		e^{i\mathcal{K}t}e^{i(\mu_L\mathcal{N}_L+\mu_R\mathcal{N}_R)t}
		\mathcal{X}e^{-i(\mu_L\mathcal{N}_L+\mu_R\mathcal{N}_R)t}
		e^{-i\mathcal{K}t}\\
		=e^{-i(\mu_L-\mu_R)t}e^{i\mathcal{K}t}\mathcal{X}e^{-i\mathcal{K}t}
		\equiv e^{-ieVt}\mathcal{X}(t),
	\end{multline}
where $V$ is the applied bias. The last line of Eq.~(\ref{eq_evolution_X}) is
obtained using $e^{\mathcal{N}}\psi^{\dagger}=\psi^{\dagger}e^{\mathcal{N}+1}$
and $e^{\mathcal{N}}\psi=\psi e^{\mathcal{N}-1}$. We then have
	\begin{subequations}\label{eq_time_evolution}\begin{eqnarray}
		i\dot{\mathcal{N}}_R(t) &=& e^{-ieVt}\mathcal{X}(t)-
			e^{ieVt}\mathcal{X}^{\dagger}(t)\\
		\mathcal{H}_T(t') &=& e^{-ieVt'}\mathcal{X}(t')+e^{ieVt'}
		\mathcal{X}^{\dagger}(t'),
	\end{eqnarray}\end{subequations}
where the time evolution of $\mathcal{X}$ and $\mathcal{X}^{\dagger}$ is now
governed by $\mathcal{K}$ as in Eq.~(\ref{eq_evolution_X}). Inserting
(\ref{eq_time_evolution}) into (\ref{eq_linear_response}) and using
(\ref{eq_tunnel_current1}) one arrives at the expression
	\begin{eqnarray}\label{eq_tunnel_current2}
		\nonumber
		I &=& 2e\,\text{Im}\,\int_{-\infty}^{+\infty}dt'\\ [0.2em]
		\nonumber
		&&\Big\{e^{-ieV(t-t')}(-i)\theta(t-t')
			\langle[\mathcal{X}(t),\,\mathcal{X}^{\dagger}(t')]\rangle\\
		\nonumber
		&&+e^{-ieV(t+t')}(-i)\theta(t-t')
			\langle[\mathcal{X}(t),\,\mathcal{X}(t')]\rangle\Big\}\\
		&\equiv& I_s+I_J(t).
	\end{eqnarray}
The first term $I_s$ is the single-particle current and the second term is the
pair (Josephson) current. According to Eq.~(\ref{eq_tunnel_current2}) $I_s$ is
the Fourier transform, at energy $-eV$, of the retarded correlation
function $X(t-t')=-i\theta(t-t')\langle[\mathcal{X}(t),\,
\mathcal{X}^{\dagger}(t')]\rangle$ of the (bosonic) operator $\mathcal{X}$. The
latter is most easily evaluated by analytically continuing onto the
real-frequency axis the corresponding temperature propagator
\cite[][p.~149]{Mahan-1990}:
	\begin{equation}\label{eq_X_propagator1}
		\mathscr{X}(i\Omega_n) = -\int_0^{\beta}d\tau\,e^{i\Omega_n\tau}
		\langle\mathcal{T}_{\tau}\mathcal{X}(\tau)\mathcal{X}^{\dagger}(0)\rangle.
	\end{equation}
$\Omega_n=2n\pi/\beta$ denote the even Matsubara frequencies with $\beta$ the
inverse temperature. One thus has
	\begin{equation}\label{eq_tunneling_current3}
		I_s=2e\,\text{Im}\,\mathscr{X}(i\Omega_n\rightarrow-eV+i0^+).
	\end{equation}
Looking at Eqs.~(\ref{eq_X_propagator1}) and (\ref{eq_X_operator}) we see that
$\mathscr{X}(i\Omega_n)$ is made of the terms
	\begin{multline}\label{eq_decoupling}
		-\langle\mathcal{T}_{\tau}\psi^{\dagger}(\vec{r}_1,\,\tau)
		\psi(\vec{l}_1,\,\tau)\psi^{\dagger}(\vec{l}_2,\,0)
		\psi(\vec{r}_2,\,0)\rangle\\
		=\langle\mathcal{T}_{\tau}\psi(\vec{l}_1,\,\tau)
		\psi^{\dagger}(\vec{l}_2,\,0)\rangle\,
		\langle\mathcal{T}_{\tau}\psi(\vec{r}_2,\,0)
		\psi^{\dagger}(\vec{r}_1,\,\tau)\rangle\\
		=\mathscr{G}(\vec{l}_1,\,\vec{l}_2,\,\tau)
		\mathscr{G}(\vec{r}_2,\,\vec{r}_1,\,-\tau).
	\end{multline}
We have introduced the temperature Green's functions on both sides of the
junction, with the usual definitions:
	\begin{subequations}\label{eq_temperature_Greens_function}\begin{eqnarray}
		\mathscr{G}(\vec{x}_1,\,\vec{x}_2,\,\tau)&=&-\langle\mathcal{T}_{\tau}
		\psi(\vec{x}_1,\,\tau)\psi^{\dagger}(\vec{x}_2,\,0)\rangle\\
		\label{eq_temperature_Greens_functionb}
		&=&\frac{1}{\beta}\sum_n e^{-i\omega_n\tau}
		\mathscr{G}(\vec{x}_1,\,\vec{x}_2,\,i\omega_n)\qquad
	\end{eqnarray}\end{subequations}
with $\omega_n=(2n+1)\pi/\beta$ the odd Matsubara frequencies. It is then
straightforward to evaluate $\mathscr{X}(i\Omega_n)$ in
Eq.~(\ref{eq_X_propagator1}) using (\ref{eq_decoupling}) and
(\ref{eq_temperature_Greens_functionb}):
	\begin{multline*}
		\mathscr{X}(i\Omega_n)=\int d\vec{l}_1d\vec{r}_1d\vec{l}_2d\vec{r}_2\,
		T(\vec{l}_1,\,\vec{r}_1)T^*(\vec{l}_2,\,\vec{r}_2)\times\\
		\frac{1}{\beta}\sum_m\mathscr{G}(\vec{l}_1,\,\vec{l}_2,\,i\omega_m)
		\mathscr{G}(\vec{r}_2,\,\vec{r}_1,\,i\omega_m-i\Omega_n).
	\end{multline*}
The frequency sum is easily performed within the spectral representation,
	\begin{equation}\label{eq_spectral_representation}
		\mathscr{G}(\vec{x}_1,\,\vec{x}_2,\,i\omega_m)=\int_{-\infty}^{+\infty}
		d\omega\,\frac{A(\vec{x}_1,\,\vec{x}_2,\,\omega)}{i\omega_m-\omega},
	\end{equation}
and using the usual trick for frequency summation \cite[p.~251]{Mattuck-1992}.
We have, letting $i\Omega_n\rightarrow-eV+i0^+$:
	\begin{multline*}
		\mathscr{X}(-eV+i0^+)=\int d\vec{l}_1d\vec{r}_1d\vec{l}_2d\vec{r}_2\,
		T(\vec{l}_1,\,\vec{r}_1)T^*(\vec{l}_2,\,\vec{r}_2)\times\\
		\int d\omega d\omega'\,
		\frac{A(\vec{l}_1,\,\vec{l}_2,\,\omega)A(\vec{r}_2,\,\vec{r}_1,\,\omega')
		\left[f(\omega)-f(\omega')\right]}{\omega-\omega'+eV-i0^+}.
	\end{multline*}
Taking the imaginary part, inserting into (\ref{eq_tunneling_current3}), and
reintroducing a dimensional factor $\hbar^{-1}$, we finally get
	\begin{multline}\label{eq_tunneling_current_r}
		I_s=\frac{2\pi e}{\hbar}\int d\omega\,
		\left[f(\omega-eV)-f(\omega)\right]\times\\
		\int d\vec{l}_1d\vec{r}_1d\vec{l}_2d\vec{r}_2\,
		T(\vec{l}_1,\,\vec{r}_1)T^*(\vec{l}_2,\,\vec{r}_2)\times\\
		A(\vec{l}_1,\,\vec{l}_2,\,\omega-eV)
		A(\vec{r}_2,\,\vec{r}_1,\,\omega).
	\end{multline}
Eq.~(\ref{eq_tunneling_current}) results immediately from
Eq.~(\ref{eq_tunneling_current_r}) by a change of representation, \textit{i.e.}
substituting the matrix element with Eq.~(\ref{eq_matrix_element_r}) and
choosing the single-particle states which diagonalize the spectral function:
	\[
		\int d\vec{x}_1d\vec{x}_2\,\varphi_i^*(\vec{x}_1)
		A(\vec{x}_1,\,\vec{x}_2,\,\omega)\varphi_j^{Ê}(\vec{x}_2)=\delta_{ij}
		A_i(\omega).
	\]
General relations about the real-space Green's function and spectral function
are given by \citet{Hedin-1969}. Of particular interest is the relation between
the spectral function and the local density of states:
$N(\vec{x},\,\omega)=A(\vec{x},\,\vec{x},\,\omega)$.

The calculation of the Josephson current proceeds along the same lines, by
considering the anomalous correlation function of the operator $\mathcal{X}$:
	\begin{eqnarray*}
		I_J(t)&=&2e\,\text{Im}\,e^{-2ieVt}
		\mathscr{Y}(i\Omega_n\rightarrow eV+i0^+)\qquad\\
		\nonumber
		\mathscr{Y}(i\Omega_n)&=&-\int_0^{\beta}d\tau\,e^{i\Omega_n\tau}
		\langle\mathcal{T}_{\tau}\mathcal{X}(\tau)\mathcal{X}(0)\rangle.
	\end{eqnarray*}
We refer the reader to the work of \citet{Josephson-1969} for a detailed
discussion of this term.

\subsection{STM and the local density of states}
\label{app_Tersoff}

Starting from Eq.~(\ref{eq_tunneling_current_r}) it is easy to see how the
\citet{Tersoff-1983} matrix element Eq.~(\ref{eq_T_Tersoff}) leads to
Eq.~(\ref{eq_current_Tersoff}) for the tunneling current. Inserting
$T_{\lambda\rho}\propto\varphi_{\rho}^*(\vec{x})$ into
Eq.~(\ref{eq_matrix_element_r}) one gets, owing to the completion property of
the functions $\varphi_{\rho}$:
	\begin{equation}\label{eq_T_Tersoff_r}
		T(\vec{l},\,\vec{r})\propto\delta(\vec{r}-\vec{x})
		\sum_{\lambda}\varphi_{\lambda}^*(\vec{l}).
	\end{equation}
The delta function in the matrix element (\ref{eq_T_Tersoff_r}) substitutes
$A(\vec{r}_2,\,\vec{r}_1,\,\omega)$ in Eq.~(\ref{eq_tunneling_current_r}) with
$A(\vec{x},\,\vec{x},\,\omega)=N_R(\vec{x},\,\omega)$. On the other hand, if
the tip material is a simple metal its spectral function takes the form
\cite{Hedin-1969}
	\begin{equation}\label{eq_spectral_function_r}
		A(\vec{l}_1,\,\vec{l}_2,\,\omega)=\sum_{\lambda}
		\varphi^{Ê}_{\lambda}(\vec{l}_1)\varphi^*_{\lambda}(\vec{l}_2)
		\delta(\omega-\varepsilon_{\lambda}).
	\end{equation}
Then the $\vec{l}$ integrals in Eq.~(\ref{eq_tunneling_current_r}) cancel the
$\lambda$ sums of Eq.~(\ref{eq_T_Tersoff_r}) due to the orthogonality of the
functions $\varphi_{\lambda}$. Thus Eqs.~(\ref{eq_tunneling_current_r}),
(\ref{eq_T_Tersoff_r}), and (\ref{eq_spectral_function_r}) lead to
	\[
		I_s\propto\int d\omega\,\left[f(\omega-eV)-f(\omega)\right]
		N_L(\omega-eV)N_R(\vec{x},\,\omega).
	\]
In the calculation of the matrix element Eq.~(\ref{eq_T_Tersoff}),
\citeauthor{Tersoff-1983} assumed that the sample electronic structure can be
represented by independent electrons, so that $\varphi_{\rho}$ was meant as the
true wave function in the sample. Nevertheless, we implicitly considered in our
derivation that the validity of Eq.~(\ref{eq_T_Tersoff}) carries over to the
case where the functions $\varphi_{\rho}$ are general basis functions. In this
way the validity of Eq.~(\ref{eq_current_Tersoff}) could be extended to
correlated systems. We believe that this procedure is correct, since the
\citeauthor{Tersoff-1983} calculation does not depend on the detailed form and
nature of $\varphi_{\rho}$.

\subsection{The case of a non-local matrix element}
\label{app_Mk}

If the relation between the matrix element and the sample wave function is
non-local, like in Eq.~(\ref{eq_M_r}), then using
Eq.~(\ref{eq_matrix_element_r}) the form Eq.~(\ref{eq_T_Tersoff_r}) is
generalized to
	\begin{equation}
		T(\vec{l},\,\vec{r})\propto M(\vec{r}-\vec{x})
		\sum_{\lambda}\varphi_{\lambda}^*(\vec{l}).
	\end{equation}
The resulting tunneling conductance is, using Eq.~(\ref{eq_tunneling_current_r})
and the common approximations for the tip:
	\begin{multline}\label{eq_conductance_matrix_element}
		\sigma(\vec{x},\,V)\propto\int d\omega\,\left[-f'(\omega-eV)\right]
		\int d\vec{r}_1d\vec{r}_2\\
		M(\vec{r}_1-\vec{x})M^*(\vec{r}_2-\vec{x})
		A(\vec{r}_2,\,\vec{r}_1,\,\omega).\qquad
	\end{multline}
If the sample is translation invariant,
$A(\vec{r}_2,\,\vec{r}_1,\,\omega)=A(\vec{r}_2-\vec{r}_1,\,\omega)$, this
expression takes a simpler form in reciprocal space. Then the conductance no
longer depends on $\vec{x}$ and is just Eq.~(\ref{eq_conductance_Mk}).

\section{HTS gaps measured by STS}
\label{app_gap_tables}

Here we provide a list of superconducting gap values of high-$T_c$ cuprates
which have been obtained by scanning tunneling spectroscopy. The superconducting
gap is defined by the energy $\pm\Delta_p$ of the superconducting coherence
peaks. All collected data where obtained on SIN vacuum tunnel junctions at low
temperature and on single crystals with the tip perpendicular to the (001)
surface plane ($I\parallel[100]$), unless stated differently. The data are
listed as a function of increasing doping level for underdoped (UD),
optimally-doped (OP), and overdoped (OD) materials.

%
%
\def\gaptable#1#2#3#4#5#6{%
\setlength{\LTcapwidth}{\columnwidth}
\begin{longtable}{@{\extracolsep{#1}}lccdl}
\caption{\label{#2}#3\\ \mbox{\hspace*{\columnwidth}}}
\\ [-1.75em] \hline\hline\\ [-0.8em]
$T_c$ (K) & $p$ & $\Delta_p$ (meV) &
\multicolumn{1}{c}{$\frac{2\Delta_p}{k_{\text{B}}T_c}$} & Reference
\\ [0.3em] \hline\\ [-0.9em]\endfirsthead
\caption[]{#4\\ \mbox{\hspace*{\columnwidth}}}
\\ [-1.75em] \hline\hline\\ [-0.8em]
$T_c$ (K) & $p$ & $\Delta_p$ (meV) &
\multicolumn{1}{c}{$\frac{2\Delta_p}{k_{\text{B}}T_c}$} & Reference
\\ [0.3em] \hline\\ [-0.9em]\endhead\hline\endfoot\endlastfoot
#5 
[0.3em] \hline\hline \\ [-0.6em]
\multicolumn{5}{l}{~\parbox[t]{8.4cm}{#6}}
\end{longtable}
}%
\def\notegt#1#2{$^#1$\parbox[t]{8.3cm}{\footnotesize #2}\\ [-0.1em]}
\def\ddc{\multicolumn{1}{c}{\text{--}}}


\gaptable
{0.3em}
{tab_Bi2201}
{Gap values for Bi$_2$Sr$_2$CuO$_{6+\delta}$ (Bi2201).}
{Bi2201 (continued)}
{
$<4^a$  & 0.160 & $16\pm3$   & \ddc & Kugler \textit{et al.} (Sec.~\ref{sect_pseudogap_TdepBi2201})\\
10$^b$  & 0.180 & $12\pm3$   & 27.9 & \citet{Kugler-2001}\\
3.5$^b$ & 0.185 & $13\pm3$   & 86.2 & \citet{Kugler-2000}\\
\hline\\ [-0.8em]
\multicolumn{5}{l}{With La or Pb substitutions:}\\ [0.2em]
29$^c$  & 0.145 & $14.5\pm3$ & 11.6 & \citet{Kugler-2000}\\
10$^d$  & \ddc  & $25\pm5$   & 58.0 & \citet{Mashima-2003}\\
}{
\notegt{a}{Bi$_{2.2}$Sr$_{1.8}$CuO$_{6+\delta}$}
\notegt{b}{Bi$_{2.1}$Sr$_{1.9}$CuO$_{6+\delta}$}
\notegt{c}{Bi$_{1.8}$Sr$_{1.7}$La$_{0.5}$CuO$_{6+\delta}$}
\notegt{d}{Bi$_{2.1}$Pb$_{0.37}$Sr$_{1.91}$CuO$_{6+\delta}$}
}


\vspace*{-3mm}
\gaptable
{-0.5em}
{tab_Bi2212}
{
Gap values for Bi$_2$Sr$_2$CaCu$_2$O$_{8+\delta}$ (Bi2212). Values indicated
without parentheses are given explicitly in the publications. Values within
parentheses were estimated using the generic formula
$p=0.16\pm\sqrt{(1-T_c/T_c^{\text{max}})/82.6}$ \cite{Presland-1991} with
$T_c^{\text{max}}=92$~K. Where the sample is claimed to be optimally doped
although $T_c\neq T_c^{\text{max}}$, we set $p$ to [0.16].
}
{Bi2212 (continued)}
{
63        & (0.10) &  40          &    14.7 & \citet{Nakano-1998}\\
60        &  0.10  & $36\pm2$     &    13.9 & \citet{Oda-1997}\\
65        &  0.11  &  62          &    22.1 & \citet{McElroy-2004a,McElroy-2005a}\\
67        &  0.11  & $55\pm15$    &    19.1 & \citet{Matsuda-2003}\\
78        & (0.12) &  50.2        &    14.9 & \citet{Hoffman-2002}\\
80        & (0.12) & $42\pm2$     &    12.2 & \citet{Howald-2001}\\
81        & (0.12) &  50          &    14.3 & \citet{Matsuda-1999a,Matsuda-1999b}\\
81        & (0.12) &  40          &    11.5 & \citet{Murakami-1995}\\
75        &  0.13  & $48\pm1$     &    14.9 & \citet{McElroy-2004a,McElroy-2005a}\\
86        & (0.13) &  52          &    14.0 & \citet{Matsuda-1999a,Matsuda-1999b}\\
83        & (0.13) & $47\pm4$     &    13.1 & \citet{Matsuba-2003}\\
(85)      &  0.13  & $45\pm12$    &    12.3 & \citet{Matsuda-2003}\\
85        & (0.13) &  33          &     9.0 & \citet{Nakano-1998}\\
83        & (0.13) &  44          &    12.3 & \citet{Renner-1998a}\\
86        & (0.13) &  45          &    12.1 & \citet{Sakata-2003}\\
79        &  0.14  &  50          &    14.7 & \citet{Lang-2002}\\
88        & (0.14) &  50          &    13.2 & \citet{Matsuura-1998}\\
82        &  0.14  & $34\pm2$     &     9.6 & \citet{Oda-1997}\\
79        &  0.15  & $43\pm1$     &    12.6 & \citet{McElroy-2004a,McElroy-2005a}\\
80        &  0.15  &  38          &    11.0 & \citet{Oda-2000}\\
92        & (0.16) &  35          &     8.8 & \citet{DeWilde-1998}\\
92        & (0.16) &  43.7        &    11.0 & \citet{Hoffman-2002}\\
92        & [0.16] & $30\pm2$     &     7.6 & \citet{Iavarone-1998}\\
84$^a$    & [0.16] &  40          &    11.0 & \citet{Misra-2002a}\\
93$^{ab}$ & (0.16) & $25$--$55$   &\sim10.0 & \citet{Cren-2000}\\
90        & [0.16] &  54          &    13.9 & \citet{Matsuda-1999a,Matsuda-1999b}\\
89        &  0.16  & $40\pm10$    &    10.4 & \citet{Matsuda-2003}\\
88        & [0.16] & $32\pm2$     &     8.4 & \citet{Oda-1997}\\
92.3      & [0.16] & $29\pm4$     &     7.3 & \citet{Renner-1995}\\
92.2      & [0.16] &  41.5        &    10.4 & \citet{Renner-1998a}\\
90        & [0.16] &  39          &    10.1 & \citet{Wolf-1994}\\
86.5$^c$  & (0.18) & $45\pm15$    &    12.1 & \citet{Howald-2003,Howald-2003b}\\
88        & (0.18) & $28\pm2$     &     7.4 & \citet{Kitazawa-1997}\\
(89)      &  0.18  &  35.6        &     9.3 & \citet{Lang-2002}\\
(89)      &  0.18  & $36\pm1$     &     9.4 & \citet{McElroy-2004a,McElroy-2005a}\\
(89)      &  0.18  & $35\pm7$     &     9.1 & \citet{Matsuda-2003}\\
85        & (0.19) &  25          &     6.8 & \citet{Chen-1992}\\
85        & (0.19) &  36.7        &    10.0 & \citet{Hoffman-2002}\\
87        & (0.19) &  32          &     8.5 & \citet{Hudson-1999}\\
87        & (0.19) &  26          &     7.4 & \citet{Ichimura-1993}\\
86        & (0.19) & $35\pm5$     &     9.4 & \citet{Kaneko-1999}\\
89        &  0.19  & $33\pm1$     &     8.6 & \citet{McElroy-2004a,McElroy-2005a}\\
85        & (0.19) &  35          &     9.6 & \citet{Maki-2001}\\
84        & (0.19) &  50          &    13.8 & \citet{Matsuda-1999a,Matsuda-1999b}\\
85        & (0.19) &  26          &     7.1 & \citet{Nakano-1998}\\
84        & (0.19) &  32          &     8.8 & \citet{Pan-2000b}\\
87        & (0.19) &  17$^d$      &     4.5 & \citet{Suzuki-1999c}\\
87        & (0.19) & $31$--$40^e$ & \sim9.3 & \citet{Suzuki-1999b}\\
\;--      &  \ddc  &  40          &    \ddc & \citet{Hudson-2000}\\
76.4      & (0.20) & $47\pm13$    &    12.2 & \citet{Hoogenboom-2003a}\\
77        & (0.20) & $30\pm1$     &     9.0 & \citet{Kugler-2000a}\\
77        & (0.20) & $40\pm10$    &    12.1 & \citet{Hoogenboom-2000b}\\
77        & (0.20) & $45\pm5$     &    13.6 & \citet{Hoogenboom-2000a}\\
80.7      & (0.20) & $36\pm2$     &    10.4 & \citet{Hoogenboom-2003a}\\
81        & (0.20) &  43          &    12.3 & \citet{Matsuda-1999a,Matsuda-1999b}\\
82        & (0.20) & $34\pm1$     &     9.6 & \citet{Oda-1996}\\
72        & (0.21) &  30          &     9.7 & \citet{DeWilde-1998}\\
72        & (0.21) & $22\pm2$     &     7.1 & \citet{Iavarone-1998}\\
\;--      &  \ddc  & $29\pm5$     &    \ddc & \citet{Mallet-1996}\\
81        &  0.21  & $27\pm2$     &     7.7 & \citet{Oda-1997}\\
80        &  0.21  &  24          &     7.0 & \citet{Oda-2000}\\
74.3      & (0.21) &  34          &    10.6 & \citet{Renner-1998a}\\
74        & (0.21) & $32\pm2$     &    10.0 & \citet{Yazdani-1999}\\
(64)      &  0.22  & $22\pm5$     &     8.0 & \citet{Matsuda-2003}\\
56        & (0.23) &  21          &     8.7 & \citet{Renner-1998a}\\
\hline\\ [-0.8em]
\multicolumn{5}{l}{With Zn, Ni, Co or Pb substitutions:}\\ [0.3em]
84$^f$    & (0.13) &  45          &    12.4 & \citet{Pan-2000a}\\
85$^g$    & (0.19) &  20          &     5.5 & \citet{Hudson-2003}\\
\;--\;$^h$&  \ddc  &  50$^i$      &    \ddc & \citet{Zhao-2000}\\
\;--\;$^j$&  \ddc  &  50          &    \ddc & \citet{Cren-2001}\\
68$^k$    & (0.22) & $40\pm20$    &    13.7 & \citet{Kinoda-2003}\\
}{
\notegt{a}{Thin film}
\notegt{b}{Bi$_2$Sr$_{1.98}$Ca$_{1.38}$Cu$_{2.28}$O$_{8+\delta}$, $T_c^{\text{onset}}=93$~K by dc-resistivity and 55~K by ac-susceptibility}
\notegt{c}{Bi$_{2.1}$Sr$_{1.9}$CaCu$_2$O$_{8+\delta}$}
\notegt{d}{Tunneling direction in $ab$-plane varies between [110] and [100]}
\notegt{e}{$I\parallel[110]$}
\notegt{f}{Bi$_2$Sr$_2$Ca(Cu$_{1-x}$Zn$_x$)$_2$O$_{8+\delta}$}
\notegt{g}{Bi$_2$Sr$_2$Ca(Cu$_{1-x}$Ni$_x$)$_2$O$_{8+\delta}$}
\notegt{h}{Bi$_{1.83}$Sr$_{1.8}$Ca(Cu$_{1-x}$Co$_x$)$_2$O$_{8+\delta}$}
\notegt{i}{Measured at $T=66$~K}
\notegt{j}{Bi$_{1.95}$Pb$_{0.5}$Sr$_2$CaCu$_2$O$_{8+\delta}$}
\notegt{k}{Bi$_{1.4}$Pb$_{0.6}$Sr$_2$CaCu$_2$O$_{8+\delta}$}
}


\gaptable
{1.2em}
{tab_Bi2223}
{Gap values for Bi$_2$Sr$_2$Ca$_2$Cu$_3$O$_{10+\delta}$ (Bi2223).}
{Bi2223 (continued)}
{
109 & UD & $60\pm3$ & 12.8 & \citet{Kugler-2006}\\
111 & OP & $45\pm7$ &  9.4 & \citet{Kugler-2006}\\
}
{}


\gaptable
{1.2em}
{tab_HBCCO}
{Gap values for HgBa$_2$Ca$_{n-1}$Cu$_n$O$_{2n+2+\delta}$.}
{HBCCO (continued)}
{
\\ [-0.8em] \multicolumn{5}{l}{Hg1201~$(n=1)$}\\ [0.2em]
97$^a$  &  OP  &      33    &  7.9 & \citet{Wei-1998}\\
\hline\\ [-0.3cm] \multicolumn{5}{l}{Hg1212~$(n=2)$}\\ [0.2em]
123$^b$ &  OP  &      50    &  9.4 & \citet{Wei-1998}\\
\hline\\ [-0.3cm] \multicolumn{5}{l}{Hg1223~$(n=3)$}\\ [0.2em]
125$^c$ & (OP) & $\sim24^d$ &  4.5 & \citet{Rossel-1994}\\
133$^a$ & (OP) & $\sim38^e$ &  6.6 & \citet{Rossel-1994}\\
135$^a$ &  OP  &      75    & 12.7 & \citet{Wei-1998}\\
}
{
\notegt{a}{Polycrystal}
\notegt{b}{Thin film}
\notegt{c}{Melted sample}
\notegt{d}{$\Delta=15$~meV using Dynes fit}
\notegt{e}{$\Delta=24\pm2$~meV using Dynes fit}
}


\gaptable
{0.5em}
{tab_La214}
{
Gap values for La$_{2-x}$Sr$_x$CuO$_4$ (La214). The doping level was determined
directly by the Sr stoichiometry ($p=x$).
}
{La214 (continued)}
{
   29.5  & 0.10 & 7.5     &  5.9 & \citet{Tanaka-1995,Tanaka-1996a}\\
$\sim40$ & 0.14 & 16      &  9.3 & \citet{Nakano-1998}\\
$\sim35$ & 0.14 & 17      & 11.3 & \citet{Oda-2000}\\
   33    & 0.15 & 7       &  4.9 & \citet{Kirtley-1987}\\
   35.5  & 0.15 & 5.5     &  3.6 & \citet{Tanaka-1995,Tanaka-1996a}\\
   39    & 0.16 & $8\pm2$ &  4.8 & \citet{Kato-2003}\\
$\sim40$ & 0.18 & 10      &  5.8 & \citet{Nakano-1998}\\
$\sim35$ & 0.18 & 9       &  5   & \citet{Oda-2000}\\
}
{}


\gaptable
{0.3em}
{tab_Y123}
{Gap values for YBa$_2$Cu$_3$O$_{7-\delta}$ (Y123).}
{Y123 (continued)}
{
60        &  UD  & $\sim20$     & 7.7 & \citet{Yeh-2001}\\
85$^a$    &  UD  & 19           & 5.2 & \citet{Yeh-2001}\\
89        &  UD  & 20           & 5.2 & \citet{Born-2002}\\
82$^b$    &  OP  & 18           & 5.1 & \citet{Yeh-2001}\\
85$^c$    & (OP) & $18$         & 4.9 & \citet{Kirtley-1987}\\
85$^c$    & (OP) & $15^d$       & 4.1 & \citet{Kirtley-1987}\\
90$^a$    &  ?   & $14\pm2^e$   & 3.6 & \citet{Hoevers-1988}\\
90        &  OP  & 28$^f$       & 7.2 & \citet{Tanaka-1994}\\
90$^a$    &  OP  & $\sim20$     & 5.2 & \citet{Ueno-2001,Ueno-2003}\\
90        &  OP  & $28$         & 7.2 & \citet{Wan-1988}\\
$>90^a$   & (OP) & $\sim17.5^e$ & 4.5 & \citet{Miller-1993}\\
91        & (OP) & $24$--$32$   & 7.1 & \citet{Murakami-2001}\\
91        & (OP) & $\sim25^g$   & 6.4 & \citet{Murakami-2001}\\
92$^a$    &  OP  & $20^h$       & 5.0 & \citet{Alff-1997}\\
92$^a$    &  OP  & 20           & 5.0 & \citet{Born-2002}\\
92        &  OP  & $30\pm8$     & 7.6 & \citet{Edwards-1992}\\
92        &  OP  & $20$         & 5.0 & \citet{Edwards-1995}\\
92$^a$    &  OP  & $17.5^e$     & 4.4 & \citet{Koinuma-1993}\\
92$^a$    &  OP  & $\sim21^h$   & 5.3 & \citet{Koyanagi-1995}\\
92        &  OP  & $28\pm2$     & 7.1 & \citet{Kugler-2000a}\\
92        &  OP  & $20\pm2$     & 5.0 & \citet{Maggio-Aprile-1995}\\
92        &  OP  & $18$         & 4.5 & \citet{Maggio-Aprile-2000}\\
92        & (OP) & 18           & 4.5 & \citet{Maki-2001}\\
92$^a$    &  OP  & $20^e$       & 5.0 & \citet{Nantoh-1994a}\\
92$^a$    &  OP  & $30\pm10^h$  & 7.6 & \citet{Nantoh-1995}\\
92        &  OP  & $19\pm4$     & 4.8 & \citet{Wei-1998b}\\
92        &  OP  & $29\pm3^e$   & 7.3 & \citet{Wei-1998b}\\
92        &  OP  & $27\pm4^h$   & 6.8 & \citet{Wei-1998b}\\
92.9      &  OP  & 18           & 4.5 & \citet{Yeh-2001}\\
90        &  OD  & 20           & 5.2 & \citet{Shibata-2003a,Shibata-2003b}\\
78$^{ai}$ &  OD  & 19           & 5.7 & \citet{Yeh-2001}\\
}
{
\notegt{a}{Thin film}
\notegt{b}{YBa$_{2}$Cu$_{0.9934}$Zn$_{0.0026}$Mg$_{0.004}$$_{3}$O$_{6.9}$}
\notegt{c}{Measured by high-field magnetic susceptibility, $T_c=92$~K by resistivity}
\notegt{d}{$I\parallel ab$-plane}
\notegt{e}{Tunneling on (100) plane}
\notegt{f}{Tunneling on (110) and (100) planes}
\notegt{g}{Tunneling on electrical field etched surface}
\notegt{h}{Tunneling on (110) plane}
\notegt{i}{Y$_{0.7}$Ca$_{0.3}$Ba$_{2}$Cu$_{3}$O$_{7-\delta}$}
}


\gaptable
{0.9em}
{tab_NBCO}
{
Gap values for NdBa$_2$Cu$_3$O$_{7-\delta}$ (Nd123).}
{Nd123 (continued)}
{
95$^a$  & OP & 30     & 7.3 & \citet{Nishiyama-2002}\\
}
{
\notegt{a}{Measured by SQUID}
}


\gaptable
{0.7em}
{tab_NCCO}
{Gap values for Nd$_{2-x}$Ce$_x$CuO$_{4-\delta}$, an electron-doped cuprate.}
{NCCO (continued)}
{
18   & (OP) & $4.5^a$     & 5.8 & \citet{Kashiwaya-1997}\\
17.5 &  OP  & $4\pm1.5$   & 5.3 & \citet{Kashiwaya-1998}\\
18   & (OP) & $>5$        & 6.4 & \citet{Kashiwaya-1997}\\
20.5 &  OP  & $\sim4.5^b$ & 5.1 & \citet{Hayashi-1998b}\\
}
{
\notegt{a}{Tunneling on (100) plane}
\notegt{b}{$I\parallel[110]$ and [100]}
}

\end{document}